\documentclass{ar-1col}

\usepackage[comma]{natbib}
\usepackage{graphicx}	
\usepackage{amsmath}	
\usepackage{amssymb}	
\usepackage{url}
\usepackage{verbatim}
\usepackage{color}
\usepackage{xcolor}
\usepackage{subcaption}

\newcommand{\apjs}{ApJS}
\newcommand{\apjl}{ApJL}

\newcommand{\aap}{A\&A}

\def\gsim{\;\rlap{\lower 2.5pt
 \hbox{$\sim$}}\raise 1.5pt\hbox{$>$}\;}
\def\lsim{\;\rlap{\lower 2.5pt
   \hbox{$\sim$}}\raise 1.5pt\hbox{$<$}\;}
\newcommand{\Da}{Damk{\"o}hler }

\newcommand{\Msun}{${\rm M}_{\odot}$}

\newcommand{\mhalo}{M_{\rm h}}

\newcommand{\rcloudlet}{$\ell_{\rm cloudlet} \,$}

\jname{Xxxx. Xxx. Xxx. Xxx.}
\jvol{AA}
\jyear{YYYY}
\doi{10.1146/((please add article doi))}

\begin{document}

\markboth{Faucher-Gigu\`ere \& Oh}{Physical Processes in the CGM}

\title{Key Physical Processes in the Circumgalactic Medium}

\author{Claude-Andr\'e Faucher-Gigu\`ere$^1$ \\ S. Peng Oh$^2$
\affil{$^1$CIERA and Department of Physics and Astronomy, Northwestern University, 1800 Sherman Ave, Evanston, IL 60201, USA; email: cgiguere@northwestern.edu}
\affil{$^2$Department of Physics, University of California, Santa Barbara, CA 93106, USA; email: peng@physics.ucsb.edu}}

\begin{abstract}
Spurred by rich, multi-wavelength observations and enabled by new simulations, ranging from cosmological to sub-pc scales, the last decade has seen major theoretical progress in our understanding of the circumgalactic medium. We review key physical processes in the CGM. Our conclusions include:

\vspace{0.1in}
\begin{minipage}[l]{0.75\textwidth}
\begin{itemize}
\item[{\scriptsize$\blacksquare$}] The properties of the CGM depend on a competition between gravity-driven infall and gas cooling. When cooling is slow relative to free fall, the gas is hot (roughly virial temperature) whereas the gas is cold ($T \sim 10^4$ K) when cooling is rapid.
\item[{\scriptsize$\blacksquare$}] Gas inflows and outflows play crucial roles, as does the cosmological environment. Large-scale structure collimates cold streams and provides angular momentum. Satellite galaxies contribute to the CGM through winds and gas stripping.
\item[{\scriptsize$\blacksquare$}] In multiphase gas, the hot and cold phases continuously exchange mass, energy and momentum. The interaction between turbulent mixing and radiative cooling is critical. A broad spectrum of cold gas structures, going down to sub-pc scales, arises from fragmentation, coagulation, and condensation onto gas clouds.
\item[{\scriptsize$\blacksquare$}] Magnetic fields, thermal conduction and cosmic rays can substantially modify how the cold and hot phases interact, although microphysical uncertainties are presently large.

\end{itemize}
\end{minipage}
\vspace{0.1in}

Key open questions for future work include the mutual interplay between small-scale structure and large-scale dynamics, and how the CGM affects the evolution of galaxies.
\end{abstract}

\maketitle

\tableofcontents

\section{INTRODUCTION}
Observations indicate that galaxies from dwarfs to massive ellipticals are enclosed by massive gaseous atmospheres, known as the circumgalactic medium (CGM). 
The total gas mass and the metal mass in the CGM can exceed the corresponding masses in galaxies \citep[e.g.][]{tumlinson17}. 
Moreover, it has become clear in recent years that the CGM crucially affects the evolution of galaxies by mediating interactions between galaxies and the larger-scale intergalactic medium (IGM). 
Gas inflows from the cosmic web are necessary to sustain star formation in galaxies over cosmological timescales, while galactic winds play a critical role in regulating star formation rates (SFRs). 
Studies of the CGM therefore constrain, or make predictions for, the mass distribution, kinematics, thermodynamics, and chemical abundances of the gas flows that regulate galaxy formation.
\begin{figure}[h]
\includegraphics[width=0.98\textwidth]{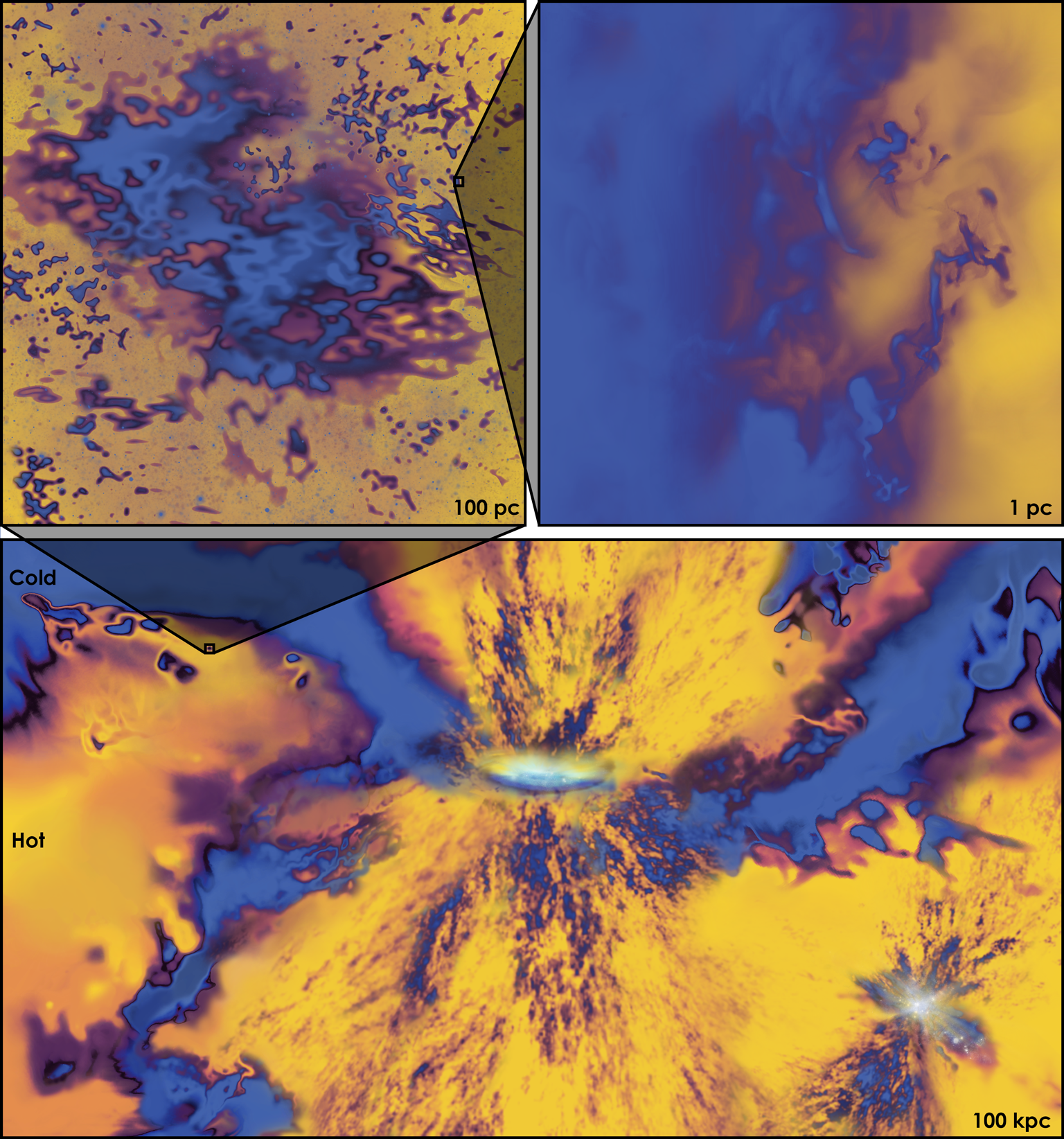}
\caption{\small Schematic illustration of important properties of the CGM. 
The CGM is a complex system involving physics on an enormous dynamic range, from the dark matter halo on scales $\gtrsim 100$ kpc to sub-parsec structure in cold gas. 
Key concepts include the interplay between gas cooling, heating, and gravity in the halo and how the hot and cold phases exchange mass, energy, and momentum. 
The bottom panel shows a central galaxy whose star formation is fueled by a mixture of cold (blue) and hot (yellow) accreting gas and which powers a multiphase galactic wind, while the top two panels zoom onto a highly structured cold cloud complex (left) and a turbulent mixing layer (right).
A complete understanding of the CGM and accurate observational predictions require consideration of various additional processes, including the effects of satellite galaxies (as in the bottom right) as well as magnetic fields, thermal conduction, cosmic rays, and feedback from accreting black holes (not shown here). 
Figure credit: Aaron M. Geller (Northwestern University/CIERA/IT Research Computing Services). \copyright 2022 Aaron M. Geller.
} 
\label{fig:multiscale_cartoon}
\end{figure}

An earlier Annual Reviews article by \cite{tumlinson17} concluded that the CGM contents are now reasonably well characterized observationally and that key questions going forward include the physics that govern the CGM and how it interacts with galaxies. 
In the last decade, there have been a number of theoretical developments directly relevant to answering these physics questions. 
On large scales, the major advances include cosmological hydrodynamic simulations that now produce broadly realistic galaxy populations and which have been used to analyze CGM gas flows on large scales \citep[for a recent review of cosmological simulations of galaxy formation, see][]{Vogelsberger2020_review}. 
On small scales, there has been similarly important progress studying processes that are not well resolved in cosmological models, including the microphysics of how cold and hot gas phases exchange mass, momentum, and energy, and the inclusion of physics beyond ideal hydrodynamics, such as magnetic fields, thermal conduction, and/or cosmic rays.

Our goal in this article is to draw on these recent advances and summarize our current understanding of the key physical processes that operate in the CGM. 
Our point of view is primarily theoretical, but our choices of topics are in many instances motivated by observations. 
An understanding of CGM physical processes is relevant to several key questions, including: How does gas flow into CGM and accrete onto galaxies? 
How do galactic winds affect the CGM? 
How does the CGM affect the formation and evolution of galaxies? 
How is the multiphase structure of the CGM, which is critical to the interpretation of many observations, produced?    
What are the important physical scales in the CGM and what are the requirements to produce realistic simulations?
We envision that our audience could range from new graduate students entering the field to more experienced CGM researchers interested in a summary of recent theoretical developments. 

By CGM, we typically refer to the gas within one virial radius $R_{\rm vir}$ of dark matter halos, but outside galaxies. 
We stress, however, that CGM processes such as galactic outflows can reach larger radii and we do not exclude such gas only because it has crossed the somewhat arbitrary virial-radius boundary. 
We focus mainly on the CGM around isolated galaxies, in halos of total mass up to a few times $10^{12}$~\Msun, corresponding to central galaxies of order the mass of the Milky Way (the $L_{\star}$ mass scale), or a few times this value. 
The most massive dark matter halos ($\mhalo \gtrsim10^{13}$~\Msun) correspond to galaxy groups and clusters of galaxies and they host 
the intra-group medium (IGrM) and the intracluster medium (ICM), respectively. 
While some physical processes are common to the CGM and the IGrM/ICM, we generally avoid discussing processes that are specific to group or cluster environments. 
There are other review articles which focus on groups and clusters  \citep[e.g.][]{Kravtsov12_ARAA_clusters,DV22_review}.

We organize our review into two main parts: one on cosmological processes (\S \ref{sec:cosmo_processes}) and one on small-scale processes (\S \ref{sec:small_scales}). 
The main theme of the section on cosmological processes are the properties and physics of gas flows on the scale of the halo, which connect the IGM to galaxies. 
The main theme of the section on small-scale processes is physical processes that arise in a multiphase medium, particularly how gas flows between different phases. 
A key emphasis is on processes that drive the formation, destruction, and structure of cold gas. 
The properties of this cold gas are important because common observational techniques, including as rest-UV absorption and emission, are sensitive to the cold gas phase. 
Section \ref{sec:cosmo_resolving} combines cosmological and small scales in a discussion of the requirements for resolving cold gas and of the prospects for modeling cold gas in cosmological simulations. 
We summarize our outlook and outline key areas for future research in \S \ref{sec:outlook}.

The artist's conception of the CGM in Figure \ref{fig:multiscale_cartoon} previews some key themes. 
The figure illustrates a huge dynamic range and the importance of hot and cold phases. 
This illustration contains several of the same concepts as \cite{tumlinson17}'s Figure 1, which has been widely used to summarize key CGM processes, including filamentary accretion and outflows. 
We have produced a new cartoon picture to emphasize some of the complexity expected in the CGM, including messy structure on scales ranging from the halo ($\gtrsim100$ kpc) to turbulent mixing layers ($<1$ pc) and interactions with companion  galaxies. 
These are aspects which have seen significant theoretical progress in recent years.

We refer to other recent reviews for complementary information on the CGM. 
\cite{tumlinson17} provide an excellent overview, but with more emphasis on observational properties. 
\cite{PerouxHowk20_ARAA} cover the baryon cycle, but also with a focus on observations. 
The recent article by \cite{DV22_review} covers both observations and theory, but focuses on the more massive halos. 
Another perspective is provided by \cite{putman12}, who review the observational properties of gaseous halos around the Milky Way and other low-redshift spiral galaxies. 
Readers interested in the physics of the IGM on larger scales can refer to the excellent review articles by \cite{Meiksin09_IGM_review} and \cite{McQuinn16_ARAA}.

Galactic winds are critical in shaping the CGM but they are a large subject by themselves, so our treatment of winds will be limited in this article. 
For more on galactic winds, readers can refer to \cite{Veilleux2020_review}. 
Also mostly beyond the scope of this review is feedback from active galactic nuclei \citep[AGN; e.g.][]{Fabian12_AGN}. 
AGN feedback may have large effects on the CGM, but most of the work on AGN feedback has focused on massive ($>L_{\star}$) halos and the results typically depend on highly uncertain black hole physics assumptions. 
This is a fascinating topic which would merit much more comprehensive discussion than we are able to fit in this article. 
Among the open questions is whether AGN feedback could have important effects on the CGM across a wider range of halos than is often assumed. 
Observations of AGN-driven outflows in dwarf galaxies \citep[e.g.,][]{MK19} and of the Fermi bubbles in the Milky Way \citep[][]{Su10_Fermi, Predehl20} suggest this is a possibility worthy of serious consideration for future research. 

\section{COSMOLOGICAL PROCESSES}
\label{sec:cosmo_processes}
\begin{figure}[h]
\includegraphics[width=0.99\textwidth]{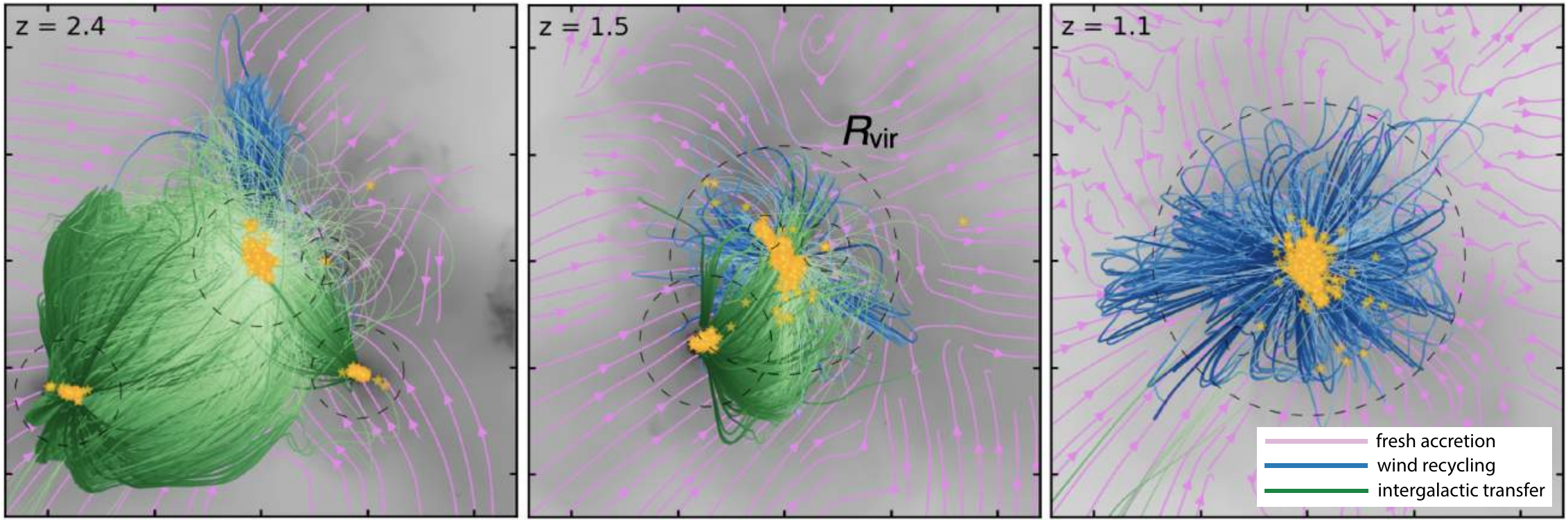}
\caption{Different types of circumgalactic gas flows in a cosmological zoom-in simulation of a main galaxy with halo mass $\mhalo \approx 1.4\times10^{11}$ \Msun~at $z=0$. 
The background gray scale shows projected gas density (logarithmically scaled) in a (240 kpc)$^3$ volume (physical), while the orange stars represent the stellar distribution in different galaxies (the virial radius of each dark matter halo is indicated by a dashed circle). 
The different panels show the system at different redshifts. 
Purple lines indicate streamlines of ``fresh accretion,'' i.e. gas that will accrete onto the main galaxy directly from the IGM by $z=0$. 
Blue lines indicate the future trajectories of gas particles ejected from the central galaxy that will accrete back as part of the ``wind recycling'' component. 
Green lines indicate the future trajectories of gas particles removed from the ISM of another galaxy that will later smoothly accrete onto the central galaxy as part of the ``intergalactic transfer'' component. 
Not shown in this visualization are galactic winds that escape galaxies and contribute to CGM mass but do not re-accrete onto the central galaxy. 
Adapted from \cite{AA17_cycle}. 
}
\label{fig:AA17_tracks}
\end{figure}

Figure \ref{fig:AA17_tracks} sets the stage with an illustration of different physical physical processes found in a cosmological simulation following the formation of a main galaxy which by $z=0$ will have a dark matter halo of mass $M_{\rm h} \approx 1.4\times10^{11}$ \Msun \citep[][]{AA17_cycle}. 
It is clear that the CGM is a highly dynamic environment intimately tied to the assembly of galaxies. 
Gas flows into dark halos from the IGM that pervades the large-scale structure of the Universe. 
Some of these inflows accrete directly onto the central galaxy, where a fraction of the baryons form stars or build up the interstellar medium (ISM). 
Of the gas accreted by galaxies, some or even most is ejected back into the CGM through galactic winds before it has time to form stars. 
These winds can later re-accrete and thus ``recycle,'' potentially a large number of times (more on this in \S \ref{sec:recycling}). 
Moreover, galaxies do not form in isolation but rather frequently have satellite galaxies. 
These satellites can strongly affect the CGM of the central galaxy, for example by losing their ISM through tidal stripping, ram pressure stripping, or galactic winds of their own (more on this in \S \ref{sec:companions}). 
The fractions of the total CGM mass originating from fresh accretion, galactic winds, and companion galaxies vary as a function of halo mass and redshift, and depend on feedback details, but can be comparable to each other especially around $\sim L_{\star}$ galaxies \citep[][]{Hafen19_origins}. 
Thus, all these processes are important to consider in general.

With the main CGM components identified, the rest of this section reviews in more detail our current knowledge of the cosmological processes that shape the CGM.

\subsection{Gas accretion}
\label{sec:gas_accretion}

\subsubsection{Cold vs. hot accretion}
\label{sec:cool_vs_hot}
Before considering the full complexity of the CGM, it is useful to examine the cooling physics of gas in dark matter halos in the idealized approximation of spherical symmetry and neglecting feedback processes.
Three timescales are important: the Hubble time $t_{\rm H}=1/H$ (where $H$ is the redshift-dependent Hubble parameter), the free-fall time in the gravitational potential $t_{\rm ff}$ and the cooling time of the gas $t_{\rm cool}$. 
As dark matter halos form from gravitational clustering, gas is dragged inward. 
In the lowest-mass halos the gas inflows remain subsonic owing to heating by photoionization by the cosmic ionizing background; in those small halos galaxy formation is suppressed \citep[e.g.,][]{Efstathiou92, Noh2014}. 
In more massive halos, inflows reach supersonic velocities and are shock-heated to a temperature of order the virial temperature, $T_{\rm vir}=(\mu m_{\rm p}/2 k) (G M_{\rm h} /R_{\rm vir})$, where $\mu$ is the mean molecular weight ($\approx 0.6$ for an ionized cosmic plasma) and $m_{\rm p}$ is the proton mass. 
For a halo of mass $M_{\rm h}=10^{12}$ M$_{\odot}$ at $z=0$ (similar to the Milky Way), the virial radius $R_{\rm vir}\approx 260$ kpc and the virial temperature $T_{\rm vir} \approx 6\times10^{5}$ K \citep[][]{barkana01}. 
Since $R_{\rm vir} \propto M_{\rm h}^{1/3}$, the virial temperature $T_{\rm vir} \propto M_{\rm h}^{2/3}$; for $\gtrsim L_{\star}$ halos this gas emits in X-rays. 
The character of gas accretion onto the central galaxy (and of the CGM) depends on whether the cooling of the shocked gas is rapid or slow relative to the free-fall time. 

{\bf Cold accretion.} When $t_{\rm cool}<t_{\rm ff}$, the shocked gas rapidly cools and loses its thermal pressure support.  
The cold $T\sim 10^{4}$ K gas that results tends to fragment and clump, and can also form narrow filaments known as ``cold flows'' or ``cold streams'' (see \S \ref{sec:cold_streams} on cold streams and more in \S \ref{sec:small_scales} about the small-scale properties of cold gas). 
If unimpeded, e.g. by feedback or angular momentum, the cold gas can accrete onto the central galaxy on a free-fall time. 
Since the infall of the cold gas is highly supersonic (relative its internal sound speed), a strong shock can form on impact with the central galaxy. 

{\bf Hot accretion.} When $t_{\rm cool}>t_{\rm ff}$, gas cooling becomes a rate-limiting step. 
Shock-heated gas can be supported for an extended period of time $\sim t_{\rm cool}$ in the halo potential by thermal pressure. 
In the inner regions, within the ``cooling radius'' where $t_{\rm cool} < t_{\rm H}$, there is sufficient time for the hot gas to cool and accrete smoothly onto the central galaxy.  
Absent feedback, these cooling regions tend to a steady-state ``cooling flow'' in which compressional heating in the inflowing gas balances radiative losses and $t_{\rm cool} \approx t_{\rm ff}$ \citep[e.g.,][]{Fabian84}, though in practice feedback processes can modify the flow. 

The different limits corresponding to different regimes of $t_{\rm cool}/t_{\rm ff}$ are core ingredients of theories of galaxy formation, starting from influential analytic models from the 1970s \citep[][]{Binney77, Silk77, RO77, WR78}. 
The implications of these limits for galaxy formation as well as the CGM have been the subject of extensive investigation since, using analytic and semi-analytic techniques \citep[e.g.,][]{WF91, SP99, DB06}, idealized numerical simulations \citep[e.g.,][]{birnboim03, Fielding17, Stern20_CF2}, and detailed cosmological simulations \citep[e.g.,][]{Keres05, Keres09, FG11, vdVoort11, Nelson13}. 
Some ideas are summarized in \S \ref{sec:ICV_implications} though this is still an active area of research and (perhaps surprisingly) there is not yet agreement on the effects of cold vs. hot accretion for galaxy formation and evolution.

In the above sketch, we have deliberately been ambiguous about where the cooling and free-fall times are evaluated. 
Modern hydrodynamic simulations as well as observations indicate that the CGM can be highly inhomogeneous and consist of multiple phases. 
Therefore, different $t_{\rm cool}/t_{\rm ff}$ limits can be realized in different regions. 
The physical picture is further complicated by outflows from stars and black holes (\S \ref{sec:outflows}), as well as additional physics such as magnetic fields, thermal conduction, and cosmic rays (\S \ref{sec:small_scales}), which imply there is in general much more to the CGM than just cooling and gravity.

\subsubsection{Maximum hot gas accretion}
\label{sec:hot_max}
To gain further insight into the different modes of gas accretion in halos, we consider some analytic results regarding the maximum rate of hot gas accretion.
Our treatment here follows \cite{Stern19_CF1} and \cite{Stern20_CF2}, who analyzed the physics of cooling flows in galaxy-scale halos. 
Although real halos can be much more complex and dynamic than idealized cooling flows, this simplified setup allows us to develop analytic insights that apply in regions where the gas dynamics is dominated by gravity and cooling.  
These results build on and extend previous on work on cooling flows in clusters of galaxies \citep[][]{MB78, Fabian84}. 
In clusters it is well known that cooling flow models fail to explain the X-ray properties of the ICM. 
The jury is still out as to whether pure cooling flow models can adequately model the CGM of some lower-mass systems, since X-ray observations can currently only barely probe the hot gas in such halos.\footnote{We note this is plausible since e.g. stellar feedback can in principle act very differently on galaxy scales than AGN feedback acts on cluster scales. Moreover, outflows appear to be relatively weak around low-redshift $\sim L_{\star}$ galaxies such as the Milky Way, so their hot CGM may be reasonably well approximated by pure cooling physics \citep[][]{Stern19_CF1}.} 
We do not take a position on this here, but simply use cooling flows as a useful baseline solution to gain insight into expected CGM properties before they are modified by feedback.

The setup is a spherically-symmetric dark matter halo in which there is initially a pressure-supported, steady flow of gas near the virial temperature. 
The energy conservation equation is $v_r d[ v_r^2/2+\gamma\epsilon+\Phi]/dr = -q$, where $r$ is the radius, $v_r$ is the radial velocity,  $\epsilon$ is the specific thermal energy, $\gamma$ is the adiabatic index of the gas, $\Phi$ is the gravitational potential, and $q$ is the cooling rate per unit mass. 
The sum in square brackets is the Bernoulli parameter, which is conserved along stream lines in a steady flow. 
To first approximation the first two terms can be neglected for slow inflow and for potentials that are not too far from isothermal (such that the specific thermal energy gradient is small), so that $d\Phi/dr \approx -q/v_{r} = -n_{\rm H}^2 \Lambda / \rho v_{r}$, where $\Lambda$ is the cooling function. 
Since the mass accretion rate $\dot{M}=-4 \pi r^{2} \rho v_{r}$ (the accretion rate is positive when the radial velocity is negative), we have the following expression in terms of the gas cooling rate and the radial gradient of the potential:
\begin{equation}
\dot{M} \approx \frac{4 \pi r^{2} n_{H}^{2} \Lambda}{d\Phi/dr}.
\end{equation}

At any radius in the halo, there is a maximum steady accretion rate of hot gas, which is set by the requirement that the density must be low enough that $t_{\rm cool} \gtrsim t_{\rm ff}$.
At higher densities, the rate of compressional heating in the cooling flow cannot balance the radiative cooling rate:  
the gas rapidly cools to $\ll T_{\rm vir}$.  
The maximum density can be evaluated using $t_{\rm ff} = \sqrt{2}r/v_{\rm c}$ (where $v_{\rm c}$ is the circular velocity in the potential) and $t_{\rm cool} = \epsilon / q = \rho \epsilon / n_{\rm H}^{2} \Lambda$:
\begin{equation}
\label{eq:nmax}
n_{\rm H,max}(r) \approx \frac{m_{\rm p} v_{\rm c} \epsilon}{X \Lambda r} \approx \frac{m_{\rm p} v_{\rm c}^{3}}{X \Lambda r} \approx 0.007~{\rm cm}^{-3}~v_{100}^{3} r_{10}^{-1} \Lambda_{-22}^{-1},
\end{equation}
where $X=0.75$ is the hydrogen mass fraction, $v_{100}=v_{\rm c}/(100$ km s$^{-1}$), $r_{10}=r/(10$ kpc), and $\Lambda_{-22}=\Lambda/(10^{-22}$ erg cm$^{3}$ s$^{-1}$). 
The second equality follows from $\epsilon \approx v_{\rm c}^{2}$, which is equivalent to the statement that the gas is at the virial temperature. 
Using $d\Phi/dr = v_{\rm c}^{2}/r$, the maximum density corresponds to a maximum hot gas accretion rate
\begin{equation}
\label{eq:Mdotmax}
\dot{M}_{\rm max}(r) \approx \frac{4\pi m_{\rm p}^{2} v_{\rm c}^{4} r}{X^{2} \Lambda} \approx 3~{\rm M}_{\odot}~{\rm yr}^{-1}~v_{100}^{4} r_{10} \Lambda_{-22}^{-1}.
\end{equation}

\subsubsection{Virialization of the inner CGM and the threshold halo mass}
\label{sec:threshold_mass}
Equations (\ref{eq:nmax}) and (\ref{eq:Mdotmax}) imply that the maximum hot gas density and accretion rate depend on radius. 
For gas in galaxy halos, the ratio of the cooling time to the free-fall time generally increases from the inside out \citep[e.g.,][]{Stern20_CF2}. 
Therefore, the outer parts can be hot and virialized ($t_{\rm cool}/t_{\rm ff}>1$),\footnote{By virialized, we mean that a virial-temperature phase is long-lived. Such gas can be sustained for longer than either a cooling time or a free-fall time if there is a continuous supply of gas, e.g. through accretion from the IGM, because in the cooling flow that develops (if feedback is neglected), compressional heating in the accreting gas balances radiative cooling.} while the inner parts cool rapidly and tend toward free fall ($t_{\rm cool}/t_{\rm ff}<1$). 
The fact that the inner CGM virializes last is important because this defines the time at which the boundary conditions of the central galaxy change. 

\begin{figure}[h]
\includegraphics[width=0.98\textwidth]{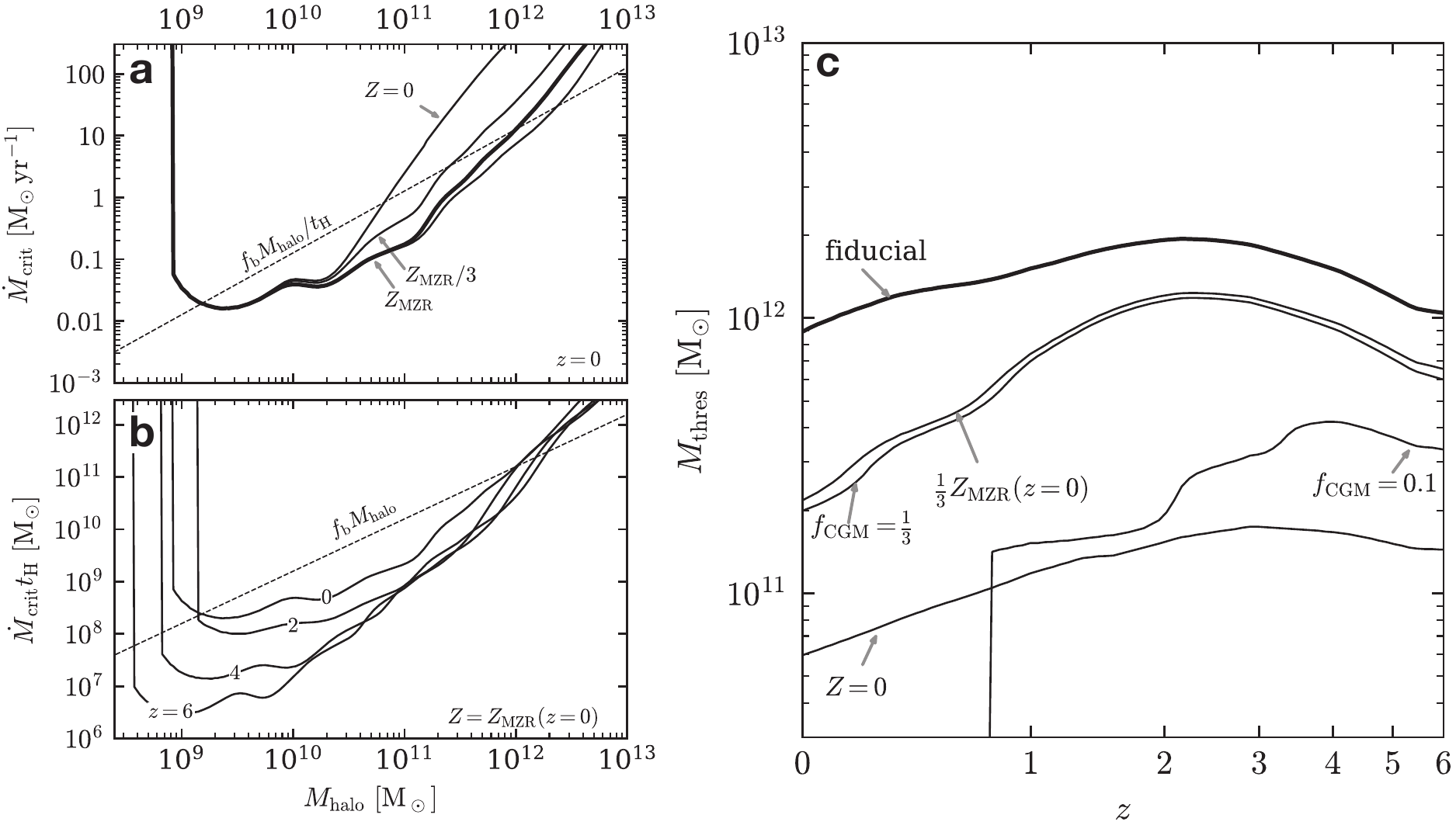}
\caption{Critical gas accretion rate for hot gas and the threshold halo mass, indicating when halos complete virialization. 
\emph{(a)} The critical accretion rate as a function of halo mass at $z=0$, for different assumed metallicities $Z$ for the gas ($Z_{\rm MZR}$ is the metallicity implied by the observed relationship between galaxy mass and ISM metallicity). 
\emph{(b)} Similar to (a), but with the critical accretion rate converted into a total gas mass by multiplying by the Hubble time. 
The curves all assume that the gas metallicity is consistent with the $z=0$ mass-metallicity relation, but show the results for different redshifts. 
In this figure, $f_{\rm b}=\Omega_{\rm b}/\Omega_{\rm m}$ is the cosmic baryon fraction and $f_{\rm CGM}$ is the fraction of the halo baryonic mass that is in CGM gas.
\emph{(c)} The threshold halo mass above which the CGM is expected to be completely hot, $M_{\rm thres}$, as a function of redshift for different assumptions. 
This corresponds to when the gas mass in the halo is equal to $\dot{M}_{\rm crit} t_{\rm H}$. 
The fiducial case corresponds to a baryon-complete CGM ($f_{\rm b}=f_{\rm CGM}=1$) on the mass-metallicity relation, and the other curves show how the threshold mass is modified when either the gas mass in the CGM or its metallicity is reduced. 
The threshold halo mass also depends on the spin of the gas (via the circularization radius), but this dependence is not shown here for simplicity. 
Adapted from Stern et al. (2020).
}
\label{fig:Mdot_crit}
\end{figure}

Cooling flow solutions also reveal an important connection between cooling and whether the flow is subsonic or supersonic. 
In the hot part of the cooling flow, where the temperature $k T_{\rm vir} \sim m_{\rm p} v_{\rm c}^{2}$, the sound speed $c_{\rm s} \sim \sqrt{P/\rho} \sim v_{\rm c}$. Thus, the free-fall time $t_{\rm ff} \sim r/c_{\rm s}$ is of order a sound crossing time. 
In this region, the inflow rate is limited by cooling, so we have $t_{\rm cool} \sim r/|v_{r}|$. 
Combining these results and defining the Mach number $\mathcal{M}=|v_{r}|/c_{\rm s}$, 
\begin{equation}
\label{eq:mach_number_rel}
\frac{t_{\rm cool}}{t_{\rm ff}} \approx \mathcal{M}^{-1}.
\end{equation}
In this expression we have omitted a prefactor $\approx1$ whose exact value  depends on the shape of the gravitational potential. 
It follows from equation (\ref{eq:mach_number_rel}) that the radius where $t_{\rm cool}/t_{\rm ff}\approx1$ coincides with the sonic radius $R_{\rm sonic}$, where $\mathcal{M}=1$. 
The flow is subsonic outside $R_{\rm sonic}$ but supersonic inside that radius. 
The transition from a subsonic to a supersonic flow has important implications for both the physics and observational properties of the CGM. 
In particular, thermal instability is inhibited in the subsonic region of a standard cooling flow whereas it can grow faster than the flow time in the supersonic region \citep[][]{balbus89}. 
In the supersonic region, large density and pressure fluctuations develop as a result of thermal instability \citep[e.g.,][]{Stern20_CF2}, which may have important implications for observational signatures as well as how galaxies interact with the CGM via inflows and outflows (see \S \ref{sec:ICV_implications}). 
We discuss thermal instability further in \S \ref{sec:formation}, where we note that if feedback keeps the hot gas close to global hydrostatic and thermal equilibrium, local thermal instability can develop if $t_{\rm cool}/t_{\rm ff} \lesssim 10$.

We now address the question of which halos are expected to be virialized. 
Since a given halo can be virialized outside its sonic radius but not inside it, the question is made more precise by considering the point at which the CGM becomes \emph{entirely} virialized, i.e. when the sonic radius becomes equal to the radius of the central galaxy. 
As stressed above, whether the CGM can sustain $t_{\rm cool}/t_{\rm ff}>1$ depends on the complexities of the baryon cycle (see Fig. \ref{fig:AA17_tracks}) and feedback in particular, as it affects $t_{\rm cool}$ by heating up the gas and by changing its density and metallicity.  
A critical halo mass can however be derived based on simplified assumptions.

The CGM can be considered to complete virialization when the accretion rate is below $\dot{M}_{\rm max}$ all the way to the circularization radius $R_{\rm circ} \approx \sqrt{2} R_{\rm vir} \lambda \approx 0.05R_{\rm vir}$, where inflowing gas becomes supported by angular momentum and which we use as a proxy for the inner boundary of the CGM (the spin parameter $\lambda$ is defined and discussed further in \S \ref{sec:accretion_angular momentum}).
For any given halo, this defines a critical gas accretion rate $\dot{M}_{\rm crit}$ equal to $\dot{M}_{\rm max}(R_{\rm circ})$. 
Approximating the cooling function as $\Lambda \propto T^{-0.7} Z^{0.9}$, valid for $T\sim 10^{5}-10^{7}$ K and metallicities $Z \gtrsim 0.3 Z_{\odot}$ \citep[][]{Wiersma09}, \cite{Stern20_CF2} obtained
\begin{equation}
\label{eq:Mdotcrit}
\dot{M}_{\rm crit} \approx 0.7~{\rm M}_{\odot}~{\rm yr}^{-1} \, v_{100}^{5.4} R_{10} Z_{0.3}^{-0.9},
\end{equation}
where $v_{100}=v_{\rm c}/({\rm 100~km~s^{-1}})$ is the circular velocity, $R_{10}=R_{\rm circ}/(10~{\rm kpc})$, and $Z_{0.3}=Z/(0.3Z_{\odot})$. The circular velocity and gas metallicity are evaluated at $R_{\rm circ}$. 
The value of $\dot{M}_{\rm crit}$ as a function of halo mass is plotted in the top left panel of Figure \ref{fig:Mdot_crit} for different metallicities at $z=0$. 
The bottom left panel of Figure \ref{fig:Mdot_crit} shows $\dot{M}_{\rm crit} t_{\rm H}$ as a function of halo mass for different redshifts, assuming a mass-dependent metallicity consistent with the observed mass-metallicity relation for galaxies.

The critical accretion rate can be translated into a threshold halo mass $M_{\rm thres}$ by setting the total gas mass in the halo $M_{\rm gas}=f_{\rm CGM} f_{\rm b} M_{\rm h} $ (where $f_{\rm b}=\Omega_{\rm b}/\Omega_{\rm m}\approx0.16$ is the cosmic baryon budget and $f_{\rm CGM}$ is the fraction of this budget in CGM gas) to $\dot{M}_{\rm crit} t_{\rm H}$. 
The idea is that $\dot{M}_{\rm crit} t_{\rm H}$ is an estimate of the hot gas mass in the halo when virialization completes, so the CGM will be fully virialized only for $M_{\rm h} \geq M_{\rm thres}$.  
The right panel in Figure \ref{fig:Mdot_crit} shows $M_{\rm thres}$ as a function of redshift. 
The solid curve in this panel shows the result for a baryon-complete CGM ($f_{\rm CGM}=1$) and other fiducial assumptions. 
Interestingly, $M_{\rm thres}$ is roughly independent of redshift, staying in the range $\approx(1-2)\times10^{12}$ M$_{\odot}$ from $z=0$ to $z=6$. 
To see why, note that at fixed metallicity, equation (\ref{eq:Mdotcrit}) implies $\dot{M}_{\rm crit} \propto v_{\rm c}^{5.4} R_{\rm circ}$. 
For a matter-dominated universe, at fixed $M_{\rm h}$, $R_{\rm circ} \propto R_{\rm vir} \propto 1/(1+z)$, $v_{\rm c} \sim v_{\rm vir} = \sqrt{G M_{\rm h}/R_{\rm vir}} \propto (1+z)^{1/2}$ ($v_{\rm vir}$ is the virial velocity), and $t_{\rm H} \propto (1+z)^{-3/2}$. 
Therefore, $\dot{M}_{\rm crit} t_{\rm H} \propto (1+z)^{0.2}$, which depends weakly on redshift.\footnote{The bottom left panel of Figure \ref{fig:Mdot_crit} shows that $\dot{M}_{\rm crit} t_{\rm H}$ depends more strongly on redshift for low-mass halos. This is because the weak redshift scaling depends on the $\Lambda \propto T^{-0.7}$ temperature scaling, which is only a valid approximation for $T\sim 10^{5}-10^{7}$ K, where metal lines dominate the cooling rate. For lower mass halos, cooling by H and He is important and $\Lambda$ has a different temperature scaling.}

We note that halos can be virialized substantially below the fiducial threshold mass $M_{\rm h} \approx 10^{12}$ \Msun, for example if the gas metallicity is lower than assumed or if the CGM density is below that implied by the cosmic baryon budget. 
Strong stellar feedback may indeed deplete the CGM by large factors in low-mass halos \citep[e.g.][]{Hafen19_origins}. 
In these limits, the entire CGM can potentially be virialized in halos of mass as low as $\sim 10^{11}$ M$_{\odot}$, or even less.

\begin{figure}[h]
\includegraphics[width=\textwidth]{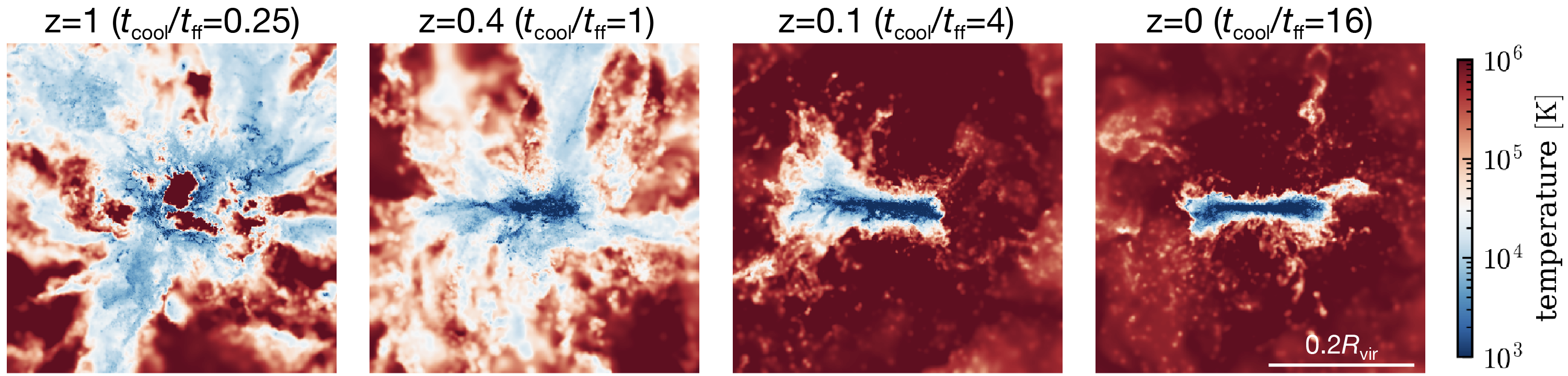}
\caption{\small Time sequence, from $z=1$ to $z=0$, showing how the CGM completes virialization in a cosmological zoom-in simulation with realistic stellar feedback of a dark matter halo of final mass $M_{\rm h}\sim 10^{12}$ M$_{\odot}$. 
Each panel shows gas temperature for a slice through the halo center and is labeled by the ratio of the cooling time of the hot, shocked gas to the free-fall time evaluated at 10\% of the virial radius. 
The virialization of the CGM completes when $t_{\rm cool}/t_{\rm ff} \gtrsim 1$ in the inner regions, but hot gas is present at larger radii earlier. 
Before the inner CGM virializes, the central galaxy is surrounded by a highly inhomogeneous mixture of cold and hot gas. 
Following virialization of the inner CGM, the halo is filled with hot, relatively uniform gas extending to the boundary of the central disk galaxy. 
Adapted from Stern et al. (2021).}
\label{fig:ICV_FIRE}
\end{figure}
The threshold mass derived above based on cooling-flow arguments is similar to threshold mass previously derived based on the stability of virial shocks \citep[][see Fig. \ref{fig:db06_nelson16}]{birnboim03, DB06}. 
In these derivations, one considers cool gas accreting supersonically into halos and shocking as the central galaxy is approached in the inner regions. 
The shock is considered stable when the cooling time of the shocked gas is sufficiently long for its thermal pressure to drive outward expansion of the accretion shock. 
The threshold halo mass derived in this way roughly matches the one derived based on cooling flows because both follow from a comparison of similar cooling and flow timescales. 
The cooling-flow derivation has the advantage of highlighting the fact that the inner parts of the CGM ``stay hot and virialized'' last, which is opposite to the inside-out direction in which accretion shocks propagate. 
The key reason for this difference is that, once hot gas is created, whether it will stay hot or rapidly cool in a given region of the CGM is a function of the local $t_{\rm cool}/t_{\rm ff}$ ratio, regardless of the directionality the shock that originally heated the gas. 
On average this ratio increases from the inside out.

We stress that the cooling flow and virial shock stability treatments are two idealized models for gas virialization in halos. 
The two models provide complementary insights, but we do not expect either to perfectly describe the dynamics of the real CGM, which are more complex due to time-variable inflows and outflows, as well as strong departures from spherical symmetry. 
In cosmological simulations including realistic feedback, such as the FIRE zoom-in simulation of a Milky Way-mass halo shown in Figure \ref{fig:ICV_FIRE}, it is found that the CGM is often first heated out to large radii by shocks due to star formation-driven galactic winds, before the theory predicts that pure accretion-driven shocks should be stable.\footnote{Note that there is evidence that star formation-driven outflows have typical velocities $\sim v_{\rm c}$ (see \ref{sec:primary_outflows}), so it can be difficult to distinguish gas that has been shocked-heated by gravitational vs. feedback processes, especially after mixing.} 
As a result, the outer parts of low-mass halos can be hot well before cooling times in the inner CGM become long enough to sustain a virialized CGM throughout the halo. 
Although this fact has seldom been emphasized in the literature so far, other simulations also find that the outer CGM is typically heated to $\sim T_{\rm vir}$ before the inner CGM is able to virialize (for example, this is apparent in temperature profiles of halos from the EAGLE simulations analyzed in Correa et al. 2018 and Wijers et al. 2020).\nocite{Correa2018, Wijers2020}

While it is beyond the scope of this review to discuss detailed observational predictions, we note here some possible observational implications of outside-in virialization: (i) The cool inner CGM should give rise to a high incidence of strong low-ionization absorbers, such as Mg II, at small impact parameters. (ii) Since most sight lines to background quasars intersect the outer CGM rather than the inner GM (due to area weighting), the presence of hot gas at large radii may contribute to the prevalence of multiphase gas inferred in observations across a wide range of halo mass. 
These expectations appear broadly consistent with the existing quasar absorption line data \citep[e.g.,][]{tumlinson17}.
Outside-in virialization would also be good news for observations that aim to detect low surface brightness rest-UV emission from the CGM, since emission is most sensitive to the inner CGM (luminosity $L \propto n^{2}$) and these wavelengths probe cool gas \citep[][]{Morrissey2018}. 
These observations can potentially test models of CGM virialization, but the observational signatures in emission have yet to be worked out quantitatively.

\subsubsection{Cold streams and where we expect them}
\label{sec:cold_streams}
\begin{figure}[h]
\includegraphics[width=0.98\textwidth]{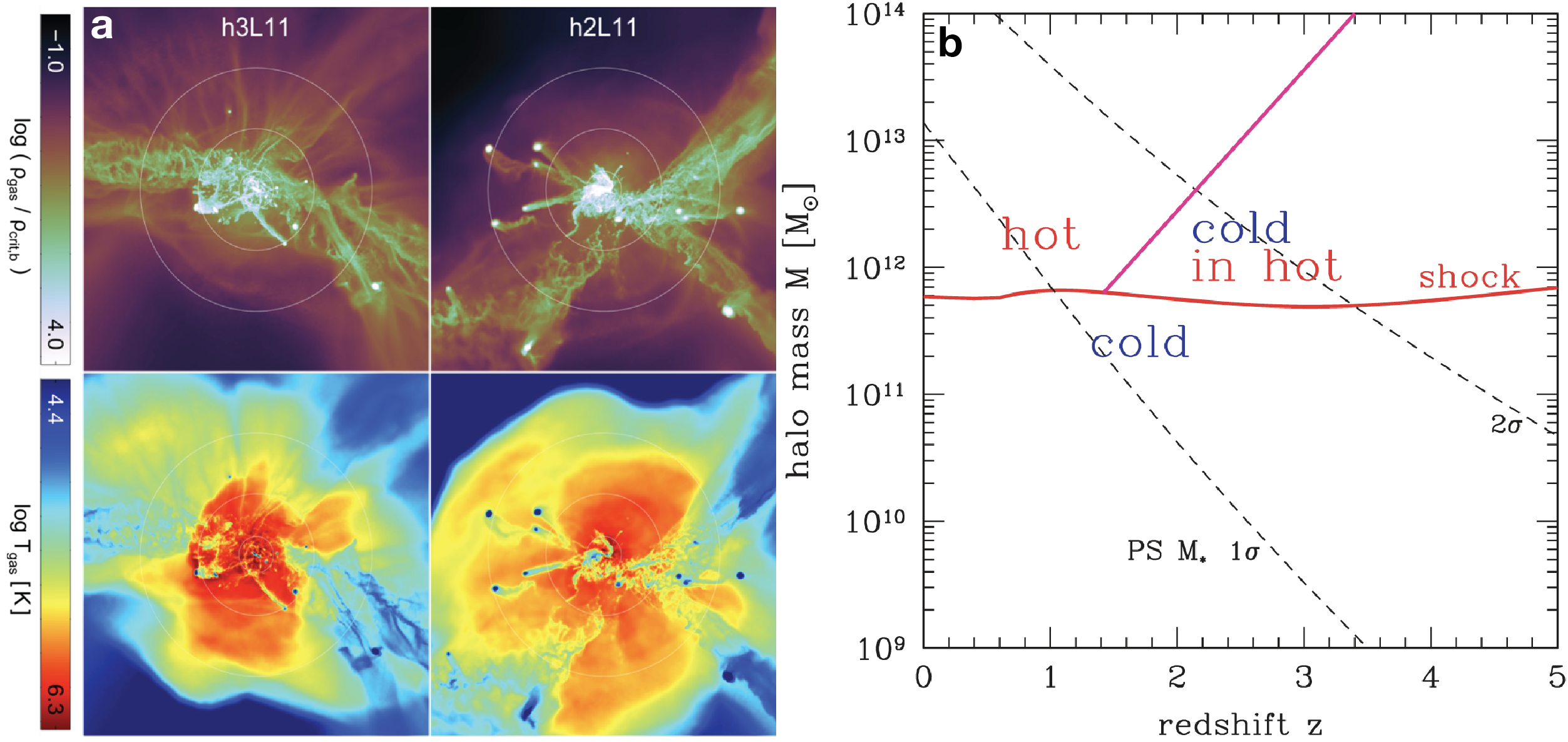}
\caption{\small \emph{(a)} Renderings of gas density and temperature in halos of total mass $\approx 10^{12}$ M$_{\odot}$ at $z=2$ in cosmological moving-mesh, zoom-in simulations. 
These simulations neglect galactic winds, so the CGM structure is primarily set by the physics of gas accretion. 
In this regime, a hot virialized CGM fills most of the volume out to beyond the virial radius (outermost circles) but dense, clumpy cold streams penetrate deep into the inner halo. 
Adapted from Nelson et al. (2016).
\emph{(b)} Analytic theory for the threshold halo mass below which accretion onto galaxies is cold and above which it is hot, assuming a post-shock gas metallicity of $0.1Z_{\odot}$ (solid red). 
This is similar to the $M_{\rm thres}$ mass shown on the right in Figure \ref{fig:Mdot_crit}, but is derived from a virial shock stability argument rather than cooling flow physics. 
The threshold halo mass is nearly constant with redshift. 
The inclined purple line shows a model for the maximum halo mass $M_{\rm stream}$ below which cold streams can persist in a hot CGM. 
The black dashed curves show the characteristic mass of newly-forming halos vs. redshift, corresponding to $1\sigma$ and $2\sigma$ fluctuations in Press-Schechter theory. From Dekel \& Birnboim (2006).}
\label{fig:db06_nelson16}
\end{figure}

Because the inflow is subsonic in virialized gas, the sound crossing time $t_{\rm s}=r/c_{\rm s}$ 
is short enough for pressure waves to smooth out density fluctuations. 
As a result, accretion of hot gas tends to proceed quasi-spherically.  
On the other hand, cold gas can clump into much finer-scale structures, as we discuss extensively in \S \ref{sec:small_scales}. 
Cosmological simulations predict that cold gas inflows often form filamentary structures known cold streams or cold flows \citep[e.g.,][]{Keres05, Keres09, dekel09, vdVoort11}. 
The panels on the left of Figure \ref{fig:db06_nelson16} shows examples of such cold filaments in $M_{\rm h} \approx 10^{12}$ M$_{\odot}$ halos at $z=2$ in cosmological zoom-in simulations evolved with the moving-mesh code Arepo \citep[][]{Nelson16_zoom}. 
The simulations shown in the figure neglect galactic winds and do not include cooling by metal lines, so these halos are significantly above the threshold mass for CGM virialization, which is lower for metal-free gas. 
In this regime, the narrow cold filaments are seen to co-exist in the CGM with a volume-filling hot phase. 
Cold streams have been the subject of much attention because in some regimes they could be a primary mode of gas accretion for galaxies \citep[e.g.,][]{dekel09}, although whether and when this is the case remains unclear as it depends on whether the cold gas survives all the way to the central galaxy during infall through the CGM, as well as the efficiency with which hot gas is accreted. 

How can cold streams exist above $M_{\rm thres}\sim10^{12}$ M$_{\odot}$? 
\cite{DB06} proposed an explanation in terms of the geometry of the large-scale structure, which also provides insight into why cold streams in massive halos appear to be a high-redshift phenomenon  \citep[e.g.,][]{Keres05}. 
The idea is that halos of different masses are, on average, located in different regions of the cosmic web. 
While low-mass halos tend to be embedded in large-scale filaments whose cross sections are larger than halo radii, high-mass halos tend to reside at the nodes where large-scale structure filaments meet. 
Therefore, while the environment of low-mass halos is roughly isotropic on the scale of the virial radius, high-mass halos are fed by collimated structures. 
What constitutes a `high' vs. a `low' halo mass in this context is determined by the non-linear clustering scale, $M_{\rm nl}$, i.e. the halo mass corresponding to density peaks that become exponentially rare. 
The key point is that $M_{\rm nl}$ increases with time due to the growth of structure, so it is smaller at high redshift. 
The panel on the right of Figure \ref{fig:db06_nelson16} shows the halo mass vs. redshift corresponding to $1\sigma$ and $2\sigma$ peaks in black dashes. 
The solid red curve in this panel shows that threshold mass above which virial shocks are stable according to \cite{DB06}'s analysis (similar to the $M_{\rm thres}$ based on the cooling flow argument outlined in \S \ref{sec:threshold_mass}). 
This plot shows that halos of mass $M_{\rm thres}$ are common ($<1\sigma$) and can be considered low-mass at $z\lesssim1$ but become increasingly rare ($>1\sigma$) above this redshift. 
Thus, above $z\sim1$ halos more massive than $M_{\rm thres}$ are increasingly fed by collimated large-scale structure filaments. 
The higher densities in filaments, relative to the mean densities in halos, imply shorter cooling times. 
On the other hand, the free-fall times are set primarily by the global mass distribution in halos and are mostly unchanged. 
Their short cooling times enable gas filaments to remain cold as they fall into massive halos. 
The cooling times can be further shortened by compression of the cold streams by the volume-filling hot phase. 
The oblique purple curve on the right in Figure \ref{fig:db06_nelson16} shows a simple analytic model from \cite{DB06} for the redshift-dependent maximum halo mass for which cold streams are expected in hot halos, $M_{\rm stream}$, based on a comparison of timescales taking into account the over-densities of filaments feeding massive halos. 
In this model, $M_{\rm stream} \approx (M_{\rm thres}/f M_{\rm nl}) M_{\rm thres}$, where $f\approx3$ is dimensionless factor calibrated from numerical simulations.

Although this estimate for the maximum mass of halos expected to contain cold streams is a useful guide, it neglects a number of important questions regarding the survival of cold gas, especially as it interacts with a hot phase. 
Whether cold streams survive during infall into halos depends on processes, such as shocks and fluid mixing instabilities, that are not well resolved in cosmological simulations. 
We discuss the small-scale physics of cold gas survival in much more detail in \S \ref{sec:survival}.

Some early results on cold streams using cosmological simulations were questioned because they were obtained using traditional smoothed particle hydrodynamics (SPH) methods, which were shown to suppress fluid mixing instabilities and can lead to the artificial survival of cold gas \citep[][]{Agertz2007, Sijacki12}. 
Although the detailed properties of cold streams remain uncertain because of the relatively low resolutions in cosmological simulations, there is currently a broad consensus between different modeling methodologies that the \emph{existence} of cold streams is a robust theoretical prediction. 
Cold streams are found not only in cosmological simulations evolved with modern SPH codes, which have been improved to more accurately capture mixing, but also in simulations using adaptive mesh refinement (AMR), moving mesh, and mesh-free codes \citep[for a comparison including several of these methods, see][]{Stewart17}. 
It is also noteworthy that cold streams are found in simulations that vary by orders of magnitude in resolution, ranging from large cosmological boxes to cosmological `zoom-in' simulations focusing on individual halos \citep[e.g.,][]{Nelson13, Nelson16_zoom}. 
Nevertheless, it is important to keep in mind that cosmological simulations still fall short of capturing all the physics relevant to cold gas formation and survival, so the theory of cold streams could still evolve substantially.  
Approaches that incorporate insights from small-scale studies will play an important role going forward (see \S \ref{sec:cosmo_resolving}).

On large scales, interactions with galactic winds and with satellite galaxies can also modify the properties of cold streams. 
For example, galactic winds (including winds blown by dwarf galaxies embedded in cold streams) can ``puff up'' the cold gas distribution \citep[][]{Nelson15_fbk, FG15}. 
The increased cold gas cross section in halos due to winds and galaxy interactions (see also \S \ref{sec:companions}) has important implications for observables, such as the predicted cross section for Lyman limit absorption. 

\subsubsection{Absorption and Ly$\alpha$ emission from cold streams}
\label{sec:lya}
Cold streams are of interest as observables in the CGM owing to their relatively high densities and their temperatures $T\sim 10^{4}$ K. 
In absorption, cold streams are predicted to manifest as HI absorbers with columns in the range $N_{\rm HI}\sim 10^{16}-10^{20}$ cm$^{-2}$, corresponding to Lyman limit systems (LLSs) and partial LLSs \citep[e.g.,][]{Fumagalli11, Fumagalli14, FGK11, FG15, Hafen17}. 
Cold streams may in fact dominate the incidence of these strong absorbers at most redshifts where they are observed \citep[e.g.,][]{vdV12_high_HI} and metal-poor LLSs have been interpreted as detections of cold streams infalling from the IGM which have not yet been significantly enriched by feedback processes \cite[e.g.,][]{Ribaudo11, Fumagalli11_Science_pristine}. 

In emission, cold streams 
may be important in explaining spatially extended structures known as Ly$\alpha$ halos \citep[e.g.,][]{Steidel11_Lya_halos} or the more extreme Ly$\alpha$ blobs \citep[e.g.,][]{Steidel2000_blobs, Matsuda04, cantalupo14}. 
One possibility is that gravitational energy is released as Ly$\alpha$ ``cooling radiation'' during the infall of cold streams \citep[e.g.,][]{DL09_LAB}. 
A simple estimate shows that cooling radiation could in principle be very important. 
Let $\dot{M}_{\rm gas}$ be the gas accretion rate in the halo and $\Delta \Phi$ the difference in gravitational potential as gas falls from the IGM down to the inner halo. 
Assuming an NFW potential \citep{navarro97} with concentration $c=5$ and a gas accretion rate $\dot{M}_{\rm gas} = f_{\rm b} \dot{M}_{\rm tot}$, where $\dot{M}_{\rm tot}$ is an average total mass accretion rate following \cite{neistein2008}, the cooling luminosity $L_{\alpha}^{\rm cool} \approx f_{\alpha,\rm eff} \dot{M}_{\rm gas} |\Delta \Phi| \approx 4\times10^{43}~{\rm erg~s^{-1}}~f_{\alpha,\rm eff} M_{12}^{1.8} [(1+z)/4]^{3.5}$, where $f_{\alpha,\rm eff}$ is an efficiency factor quantifying how much of the gravitational energy is released in the Ly$\alpha$ line and $M_{12} = M_{\rm h}/({\rm 10^{12}~M_{\odot}})$ \citep[see the appendix in][]{FG10_Lya}. 
This luminosity is comparable to observed Ly$\alpha$ halos. 

However, the temperature of cold streams puts them on the exponential part of the Ly$\alpha$ emissivity function. 
Namely, the Ly$\alpha$ emissivity powered by collisions is $\epsilon_{\alpha}^{\rm coll} = C_{\alpha}(T) n_{\rm HI} n_{\rm e}$, where $C_{\alpha}$ is the collisional excitation coefficient, $n_{\rm HI}$ is the neutral hydrogen number density, and $n_{\rm e}$ is the free electron number density. 
The collisional excitation coefficient scales as $C_{\alpha} \propto T^{-1/2} \exp{(-T_{\alpha}/T)}$, where $T_{\alpha}\equiv h \nu_{\alpha}/k\approx 1.2\times10^{5}$ K and $\nu_{\alpha}$ is the Ly$\alpha$ frequency. 
This exponential dependence on temperature makes theoretical predictions for cooling radiation highly uncertain \citep[][]{FG10_Lya, Rosdahl12_Lya, Mandelker2020_Lya}. 
In simulations, the predictions are sensitive to the numerical methods used to model the hydrodynamics (because of the importance of fluid mixing instabilities and weak shocks) and radiation (because it alters the ionization structure and photoionization also heats the gas). 
The structure of turbulent mixing layers at the boundaries between cold and hot gas, discussed in \S \ref{sec:TML}, is relevant as it may be where much of the energy dissipation occurs, but these layers are not resolved in cosmological simulations.

Alternatively, extended Ly$\alpha$ emission can be powered by recombinations following ionization by stars or AGN. 
These recombinations can occur either in the ISM (HII regions) or, for ionization radiation that escapes galaxies, in the CGM. 
In the case of Ly$\alpha$ photons produced within galaxies, diffuse halos can be formed by resonant scattering with neutral hydrogen in the CGM \citep[e.g.,][]{Dijkstra06, Gronke15}. 
For reference, the Ly$\alpha$ emission powered by stellar radiation in HII regions $L_{\alpha}^{\rm SF} \approx 10^{43}~{\rm erg~s^{-1}}~f_{\alpha,\rm esc} {\rm SFR}_{10}$, where $f_{\alpha,\rm esc}$ is the fraction of Ly$\alpha$ photons that avoid destruction by dust and escape the medium and ${\rm SFR}_{10}={\rm SFR}/({\rm 10~M_{\odot}~yr^{-1}})$ \citep[e.g.,][]{leitherer99}. 
This is comparable to the Ly$\alpha$ luminosity of cooling radiation, which in part explains why it has been difficult to unambiguously identify what powers observed sources \citep[scattering can in principle be tested using polarization;][]{dijkstra2008_polarization}. 
In the case of ionizing radiation that escapes galaxies, Ly$\alpha$ photons can be produced in the CGM via fluorescence, i.e. recombination emission powered by ionizing photons absorbed in the halo \citep[][]{Cantalupo05_fluo, Kollmeier10_Lya}. 
The Ly$\alpha$ emissivity from recombinations $\epsilon^{\rm rec}_{\alpha} = f_{\alpha,\rm rec} \alpha_{\rm HI}(T) n_{\rm HII} n_{\rm e}$, where $f_{\alpha,\rm rec}$ is the average number of Ly$\alpha$ photons produced per recombination ($f_{\alpha,\rm rec} \approx 0.68$), $\alpha_{\rm HI}(T) \propto T^{-0.7}$ is the recombination coefficient, and $n_{\rm HII}$ is the ionized hydrogen number density. 
While recombinations are not as sensitive to temperature as collisional excitation, the recombination emissivity is sensitive to gas clumping (the emissivity is proportional to the clumping factor $C = \langle n^{2} \rangle/\langle n \rangle ^{2}$). 
This dependence on the clumping factor has been used to infer unexpected small-scale structure in the cold gas in the halos of some luminous Ly$\alpha$ blobs \citep[][]{cantalupo14, hennawi15}. 
This has led to the proposal that the CGM could be filled with a ``fog'' or ``mist'' of tiny but high-density cold clouds; the physics of these tiny clouds is covered in \S \ref{sec:morphology}.

Galactic winds can also power extended emission by depositing mechanical energy into the CGM, which can then be radiated away \citep[][]{Taniguchi2000_wind_LABs, Sravan16}. 
Even if the ultimate energy source for extended emission (whether it be radiation or mechanical energy from stars and/or AGN) originates from galaxies, cold streams may be important to explain Ly$\alpha$ emission on halo scales. 
This is especially the case in massive halos exceeding the threshold mass $M_{\rm thres}\sim10^{12}$~\Msun, above which the volume-filling phase is expected to be hot.
If all the halo gas were hot, most of the emission would be expected to come out in X-rays. 
Cold streams and other cold gas structures in halos, such as a possible cold `fog' (\S \ref{sec:morphology}), can scatter Ly$\alpha$ photons that escape galaxies or otherwise ensure that a significant fraction of the energy deposited into the CGM is radiated in Ly$\alpha$ rather than in higher energy bands.

\subsubsection{Effects of CGM virialization and accretion mode on galaxies}
\label{sec:ICV_implications}
Much of the interest in the CGM is rooted in the presumption that the physics of gaseous halos plays an important role in the formation of galaxies. 
In particular, there is broad but indirect observational evidence that CGM virialization is important for galaxy evolution. 
The characteristic luminosity of galaxies, $L_{\star}$ (above which the galaxy stellar mass function is exponentially suppressed), corresponds to a roughly constant halo mass $M_{\rm h}\sim 10^{12}$ M$_{\odot}$, only weakly dependent on redshift. 
This is also the halo mass scale above which the fraction of galaxies that are quiescent rises above $\sim 50\%$ \citep[e.g.,][]{Behroozi19}. 
In the last few years, observations of spatially resolved galaxy kinematics have suggested that the $L_{\star}$ mass scale is consistent with the emergence of large disk galaxies \citep[e.g.,][]{Tiley2021}. 
This mass scale, termed the ``golden mass'' by some authors \citep[e.g.,][]{Dekel2019_golden}, is similar to the halo mass at which the CGM is theoretically expected to complete virialization (see Fig. \ref{fig:Mdot_crit}). 

Despite the substantial evidence that CGM virialization correlates with major changes in galaxy properties, whether and how CGM physics affect galaxy evolution remains an active area of research, with basic questions still the subject of debate. 
We summarize below some ideas that have been proposed for how CGM processes could affect galaxy evolution for $L \sim L_{\star}$ galaxies, and which in our view deserve deeper investigation:

{\bf A quasi-isotropic, hot CGM is necessary for effective preventative feedback.} There is a broad consensus that in order to explain the observed population of ``red and dead'' galaxies at the massive end, it is not sufficient for feedback to eject gas from galaxies. 
There must also be ``preventative feedback'' which prevents halo gas from cooling and raining onto galaxies at overly high rates \citep[e.g.,][]{Bower2006, Croton2006}. 
In the most massive halos, this feedback is often assumed to come from jets powered by AGN, but wider-angle winds powered by either AGN or supernovae can play a role \citep[Type Ia supernovae can be energetically important in ellipticals with old stellar populations; e.g.][]{voit15_galaxy}. 
An idea often discussed in this context is that preventative feedback only becomes important after most of the CGM has become hot and quasi-isotropic \citep[e.g.,][]{Keres09b}. 
This is because only in this limit can feedback keep the gas hot. 
In contrast, when there are massive inflows of clumpy or filamentary cool gas, the smaller geometric cross section of the inflows strongly reduces the efficiency with which feedback couples to accreting gas.

{\bf Pressure fluctuations change at the order-of-magnitude level at inner CGM virialization.} Whether the CGM is virialized or not also changes the boundary conditions of the central galaxy. 
Both idealized simulations \citep[][]{Stern19_CF1} and cosmological simulations \citep[][]{Stern20_FIRE} show that when the inner CGM virializes, there is a change from order-of-magnitude thermal pressure fluctuations in the gas around the galaxy (prior to virialization) to a roughly uniform pressure (after virialization; see Fig. \ref{fig:ICV_FIRE}). 
Large pressure fluctuations in the inner CGM create paths of least resistance through which feedback can more easily expel gas from the galaxy. 
Thus, we may expect that large-scale galactic winds will be stronger and reach farther out before the CGM virializes. 
There is some evidence from galaxy formation simulations with resolved ISM physics that star formation-driven outflows are suppressed when the inner CGM is virialized, such as around Milky Way-like galaxies at $z\sim0$ \citep[e.g.,][]{Muratov15, Stern20_FIRE}. 
Large pressure fluctuations in the inner CGM may also make it difficult for the ISM to reach a statistical steady state, which could result in highly time-variable (or ``bursty'') star formation rates \citep[e.g.,][]{Gurvich2022}.

{\bf CGM virialization changes the buoyancy of supernova-driven outflows.} \cite{Keller2016} and \cite{Bower2017} proposed a related but different effect of CGM virialization on outflows. 
These authors suggested that supernova-inflated superbubbles are buoyant in the CGM prior to virialization, so that outflows can be ``lifted'' in the CGM by buoyancy forces, but that the bubbles would cease being buoyant once a hot CGM develops. 
These authors argued that stellar feedback would therefore become ineffective at expelling gas once the CGM virializes. 
They furthermore hypothesized that this would lead to the accumulation of gas in galaxy centers, which would allow nuclear black holes to start growing more rapidly. 
If correct, this mechanism would represent another connection between CGM virialization and AGN feedback. 
Similar phenomenology regarding accelerated black hole feeding starting around $L_{\star}$, found also in other simulations, has however been attributed by other authors to changes in star formation-driven outflows either due to confinement by gravity or to the pressure fluctuations effect mentioned above \citep[e.g.,][]{Dubois2015,Byrne2022}.

{\bf Hot accretion promotes the formation of thin disks by making angular momentum (AM) coherent.} Recently, \cite{Hafen2022} reported evidence from cosmological simulations that hot-mode accretion promotes the formation of galaxies with thin disks, such as observed in low-redshift Milky Way-like galaxies. 
The basic idea is that gas from large-scale structure enters dark matter halos with a broad distribution of specific AM (sAM). 
When the gas falls in toward the galaxy as cold clumps or filaments, spatially separated gas parcels are causally disconnected. 
In this regime, the cold gas reaches the halo center supersonically with a still-broad sAM distribution and tends to form stars in irregular or thick disk morphologies. 
On the other hand, when the gas accretes onto the central galaxy in a smooth, subsonic cooling flow, the sAM distribution becomes coherent (i.e., narrow) before accretion onto the galaxy, and stars form in a thin disk configuration.\footnote{It is likely that the net result depends not only on how the gas accretes but also on feedback, because absent feedback we would expect a thin gas disk to eventually form as a result of dissipation, even if the sAM distribution is not initially coherent (as in other astrophysical settings, e.g. protoplanetary disks that form in turbulent molecular clouds).}

The role of the gas accretion mode in determining the morphology of galaxies is an example of how there is not yet a consensus on the role of CGM physics in galaxy formation. 
While the recent work mentioned in the previous paragraph highlights the role of hot mode accretion in the formation of thin disks, a substantial body of work has instead emphasized the role of cold streams in feeding massive disks at high redshift \citep[e.g.,][]{DSC09}. 
These results are not necessarily inconsistent because the disks in the massive, high-redshift regime are highly turbulent and geometrically thick. 
More work on the role of the gas accretion mode on the formation of disk galaxies will be important, including special attention to how results vary as a function of halo mass and redshift. 

Despite the plausible causal CGM mechanisms summarized above, we must stress it has proved challenging to disentangle whether CGM changes cause changes in galaxy properties, or whether changes in CGM and galaxy properties simply correlate. 
For example, analytic arguments suggest that the mass scale of CGM virialization is similar to the mass scale where supernova-driven outflows become confined by gravity \citep[e.g.,][]{Lapiner2021, Byrne2022}, so it is possible that outflows are suppressed around the same time as the CGM virializes but that neither change drives the other. 
We conclude that more research is needed to firmly establish the roles the CGM plays in galaxy evolution.

\subsection{Angular momentum}
\label{sec:accretion_angular momentum}
We now expand on what is known about the AM content and exchange processes in the CGM. 
The motivation for this is twofold. 
First, in the standard cosmological picture, galaxies inherit their AM from gas accreted from their host halos \citep[e.g.,][]{Fall80, Dalcanton97, MMW98}. 
In this picture, the AM of halos is first acquired via gravitational torques during structure formation \citep[e.g.,][]{Peebles69, White84}, although the AM of a particular halo fluctuates substantially over time due to mergers \citep[e.g.,][]{Vitvitska02}. 
While on sufficiently large scales the baryons are expected to have AM properties similar to the dark matter, hydrodynamic forces and feedback processes experienced by the baryons during galaxy formation can potentially strongly affect both the AM content of the CGM as well as of galaxies. 
Second, observations indicate that in many systems the CGM has substantial rotation \citep[e.g.,][]{Bouche13, Ho17, HodgesKluck2016} and we would like to understand these CGM observations.

A useful basis to understand the AM properties of the CGM is to start with scalings derived from dark matter-only simulations. 
Dark matter halos can be characterized by a dimensionless spin parameter $\lambda \equiv \frac{J}{\sqrt{2} M R V }$, where $J$ is the total AM inside a sphere of radius $R$ containing mass $M$, and $V = \sqrt{GM/R}$ is the circular velocity at $R$ \citep[][]{Bullock01_spin}. 
We adopt the standard choice of setting $R$ to the virial radius. 
With these definitions, numerical simulations have shown that dark matter halos have a lognormal distribution of spin parameters, nearly independent of mass and redshift, with a median $\lambda \approx 0.035$ \citep[e.g.,][]{Bett07, ZS17}. 
Defining the sAM $j \equiv J/M$ and noting that $R_{\rm vir}\propto M_{\rm h}^{1/3}/(1+z)$ (for halos defined to have constant over-density relative to the mean matter density) implies $j\propto M_{\rm h}^{2/3}/ \sqrt{1+z}$. 
Thus, at fixed redshift $j$ increases with halo mass $\propto M_{\rm h}^{2/3}$, while at fixed halo mass $j$ increases with time $\propto 1/\sqrt{1+z}$ as redshift decreases. 
While these scalings apply to dark matter-only simulations, \cite{DeFelippis20} find that the same scalings roughly describe the CGM AM trends with halo mass and redshift in the IllustrisTNG hydrodynamic simulation, which includes feedback from galaxy formation. 
Assuming that the sAM of the CGM is comparable to that of the dark matter halo (though with significant differences discussed below), the small spin parameters $\lambda \sim 0.035$ imply that AM support is negligible in most of the CGM, becoming only important at a circularization radius $R_{\rm circ}\approx \sqrt{2}\lambda R_{\rm vir} \sim0.05$ $R_{\rm vir}$. 
The small contribution of rotation to the support of halo gas has been confirmed by a systematic analysis of different support terms in EAGLE simulations \citep{Oppenheimer2018_HSE_deviations}.

\begin{figure}[h]
\includegraphics[width=0.98\textwidth]{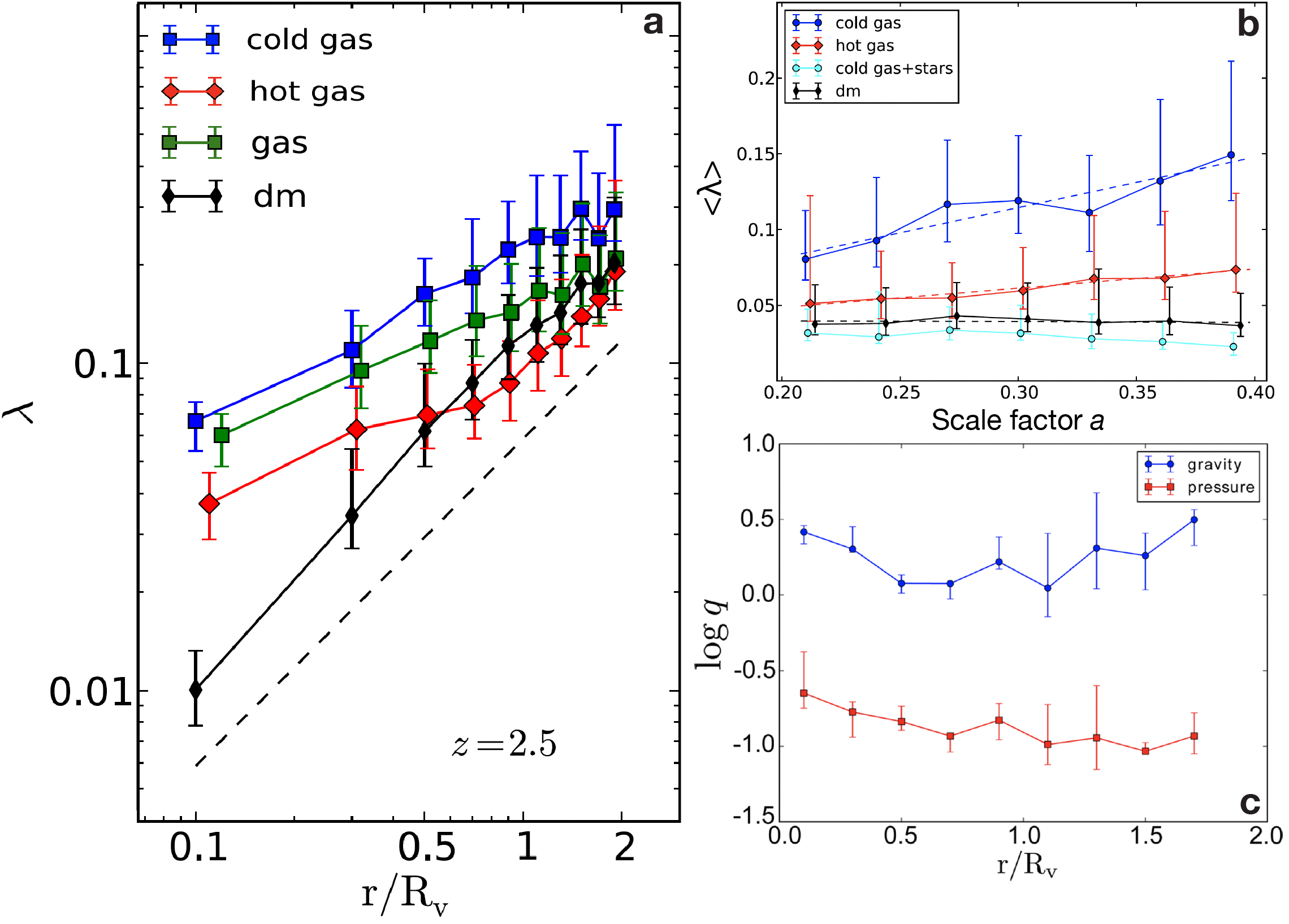}
\caption{\small \emph{(a)} Radial profiles for the spin parameter of different mass components in 29 simulated halos analyzed in Danovich et al. (2015).  
The halo mass ranges from $10^{11.5}$ to $10^{12.5}$ M$_{\odot}$ at $z\sim 2$. 
The total gas mass is divided between cold and hot components, corresponding to $T<10^{5}$ K and virial-temperature gas, respectively. 
The symbols show the mean spin in each shell, while the error bars show the standard deviation. 
The dashed line shows a slope of unity.
Several trends are apparent: (i) for all components, the mean spin parameter increases from the inside out; (ii) the spin parameter of the gas is systematically higher than that of the dark matter; and (iii) the higher spin in the gas is primarily driven by the cold gas. 
\emph{(b)} Evolution of the mean spin parameters inside the virial radius as a function of scale factor $a$. Here the cyan ``cold gas+stars'' component corresponds to $<0.1R_{\rm vir}$, a proxy for the evolution of galactic baryons. 
Averaged within the virial radius, the sAM of the cold gas exceeds that of the dark matter by a factor $\sim2-3$. 
\emph{(c)} Comparison of the radial profiles of torques on cold halo gas due to gravity vs. torques due to gas pressure, for simulation snapshots ranging from $z\sim1.6$ to $z\sim3.2$. 
The $q$ parameter is proportional to $|\pmb{\tau}|/|\pmb{l}|$, showing that gravitational torques are more important for the cold gas.   
Adapted from Danovich et al. (2015).
}
\label{fig:AM}
\end{figure}

Next, we summarize some key results concerning how the CGM AM relates to other components, including the dark matter halo and the central galaxy. We also review physical explanations for the differences found between the different components. 
We refer to Figure \ref{fig:AM} for some quantitative results on spin parameters of the gas and the dark matter around simulated galaxies with a halo mass $M_{\rm h}\sim 10^{12}$ M$_{\odot}$ at $z\sim2$. 

\nocite{Stewart17, DeFelippis20, Oppenheimer2018_HSE_deviations} 

\subsubsection{AM in the CGM vs. the dark matter halo}
Within the virial radius, CGM gas has systematically higher sAM than the dark matter. 
Interestingly, this is the case even in non-radiative simulations, so part of the difference can be attributed to hydrodynamic interactions that do not involve cooling \citep[e.g.,][]{ZS17}. 
For example, when two halos merge, ram pressure will cause the gas mass to become offset from the dark matter. 
Since the simulations also predict that the gas and dark matter spins are on average misaligned by $\sim 35^{\circ}$, mergers could on average spin up gas more than the dark matter. 

When the total CGM mass is divided between cold ($T<10^{5}$ K) and hot (virial-temperature) gas, it becomes clear that higher gas sAM relative to the dark matter is primarily driven by the higher sAM of the cold gas, which can exceed that of the dark matter by a factor $\sim2-3$ within $R_{\rm vir}$. 
This indicates an important role for gas cooling. 
\cite{Danovich15} analyzed the torques experienced by the dark matter and cold gas as they approach the virial radius of halos, and argued that the excess quadrupole moment of the cold gas relative to the dark matter could explain the additional sAM acquired by the infalling cold gas as a result of more efficient tidal torquing. 
Specifically, the elongated, thin-stream geometry of the infalling cold gas (relative to the thicker dark matter distribution) enables the cold gas to acquire AM more efficiently via tidal torques. 
In other words, cold streams are where cold gas gets extra torque.

An additional timescale effect contributes to the higher sAM of the cold gas relative to the hot gas. 
Whereas the cold gas can accrete onto the central galaxy on a free-fall time, the hot gas is supported in the halo by thermal pressure for at least a cooling time. 
Thus, while much of the cold CGM has typically only recently entered the halo, the hot CGM has been built up over a longer period in the past. 
Since the sAM of matter accreting from large-scale structure increases with time, this timescale effect alone tends to enhance the sAM of the cold gas relative to the hot gas. 
The increasing sAM of matter accreting from large scales likely explains, at least in part, why the spin parameter of the different halo components increases systematically with radius in Figure \ref{fig:AM}. 

Galaxy formation feedback can also increase the sAM of the CGM relative to the dark matter. 
Namely, the ejection of gas from galaxies by star formation or AGN-driven outflows occurs primarily from the inner parts where the baryons have relatively low sAM \citep[][]{ZS17}. 
If sufficiently strong, feedback can eject some of the low sAM gas not only from galaxies but from halos altogether. 
The preferential ejection from halos of low sAM gas also tends to enhance the sAM of the remaining CGM relative to the dark matter.

Although the detailed quantitative predictions depend on the simulation code, including the feedback model, the high sAM of the CGM relative to the dark matter (especially for the cold gas) appears robust, as similar results have been found in cosmological simulations using different codes (e.g., Stewart et al. 2017; DeFelippis 2020) and for halos in different mass ranges (e.g., Oppenheimer 2018). \nocite{Stewart17, DeFelippis20, Oppenheimer2018_HSE_deviations}
The high sAM of cold gas can produce extended rotating structures that have sometimes been called ``cold flow disks'' which may have observational signatures in low-ionization absorption systems co-rotating with central galaxies \citep[e.g.,][]{Stewart11, Stewart13}. 

\subsubsection{AM in the CGM vs. the central galaxy}
The relationship between the sAM of galaxies and that of their host halos merits some comments. 
On the one hand, observational studies \citep[e.g.,][]{Kravtsov13, Somerville18} and many numerical simulations \citep[e.g.,][]{Genel18, Rohr22} find that \emph{on average} the size of galaxies scales with the virial radius of the dark matter halo, with a normalization roughly consistent with that expected if the sAM of the galaxy is comparable to that of the dark matter halo. 
Moreover, it is found in some simulations that at fixed stellar mass, halos with larger spin parameters on average host larger galaxies \citep[][]{RG2022}. 
On the other hand, some simulations that reproduce the average trend between galaxy size and halo size indicate that on a \emph{halo-by-halo} basis, the spin parameter of the central galaxy is barely correlated with the spin parameter of the dark matter halo, when these are measured at the same final time \citep[][]{GK18_m12s, Jiang19}. 
It is also noteworthy that the AM vector of the CGM is in general misaligned with that of the stars in the central galaxy by large angles $\sim30-60^{\circ}$ \citep[][]{DeFelippis20}. 

These results could be understood if, to first order, the sAM of galaxies scales with sAM of the host dark matter halo when the galaxy is assembled but there is order-unity scatter introduced between the sAM of the baryons and of the dark matter over time. 
For example, \cite{Vitvitska02} showed that the spin parameter of a dark matter halo fluctuates by factors up to $\sim2-3$ due to halo mergers. 
Because of the very different spatial distributions of matter, we expect galaxies to be torqued differently during mergers compared to the much larger halos. 
The partial decoupling of the spin parameter of galaxies from their host halos over time is consistent with the finding of \cite{GK18_m12s} that the stellar morphology and kinematics of simulated Milky Way-mass galaxies are poorly correlated with the properties of the final dark matter halos (including spin), but that the galaxy properties correlate much better with dark matter halo properties evaluated at the time when 50\% of the stars had formed.

Processes other than mergers can also contribute to differences between the AM content of galaxies and their halos. 
One likely relevant factor is that the minority of baryons that end up in galaxies, relative to the cosmic baryon fraction, is not necessarily representative of the majority of halo baryons. 
The fraction of baryons found in galaxies peaks at $\sim0.2$ for Milky Way-mass halos and is as low as $\lesssim10^{-3}$ for dwarf galaxies and for central galaxies in massive clusters \citep[e.g.,][]{Behroozi19}. 
Another possibility is that the sAM of gas accreting through the CGM is not strictly conserved but rather experiences exchanges with other components.

\subsubsection{Gravitational vs. gas pressure torques}
We do not yet have a detailed understanding of AM transport in the CGM, but several mechanisms can contribute. 
\cite{Danovich15} decomposed the total Lagrangian torque on gas elements into three components: $\pmb{\tau}=d\pmb{l}/dt=\pmb{\tau}_{\Phi} + \pmb{\tau}_{P} + \pmb{\tau}_{s}$, where $\pmb{l}$ is the AM vector, $\pmb{\tau}_{\Phi} = -\rho \pmb{r} \times \pmb{\nabla}\Phi$ is the torque due to gravitational forces, 
$\pmb{\tau}_{P} = -\pmb{r} \times \pmb{\nabla} P$ is the torque due to pressure gradients, and $\pmb{\tau}_{s} = -\pmb{l \nabla \cdot v}$ corresponds to viscous stresses. 
The viscous stress term is negligible in the ideal hydrodynamics limit. The bottom right panel in Figure \ref{fig:AM} compares $\log{q}$ radial profiles for gravitational and pressure torques acting on cold gas for simulated halos at $z\sim1.6-3.2$ from \cite{Danovich15}, where $q_{\Phi,P} \propto |\pmb{\tau}_{\Phi,P}|/|\pmb{l}|$. 
The results indicate that the torques on cold streams are dominated by gravity rather than gas pressure. 
These gravitational torques are sourced by anisotropies in the matter distribution, ranging from large-scale structure to central disks, which tend to align the infalling gas. 
It would valuable to extend this kind of analysis to other regimes in the future. 
For example, the relative importance of gravitational torques vs. pressure torques could be very different for hot gas, which tends to be more spherical in geometry and in approximate hydrostatic equilibrium throughout the halo. 
It would also be worthwhile to quantify the effects of magnetic fields on AM transport in the CGM. 
Magnetic fields play a key role in transporting AM in accretion disks around young stars and black holes, but their effects on AM exchanges in the CGM has not yet received attention to our knowledge.

After baryons are accreted by galaxies, gravitational torques due to asymmetric features in the potential (e.g., spiral arms, bars, or massive clumps) can strongly affect the AM distribution within galaxies, such as by forming central bulges \citep[e.g.,][]{Shlosman1989}. 
This would also contribute to differences between the AM of galactic components relative to what may be expected from strict conservation of AM inherited from the halo.

Overall, AM acquisition and exchange processes in the CGM remain relatively under-studied and more work on this topic would be highly valuable.

\subsection{Galactic winds}
\label{sec:outflows}
Galactic winds are commonly observed and are an essential ingredient of modern galaxy formation theories.
In most galaxies, these outflows are understood to be primarily driven by energy and/or momentum produced by massive stars, including via supernovae \citep[e.g.,][]{Chevalier85} and/or radiation pressure \citep[e.g.,][]{Murray05}. 
In galaxies with luminous active galactic nuclei (AGN), galactic winds can also be powered by accretion onto massive black holes \citep[e.g.,][and references therein]{FGQ12}. 
Here we focus on outflows powered by star formation. 
In current models, these outflows are critical to suppress star formation in galaxies up to $\sim L_{\star}$, either by ejecting gas from the ISM before it has time to turn into stars or by preventing CGM gas from accreting onto galaxies in the first place \citep[][]{SD15_review, NO17_review}. 
However, because they originate on small scales and their driving mechanisms are not yet fully understood, the properties of galactic winds remain highly model dependent. 
Therefore, we limit our discussion below to general concepts and results which are useful to understand the impact of galactic winds on the CGM, and vice versa, rather than the detailed predictions of specific models.

\begin{figure}[h]
\includegraphics[width=1.0\textwidth]{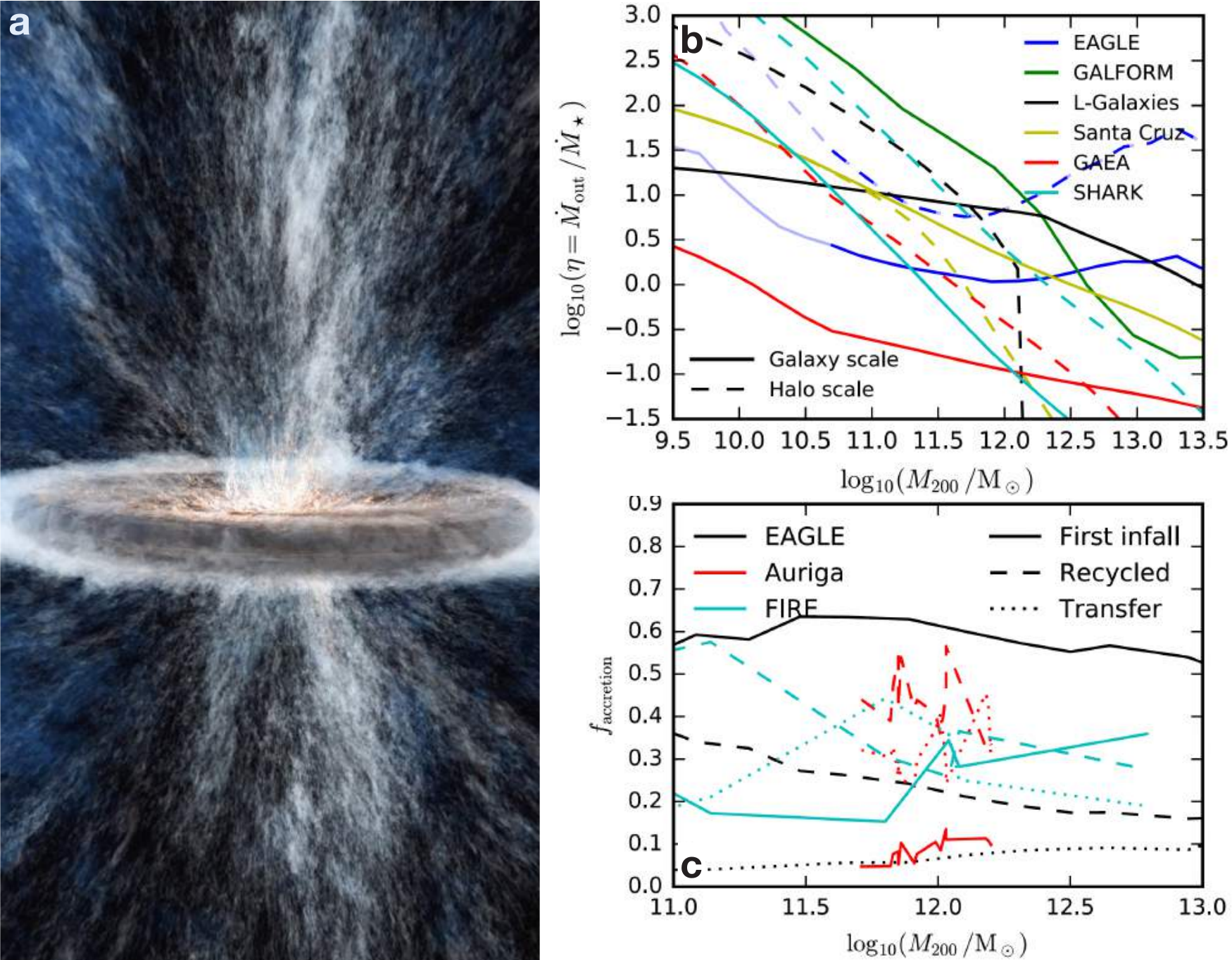}
\caption{\small \emph{(a)} Simulation of a galactic wind driven from a disk galaxy by supernovae. 
The density rendering highlights the multiphase nature of the outflow, with a low-density fluid heated by supernova explosions escaping the galaxy with a highly structured spectrum of cool clouds embedded in it. 
From Schneider et al. (2020). \copyright AAS. Reproduced with permission. 
\emph{(b)} Comparison of wind mass loading factors (defined as $\eta_{\rm M}$ in the main text) as a function of dark matter halo mass in different cosmological simulations and semi-analytic models (labeled in the legend) at $z=0$. 
The solid curves correspond to outflow rates from the galaxy, while the dashed curves correspond to outflow rates from the halo (at one virial radius). 
The outflow rate from the halo can be larger than from the ISM because of gas entrainment in the CGM. 
From Mitchell et al. (2020b). 
\emph{(c)} Fraction of the final stellar mass contributed by different accretion channels as a function of halo mass at $z=0$, for different cosmological simulations. 
The accretion channels are defined similarly as in Figure \ref{fig:AA17_tracks}, and correspond to gas that formed stars after directly accreting onto a galaxy from the IGM (``first infall''), gas that recycled in winds before turning into stars, and gas that transferred from one galaxy to another before forming stars. 
From Mitchell et al. (2020a). 
}
\label{fig:winds}
\end{figure}

Figure \ref{fig:winds} summarizes some key properties of galactic winds in simulations. 
One salient feature is that galactic winds are multiphase. 
This multiphase structure is clearly observed in the protypical example of the galactic wind driven by the M82 starburst galaxy \citep[][]{Strickland2009_M82} and is also predicted by several models. 
In the models, the multiphase structure typically consists of a hot fluid heated by supernova explosions with a spectrum of embedded cool clouds \citep[e.g.,][]{schneider20,Kim2020_outflows,fielding22}. 
However, winds accelerated by gentler processes such as radiation pressure or cosmic rays could be cooler overall \citep[e.g.,][]{Murray05,booth13}; the dominant driving mechanisms and wind properties could well vary with galaxy mass and redshift. 
Much of our discussion of the physics of multiphase gas in \S \ref{sec:small_scales}, including processes that govern cold cloud growth and survival, is relevant to galactic winds. 
We discuss a global thermal instability in winds in \S \ref{sec:wind-TI} and we discuss cloud-wind interactions extensively in \S \ref{sec:survival}. 
\nocite{schneider20, Mitchell2020_outflows, Mitchell2020_recycling}

\subsubsection{Bulk scalings}
\label{sec:primary_outflows}
Although the gas in galactic winds exhibits a range of velocities, densities, and temperatures (even in individual galaxies), there is some evidence that the mean (or median) velocity $v_{\rm w}$ scales linearly with the circular velocity of the galaxy ($v_{\rm c}$). 
This velocity scaling is predicted, for example, in FIRE simulations in which galactic winds emerge from the energy and momentum injected by multiple stellar feedback processes (including, SNe, stellar winds, and radiation pressure) on the scale of individual star-forming regions \citep[][]{Muratov15}. 
In large-volume simulations in which the generation of galactic winds is not resolved but the wind properties are instead prescribed, it is also found that a wind injection velocity proportional to $v_{\rm c}$ can produce a reasonably good fit to the observed galaxy stellar mass function \citep[e.g.,][]{Dave11, Vogelsberger14}. 
While we do not understand this scaling in detail, we can heuristically reason why it may emerge from the self-regulation of stellar feedback \citep[e.g.,][]{Murray05}. 
Namely, $v_{\rm c}$ scales with the escape velocity in the potential, so much slower outflows would quickly fall back onto galaxies, strongly suppressing their net effect. 
On the other hand, much faster outflows would easily escape halos and halt galaxy formation. 

The scaling with circular velocity can be used to derive how the mass outflow rate $\dot{M}_{\rm w}$ scales in different limits. 
When the wind is \emph{energy-driven}, the product $\dot{M}_{\rm w} v_{\rm w}^{2}$ is fixed, so the mass outflow rate scales as $\dot{M}_{\rm w} \propto 1/v_{\rm w}^{2} \propto 1/v_{\rm c}^{2}$. 
Similarly, when the wind is \emph{momentum-driven}, the fixed product is $\dot{M}_{\rm w} v_{\rm w}$, so $\dot{M}_{\rm w} \propto 1/v_{\rm w} \propto 1/v_{\rm c}$. 
Since the feedback energy scales with the star formation rate, these scalings are often expressed in terms of the mass loading factor $\eta_{\rm M} \equiv \dot{M}_{\rm w}/{\rm SFR}$. 
An example of an energy-driven wind is a hot, supernova-driven outflow in which radiative losses are negligible. 
An example of a momentum-driven wind would be one driven by radiation pressure, in which the momentum of photons is transferred to the gas but thermal energy plays a negligible role in the outflow expansion. 
The top right panel in Figure \ref{fig:winds} compares mass loading factors measured from different cosmological simulations and semi-analytic models as a function of halo mass. 

We note that, because mass loading can occur both in the ISM and in the CGM (see below), while the energy and/or momentum injection is concentrated in the galaxy, the energy and momentum loading factors $\eta_{\rm E} \equiv \dot{E}_{\rm w}/\dot{E}_{\rm feedback}$ and $\eta_{\rm p} \equiv \dot{p}_{\rm w}/\dot{p}_{\rm feedback}$ are often more robust model predictions. 
Here, the subscript `feedback' refers to the energy or momentum injected in the ISM by feedback processes while the subscript `w' refers to the energy or momentum escaping in a wind. 
The energy loading factor $\eta_{\rm E}$ can be $\ll 1$, e.g. when the majority of the energy from SNe is radiated away in the ISM before wind break out \citep[e.g.,][]{Fielding17_SNe}.

\subsubsection{Entrainment of CGM gas by galactic winds}
\label{sec:entrainment}
The properties of outflows can change greatly as they expand into the CGM. 
As outflows expand, they are decelerated by gravity as well as by entrainment of CGM mass.\footnote{In \S \ref{sec:survival} we will discuss the entrainnment of cold clouds in hot winds. Here, the entrained CGM mass can be volume-filling hot gas as well as cold gas.} 
The entrainment of CGM gas modifies the mass outflow rate, as well as its chemical composition by mixing gas recently ejected from the galaxy with ambient CGM. 
CGM entrainment can be very important: for example, \cite{Muratov17} and \cite{Mitchell2020_outflows} showed that mass outflow rates at the virial radius can be dominated by entrained gas, in the FIRE and EAGLE simulations, respectively (for EAGLE, this is shown by `halo scale' mass loading factors that are larger than `galaxy scale' loading factors in Fig. \ref{fig:winds}). 
In other simulations, such as IllustrisTNG, the entrained CGM mass is less important relative to the gas directly ejected from the ISM \citep[][]{Nelson19_outflows}, again underscoring the model dependence of outflow results. 
Since the metallicity of the CGM is generally lower than that of the ISM, entrainment tends to dilute the outflow metallicity. 
These effects imply that it is critical to specify where outflow properties are measured (such as at what radius) when comparing model predictions to observations, or different models to one another. 

As they sweep up CGM, galactic winds can affect the properties of gas accretion in halos. 
For example, outflows may push out infalling gas and prevent some of it from accreting onto galaxies \citep[e.g.,][]{Nelson15_fbk, Tollet2019}. 
Outflows may also affect the survival of cold streams, drive turbulence or inject heat into halo gas. 
Thus, it should be borne in mind that galactic winds are likely to modify some aspects of our simplified discussion of gas accretion in halos (\S \ref{sec:gas_accretion}) in model-dependent ways.

\subsubsection{Wind recycling}
\label{sec:recycling}

An important property of galactic winds is that some or most of their mass can recycle, i.e. re-accrete onto galaxies (see Fig. \ref{fig:AA17_tracks}). 
This implies that, in an instantaneous sense, some CGM gas that is observed to be infalling onto galaxies may have been previously part of a wind \citep[e.g,][]{Hafen2020_fates}. 
A phase change may occur as winds recycle, e.g. if a hot wind cools and cold clouds rain back onto the galaxy, but this does not necessarily occur if gas is ejected cold from the galaxy, as in some momentum-driven wind models. 
Recycling has also been shown to be very important in an integrated sense in shaping the galaxy stellar mass function, as was shown for example in the pioneering study of wind recycling by \cite{Oppenheimer10_recycling}. 
The fraction of wind mass that recycles in depends on galaxy mass, redshift, and on the feedback model \citep[][]{Mitchell2020_recycling}, but can be more than half and up to $\approx 1$ in some simulations \citep[e.g.,][]{Christensen2016_gas_cycle, AA17_cycle}. 

Useful concepts to characterize wind recycling include the distribution of recycling times and the distribution of the number of recyclings. 
Long recycling times mean that, after being ejected in a wind, a gas element spends a long time outside galaxies before being re-accreted. 
A given gas element can in general be recycled many times. 
In the FIRE simulations, the star formation histories of dwarf galaxies are highly time-variable, and gas elements can be ejected then re-accreted up to $\sim 10$ times by redshift zero. 
The multiple cycles of wind ejection and re-accretion in these dwarf galaxies may be an important factor driving the burstiness of star formation predicted by the simulations \citep[][]{AA17_cycle}.

As we discuss below in \S \ref{sec:companions}, another form of wind recycling occurs when gas ejected by one galaxy re-accretes onto \emph{another} galaxy.

\subsection{Satellite galaxies}
\label{sec:companions}
In this review, we focus primarily on physical processes operating in the CGM of central galaxies, i.e. main galaxies at the center of dark matter halos. 
The CGM of satellite galaxies can be affected by additional effects and does not separate cleanly from the CGM of the central galaxy they orbit. 
In this section, we briefly list some of the ways in which satellite galaxies can affect the CGM of the central galaxy.

Most directly, gas that remains bound to satellite galaxies (e.g., satellite ISM) can give rise to strong absorption features in the spectra of background sources. If the satellite is faint, the satellite may not be detected in emission and the absorption features can be mistakenly attributed to the CGM of the central galaxy. 
Similarly, satellites could contribute to the spatially extended emission that sensitive experiments aim to detected from the CGM.

Gas originally belonging to satellites can also be lost and incorporated into the CGM of a central galaxy by several different processes. These include:
\begin{itemize}
\item {\bf Ram pressure stripping:} When a body moves with velocity $v_{\rm rel}$ relative to a background gaseous medium of density $\rho$, the body experiences a ram pressure of magnitude $\sim \rho v_{\rm rel}^{2}$. This ram pressure strips gas from satellites and this gas mixes with the CGM of the central galaxy, contributing both mass and metals. 
In dense environments, ram pressure plays an important role in quenching star formation in satellite galaxies. 
This effect has been studied extensively in the context of galaxy clusters \citep[e.g.,][]{Tonnesen2007_environment_evol} and is theorized to produce ``jellyfish'' galaxies \citep[e.g.,][]{Franchetto2021_jellyfish}. 
If the relative velocity of the satellite (or, better still, it full orbital history) is known, ram pressure can be exploited to infer the density of dilute halo gas, as has been done using observations and modeling the Large Magellanic Cloud (LMC) around the Milky Way \citep[][]{Salem2015_ram_pressure}.

\item {\bf Tidal stripping:}  Tidal forces, which arise when gravitational forces are stronger on one side of a body than on the other, can also pull gas out of satellites. 
For a body of size $\Delta R$, the tidal acceleration with which opposite parts of the body are pulled apart by the gravity of a point mass $M$ at distance $R$ is $\sim GM \Delta R/ R^{3}$. 
We note that, because tidal forces scale as $1/R^{3}$, tides between low-mass satellite galaxies that are near one another can be more important than tidal forces between a satellite and the central galaxy. For example, \cite{Besla2012} modeled the Magellanic Stream (a band of HI gas trailing the Magellanic Clouds) as due to LMC tides stripping gas from the Small Magellanic Cloud \citep[other models suggest an important role for ram pressure in addition to tides, e.g.][]{tg2019_magellanic,Lucchini2021}. 
If the Milky Way were observed externally, the Magellanic Stream would appear as an important component of its CGM, so we must presume that some observed features of the CGM of other galaxies arise from similar tidal interactions. 

\item {\bf Satellite winds:} Feedback in satellite galaxies can eject gas from the ISM of a satellite and into the CGM of the central galaxy. 
This effect is illustrated in the simulation shown in Figure \ref{fig:AA17_tracks}, where it is labeled ``intergalactic transfer'' because some of the gas ejected 
by satellites can later accrete onto the central galaxy. 
This transfer process can contribute up to $\sim 1/3$ of the baryons that end up as stars in Milky Way-mass galaxies in FIRE simulations \citep[][]{AA17_cycle}, although the importance of this mode of galaxy fueling differs in other simulations (bottom right panel in Fig. \ref{fig:winds}). 
Not all the gas ejected in satellite winds necessarily re-accretes onto galaxies. 
Winds from satellites can also affect the CGM by puffing up accreting filaments in which they are often embedded (see \S \ref{sec:cold_streams}) and by creating overdensities that promote the condensation of the cool gas via thermal instability in the CGM \citep[][]{Esmerian2021}. 
At a given time, the fraction of the total CGM mass contributed by winds from satellite galaxies can be substantial (e.g., up to $\sim 20\%$ inside the virial radius of $L_{\star}$ galaxies in the FIRE simulations analyzed by Hafen et al. 2019\nocite{Hafen19_origins}).
\end{itemize}
The different mechanisms listed above that remove gas from satellites are not mutually exclusive. 
The galaxy group containing M81 and M82 is a well known example of a system in which intra-halo, filamentary HI clouds are associated with strong galaxy interactions, and thus most likely involve tidal stripping \citep[][]{Chynoweth2008}. 
In this system, M82 is also well known for its prominent galactic wind \citep[e.g.,][]{Strickland2009_M82}, so this an example where both gas ejection in a wind and tidal interactions shape the observed CGM.

Even if gas mass losses by satellites are negligible, satellites can deposit into the diffuse CGM the gravitational potential energy they lose as they fall into, or orbit within, the halo. 
Ram pressure acts as an effective friction force and removes energy from the orbit, which can in principle go into heating the CGM (e.g., as wakes dissipate). 
Similarly, dynamical friction induces wakes that can dissipate in the CGM \citep[][]{ElZant2004}. 
These processes may contribute to the excitation of disturbances or turbulence in the CGM, and operate on gas clumps that accrete in the halo even if these clumps do not correspond to satellite galaxies. 
\cite{DB2008_grav_quenching} analyzed clumpy gas accretion in massive halos and argued based on simple estimates that ``gravitational heating'' by cosmological accretion delivered to the hot gas in the inner halo could potentially maintain star formation quenching in the long term. 
More recent high-resolution simulations that self-consistently include gravitational heating indicate that this process is not sufficient to quench galaxies \citep[e.g.,][]{Su2019_failure}. 
However, more work is needed to understand whether and where these processes may have interesting effects on the CGM.

\section{SMALL-SCALE PROCESSES: MULTI-PHASE GAS}
\label{sec:small_scales}

\begin{figure*}
  \vspace*{-\baselineskip}
  \centering
  \includegraphics[width=1.0\textwidth]{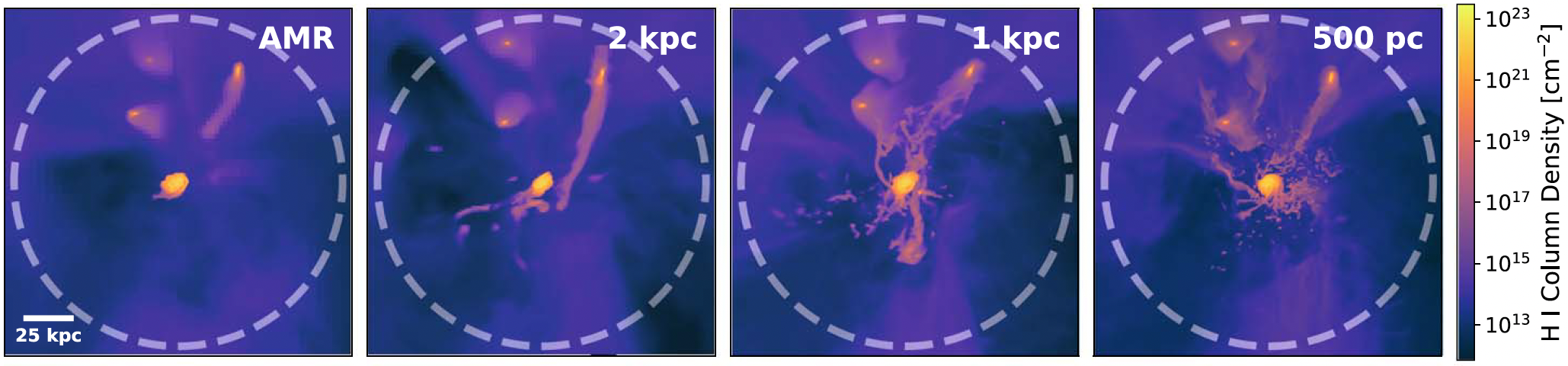}
  \caption{\small{{Cosmological simulations are unconverged in the CGM.} Shown are column density projections of a simulated $L_{*}$ galaxy with  different levels of spatial resolution. Enhanced halo resolution leads to enhanced neutral hydrogen content, with no sign of convergence. Adapted from \citet{hummels19}. \copyright AAS. Reproduced with permission.}
  }
  \label{fig:hummels}
\end{figure*}

An outstanding problem in current large-scale simulations is that the amount of CGM cold gas is unconverged. It increases monotonically with resolution \citep{FG16,vandevoort19,hummels19,peeples19,suresh19}, indicating that key physical processes remain unresolved (see Fig. \ref{fig:hummels}). In this portion of the review, we will survey small scale processes in CGM gas. This frequently includes physics which is unresolved in galaxy or cosmological scale simulations, and is often the realm of idealized simulations or analytic theory. The list of relevant physical processes is vast, and similar to that in the ISM: magnetohydrodynamics, fluid instabilities, shocks, radiative cooling, anisotropic conduction and viscosity, turbulence, cosmic rays, and gravity, to name a few. The more dilute nature of CGM plasma means that collisionality can be weak, and kinetic scale plasma processes can play a role. Entire textbooks could be devoted to some of these topics; we obviously cannot do them justice in a brief review. 

To focus our discussion, we lean on the striking abundance of atomic ($T\sim 10^4\,$K) and sometimes even molecular ($T\sim 10-100\,$K) gas in the CGM \citep{tumlinson17}, even though virialized gas should be much hotter. Indeed, cold gas forms our main observational probe of the CGM: it is observed at high spectral resolution by quasar absorption line spectroscopy, and also in spatially resolved emission maps by Integral Field Spectrographs on large ground-based telecopes such as Keck and the VLT. Direct observations of the hot gas component, in X-ray and Sunyaev-Zeldovich measurements, are few and far between, except in hotter systems such as massive ellipticals, groups and clusters; dispersion measure (DM) measurements of fast radio bursts \citep[FRBs;][]{prochaska19,chawla22} could eventually improve the situation. Given its observational prominence, and the likely importance of cold mode accretion in fueling star formation, we focus on physical processes relevant to cold gas in the CGM, and how it interacts with the hot phase. Parallels with the terrestrial water cycle are reflected in terminology (precipitation, condensation, evaporation). We consider: 

\begin{itemize} 

\item{{\it Cold Gas Formation} (\S\ref{sec:formation}; \S\ref{sec:survival}). What is the origin of cold gas in the CGM? We have already discussed cooling flows which go through a sonic point (\S\ref{sec:cool_vs_hot}) and cosmological accretion via `cold streams' (\S\ref{sec:cold_streams}), both of which occur in limited halo and accretion rate regimes. Here, we will discuss three further possibilities: thermal instability of hot halo gas (`precipitation'; \S\ref{sec:precipitation}), wholesale cooling of wind gas (\S\ref{sec:wind-TI}), and mixed-induced cooling condensation onto cold gas `seeds' (\S\ref{sec:survival}; see below).}

\item{{\it Cold Gas Survival and Growth} (\S\ref{sec:survival}). Overdense cold gas cannot be supported hydrostatically. 
It must either fall under gravity, or be flung out at high velocity. The resulting shear with hot gas should destroy the cloud via Rayleigh-Taylor and Kelvin-Helmholtz instabilities \citep{klein94,zhang17}. Yet, cold gas is seen in abundance outflowing in galactic winds, and also infalling as high-velocity clouds (HVCs; \citealt{putman12}). We describe recent progress in understanding cold gas survival, and how it can even grow in mass.}

\item{{\it Cold Gas Morphology} (\S\ref{sec:morphology}). Small-scale structure ($< 50\,$pc) in CGM cold gas has been inferred from photoionization modeling \citep{hennawi15,lau16,stern16,rudie19}; it has been suggested that cold gas is a `mist' \citep{McCourt18}. What is the topology of cold gas? Does it have a characteristic scale? Or is there structure on all scales? 
Like the IMF of self-gravitating clouds in the ISM, the mass function of pressure-confined cold gas in the CGM has fundamental observational and theoretical consequences.} 

\item{{\it Cold Gas Interactions: Turbulent Mixing Layers} (\S\ref{sec:TML}). Phase boundaries are not infinitely sharp, but thickened by diffusive transport processes such as thermal conduction, viscosity, and turbulence. The physics of these mixing layers are of great consequence: they govern the transport of mass, momentum and energy between phases. What sets these transport rates? These mixing layers are usually completely unresolved, even in idealized simulations. Does this preclude numerically converged transport rates?} 

\item{{\it Cold Gas Interactions: Cosmic Rays} (\S\ref{sec:CR}). Cosmic rays have energy densities comparable to thermal gas in the ISM; an important role is entirely plausible in the CGM. They can provide non-thermal pressure support, accelerate and heat gas; in recent years, their potential role in driving galactic winds has drawn significant attention. We briefly describe CR hydrodynamics, then draw attention to how small-scale cold gas can dramatically alter CR transport, and the spatial footprint of CR momentum and energy deposition.}

\end{itemize}

\subsection{Cold Gas Formation}
\label{sec:formation}

\subsubsection{Making Multi-Phase Gas: Local Thermal Instability}
\label{sec:TI} 

How does a multi-phase medium develop? The classic mechanism is thermal instability \citep{field65}: slightly overdense gas cools faster than its surroundings, loses pressure, undergoes compression and runaway cooling, until it reaches a new equilibrium. In the words of \citet{balbus95}, ``This seems such an economical and elegant method to make cloudy media, one feels that nature would be inexcusably remiss not to have taken advantage of it at some point." In stratified galactic halos, this is often dubbed `precipitation', since the cool gas which forms rains down on the galaxy. In this review, we distinguish precipitation, discussed here, from `condensation', whereby hot gas mixes with pre-existing cold gas to form intermediate temperature gas, which subsequently cools (\S\ref{sec:survival}). Both precipitation and condensation turn hot gas into cold gas via radiative cooling. They can take place simultaneously: for instance, cold gas which forms via thermal instability (precipitation) will fall and shear against background hot gas, triggering condensation. However, their physics is quite different. Precipitation has recently been comprehensively reviewed by \citet{DV22_review}, to which we refer the reader for more detailed discussion. 

Two points are worth noting. Firstly, local thermal instability typically presumes thermal equilibrium in the background medium; some form of heating is required. Otherwise, the entropy contrast between cool/hot gas often does not develop quickly enough, and a single phase cooling flow develops instead (\S\ref{sec:cool_vs_hot}). Secondly, there is often some damping process which counteracts the fragmentation into a multi-phase medium. The dimensionless ratio $t_{\rm cool}/t_{\rm damp}$, where $t_{\rm cool}$ is the cooling time and $t_{\rm damp}$ is the damping time, determines whether the medium is single phase (high $t_{\rm cool}/t_{\rm damp}$) or multi-phase (low $t_{\rm cool}/t_{\rm damp}$). 
Care must be taken to specify if $t_{\rm cool}$ is the cooling time of hot, cool, or intermediate temperature gas; different timescales are relevant in different contexts. Examples of $t_{\rm damp}$ are the buoyancy time $t_{\rm buoy}$ in stratified environments (where $t_{\rm buoy} \sim t_{\rm ff}$; see below), and the eddy turnover time $t_{\rm turb}$ in turbulent environments.  

In a uniform medium in thermal equilibrium ($\mathcal{L} =0$), the classic criterion\footnote{A more general criterion, $[\partial (\mathcal{L}/T)/\partial S]_{\rm A} > 0$, drops the requirement of thermal equilibrium \citep{balbus86}.} for thermal instability is \citep{field65}: 
\begin{equation}
\left( \frac{\partial \mathcal{L}}{\partial S} \right)_{\rm A} > 0,
\label{eq:TI-field}
\end{equation} 
where $\mathcal{L}$ is the net loss function (cooling minus heating per unit mass), $S$ is entropy, and A is the variable held constant. The limits of when the sound crossing time is short (long) compared to the cooling time gives rise to isobaric (isochoric) thermal instability, where pressure (density) is held constant. For a cooling function $\Lambda(T)$, with a local power law slope $\alpha= d({\rm log \Lambda})/{d( {\rm log} T)}$, equation (\ref{eq:TI-field}) gives $\alpha < 2$ ($\alpha < 0$) as the criterion for isobaric (isochoric) thermal instability. For the range of temperatures relevant for thermal instability in the CGM ($10^5 < T < 10^7$K), the gas is both isobarically and isochorically unstable ($\alpha < 0$), in which case the isobaric mode has a shorter growth time, by a factor $- \gamma_g \alpha/(2 -\alpha)$ \citep{field65}, where $\gamma_g=5/3$. There are also small `pockets' in narrow temperature ranges where only the isobaric mode is unstable \citep{pfrommer13,das21}.

\subsubsection{Precipitation: Local Thermal Instability in a Stratified Medium} 
\label{sec:precipitation} 

In a gravitationally stratified medium, an overdense cooling blob will oscillate, due to buoyant restoring forces as it falls under gravity. The rapidly changing background a cooling blob experiences modifies thermal instability \citep{defouw70}. \citet{balbus89} showed that the medium should be thermally unstable if and only if it is convectively unstable, $d{\rm ln} S/d{\rm ln} r < 0$, where $S(r) = P(r)/[\rho(r)]^{5/3}$ is the radial entropy profile. Since virialized halos are convectively stable (at least in groups and clusters, where X-ray observations show entropy profiles which increase outward), this appears to rule out thermal instability in stratified halos. 

\begin{figure}
  \vspace*{-1.5cm}
  \includegraphics[width=\textwidth]{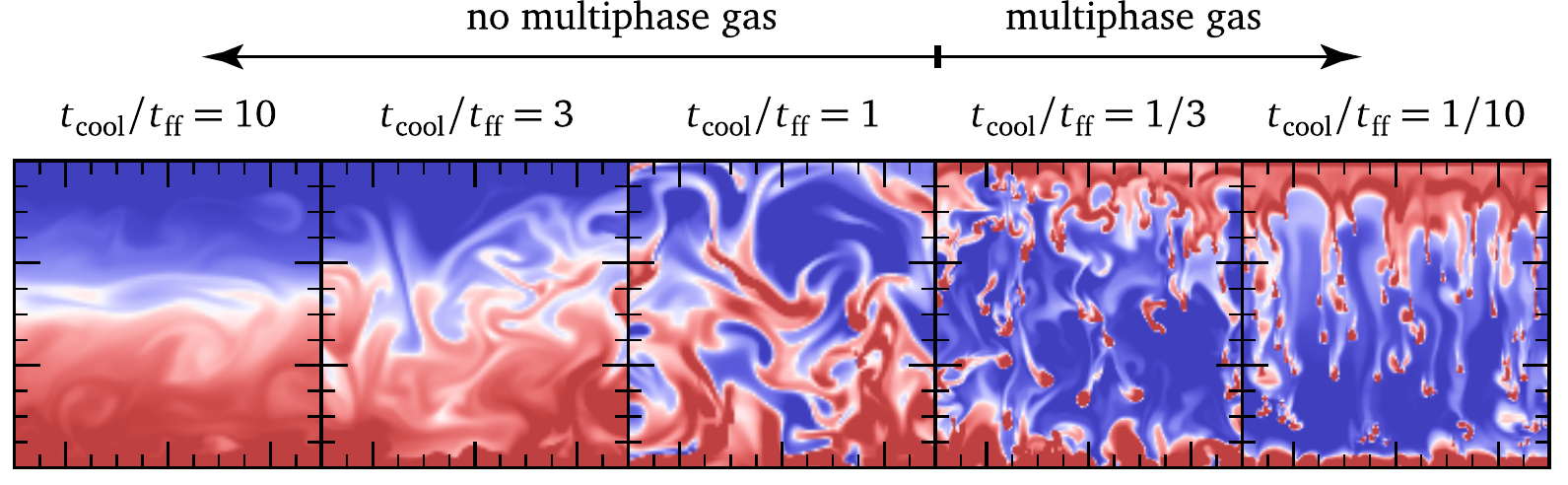}
  \caption{\small{Nonlinear saturation of thermal instability depends only on
    the ratio $t_{\text{cool}}/t_{\text{ff}}$ in hydrodynamics.  Each
    column shows images of the density in simulations with different
    values of the ratio $t_{\text{cool}}/t_{\text{ff}}$; all panels
    show the density after several cooling times, when the simulations
    have reached a steady state.  Thermal instability produces clouds
    of cold gas (with $\delta\rho/\rho{}\gtrsim{}1$) only when
    $t_{\text{cool}}\lesssim{}t_{\text{ff}}$. Adapted from \citet{mccourt12}.}  }\label{fig:thermal-instability-picture}
\vspace*{-0.5\baselineskip}
\end{figure}

In fact, \citet{balbus89} assumed that heating is a function of thermodynamic variables such as density and temperature (e.g., photoionization heating), but not an explicit function of position. Observations of AGN feedback in cluster cores means that spatial dependence is very likely; the same could be true of stellar feedback in galaxies. In that case, the \citet{balbus89} criterion does not apply. In an influential set of papers, \citet{mccourt12} and \citet{sharma12} showed that local thermal instability (commonly dubbed precipitation, since the cold gas then falls like raindrops) occurs when $t_{\rm cool}/t_{\rm ff} < 1, 10$ in planar (spherical) simulations; otherwise, the medium remains single-phase -- see Fig \ref{fig:thermal-instability-picture}. 
They also showed non-linear saturation at density amplitudes $\delta \rho/\rho \propto (t_{\rm cool}/t_{\rm ff})^{-1}$. Cooling blobs are akin to driven damped oscillators, with the ratio of driving time (the hot gas cooling time $t_{\rm cool}$) and damping time (the buoyancy time $t_{\rm buoy} \sim (d{\rm ln} S/d{\rm ln} r)^{-1/2} t_{\rm ff}$) determining the existence and saturation amplitude of thermal instability. Damping arises due to non-linear g-mode couplings, which sap energy from the oscillating blob.  
There is also considerable observational evidence in groups and clusters\footnote{For a dissenting view, see \citet{mcnamara16} and \citet{hogan17}, who argue from observational data that $t_{\rm cool}$ alone, rather than $t_{\rm cool}/t_{\rm ff}$, determines the threshold for multi-phase gas; they contend that non-linear perturbations, such as uplift of cold gas from the central galaxy, seed cold gas condensation.} that a threshold value of ${\rm min} (t_{\rm cool}/t_{\rm ff}) \lsim 10$ is required for multi-phase gas (e.g., see summary in \citealt{DV22_review}), and it has been hypothesized that massive galaxies and clusters naturally evolve, via feedback, to a state of marginal stability with respect this criteria \citep{voit17}. These considerations are also important in the vicinity of accreting black holes, which can be potentially fed by precipitating cold gas (`chaotic cold accretion'), rather than Bondi accretion of hot gas, which is inefficient and also predicts a cusp in hot gas temperature near the black hole, which is not seen \citep{gaspari13, gaspari18}. Cloud-cloud and cloud-torus collisions allow angular momentum loss. 

We have already seen the ratio $t_{\rm cool}/t_{\rm ff}$ before, which determines when halos complete virialization (\S\ref{sec:gas_accretion}). In the absence of heating, the competition between cooling and shock virialization of cosmological accretion determines {\it global} thermal stability, i.e. whether cooling flows (or stable virial shocks) develop. In the presence of heating, and background hydrostatic/thermal equilibrium, the competition between cooling and buoyancy determines {\it local} thermal stability, i.e. fragmentation into a multi-phase medium. While both virialization and buoyant oscillations share a common timescale $t_{\rm ff}$, their physics is quite different. 

For precipitation, the ratio $t_{\rm cool}/t_{\rm ff}$ is more correctly written as $t_{\rm cool}/t_{\rm buoy}$, where $t_{\rm buoy} \sim (d{\rm ln} S/d{\rm ln} r)^{-1/2} t_{\rm ff}$. Typically $(d{\rm ln} S/d{\rm ln} r) \sim {\mathcal O}(1)$ in halo gas, which is strongly stratified. However, in environments with strong turbulent mixing, entropy cores can develop, and then there are no buoyant restoring forces: $t_{\rm buoy} \rightarrow \infty$, and $t_{\rm cool}/t_{\rm ff}$ is irrelevant \citep{voit17}. While observational constraints are weak, flat hot gas entropy profiles often develop in simulated disk galaxies \citep{Esmerian2021}. Furthermore, global thermal equilibrium requires that any feedback loop which keeps the hot atmosphere stable must operate on timescales short compared to thermal timescales. This likely holds in the hot atmospheres of galaxy  groups and clusters, where the AGN duty cycle ($t \sim 10^{7}$yr) is shorter than, or comparable to the central cooling time.\footnote{Many simulations adopt idealized heating where global thermal equilibrium is enforced by fiat: at each timestep, total cooling in each shell is calculated and an equivalent amount of uniform heating is added to that shell \citep{mccourt12,sharma12}. However, simulations which incorporate more realistic AGN feedback heating obtain similar results \citep{gaspari12,li14}.} However, the situation is likely much more dynamic and out of equilibrium in galactic CGM environments (e.g., \citealt{Fielding17}). Thus, while the case for precipitation in massive halos ($M > 10^{13} \, M_{\odot}$) is strong, the extrapolation to lower masses is more uncertain, particularly given scant observational constraints on hot gas in such halos. In these environments, it is quite possible that buoyant restoring forces are secondary to other physics (e.g. turbulence). 

Indeed, the threshold value of $t_{\rm cool}/t_{\rm ff}$ can vary in different settings\footnote{It was initially thought that damping by buoyancy is geometry dependent, given different thresholds for planar and spherical simulations \citep{mccourt12,sharma12}, but subsequent work showed that this was an artifact of the how $t_{\rm cool}/t_{\rm ff}$ was defined in different setups, where $t_{\rm cool}/t_{\rm ff}$ can either increase monotonically or display a minimum. The physics of the threshold is geometry independent \citep{choudhury16, meece15}.}, or have considerable dispersion (e.g., \citealt{voit21}). A particularly important variable is the amplitude of initial density perturbations, which is presumed small ($\delta \rho/\rho \ll 1$) in most studies. In a controlled set of simulations, \citet{choudhury19_published} showed that the threshold value of $t_{\rm cool}/t_{\rm ff}$ increases (i.e., thermal instability is more easily triggered) as $\delta \rho/\rho$ increases; once $\delta \rho/\rho \sim \mathcal{O}(1)$, then buoyant forces cannot quench thermal instability, and $t_{\rm cool}/t_{\rm ff}$ becomes irrelevant. Such non-linear density perturbations could arise from shocks, or the introduction of low entropy gas into the halo via uplift from the halo center (e.g., cool gas dragged out by AGN blown bubbles), cosmological accretion, or winds from satellite galaxies. Such low entropy gas cools easily and seeds further cooling of hot gas, though in the context of this review, this falls under the rubric of `condensation' (\S\ref{sec:survival}) rather than `precipitation'. Both \citet{nelson20} and \citet{Esmerian2021} came to similar conclusions that $t_{\rm cool}/t_{\rm ff}$ is a poor predictor of thermal stability when large amplitude perturbations are present, in TNG50 and FIRE simulations respectively.   

Other physics can also influence the development and saturation of thermal instability. For instance, magnetic fields damp buoyant oscillations  (via magnetic tension) and destabilizes all scales below $l_{\rm A} \sim v_{\rm A} t_{\rm cool}$ (where $v_{\rm A} = B/\sqrt{4 \pi \rho}$ is the Alfven velocity) -- i.e., thermal instability can potentially occur anywhere in the halo, independent of $t_{\rm cool}/t_{\rm ff}$ \citep{ji18}. MHD effects appear for $\beta = P_{\rm g}/P_{\rm B} < 1000$ (where $P_g$ is gas pressure and $P_{\rm B}$ is magnetic pressure), with B-fields enhancing the amplitude of thermal instability $\delta \rho/\rho \propto \beta^{-1/2}$ even for very weak fields, surprisingly independent of field orientation\footnote{Horizontal field lines support overdense gas directly via magnetic tension, while vertical field lines confine over-pressured hot gas, which in turn supports overdense gas via pressure gradients.}. By providing rotational support against gravity, and also deflecting the descent of a condensing blob via Coriolis forces, angular momentum can also suppress buoyant damping of thermal instability \citep{sobacchi19}. Turbulence can both enhance and suppress thermal instability. Enhancement occurs via stronger density perturbations in compressive turbulence\footnote{Even for subsonic turbulence, where $\delta \rho/\rho \sim \mathcal{M} < 1$, where $\mathcal{M}$ is the Mach number, non-linear steepening can result in $\delta \rho/\rho \sim \mathcal{O}(1)$ and weak shocks.}, as well as uplift of low entropy gas, which subsequently cools \citep{voit18}. Suppression occurs via mixing of cold gas with hot gas; \citet{gaspari18} have argued that $t_{\rm cool}/t_{\rm turb} \lsim 1$ is required for perturbations to condense, which we will see is indeed the criterion for multi-phase gas in a turbulent medium (see \S\ref{sec:TML}). Cosmic rays are an interesting case. In simulations with horizontal fields perpendicular to gravity, they do not significantly change the threshold for thermal instability, although (by providing pressure support) they decrease the density, and increase the infall time of cold clouds \citep{butsky20}. However, since field-aligned CRs cannot stream down their vertical gradient, there is no background CR transport or CR heating in the horizontal field case; these only arise as perturbations. For vertical fields with CR streaming, linearly unstable modes no longer oscillate like g-modes, but instead propagate at the Alfven speed \citep{kempski20}. In the non-linear state, in some parameter regimes hydrostatic and thermal equilibrium can be maintained, but in others, a CR heated and driven wind develops -- local thermal instability triggers a global wind instability (\citet{tsung22-TI}; see \S\ref{sec:CR}).

Once dense cold gas is introduced into the halo, it cannot stay in equilibrium but either falls under gravity or is blown out by a wind. It therefore inevitably evolves as it shears against hot gas, potentially either growing (`condensation') or diminishing (destruction via hydrodynamic instabilities). Such processes are critical in regulating the cold gas content of halos. We discuss this in \S\ref{sec:survival}. 

\subsubsection{Global Thermal Instability in a Wind} 
\label{sec:wind-TI} 

Many galaxies do not have a static CGM, but exhibit outflows and inflows, where cold gas is observed. Without stable heating, global thermal instability can cool the entire CGM within a critical radius if the cooling flow goes through a sonic point (\S\ref{sec:gas_accretion}), or beyond a critical radius in a wind \citep{wang95,silich04,thompson16,scannapieco17,schneider18}. \citet{thompson16} noted that mass loaded winds always cool on large scales. This may seem counterintuitive, since in an initially adiabatic spherical wind \citep{Chevalier85}, where $v\approx$const, the density $\rho \propto r^{-2}$, and cooling rates fall rapidly. However, as the wind cools adiabatically, it moves into a temperature regime ($10^5 \, {\rm K} < T < 10^7$K) where metal line cooling dominates, $\Lambda \propto T^{-0.7}$. Since $T \propto P/\rho \propto \rho^{\gamma-1} \propto r^{-4/3}$, we have $t_{\rm cool} \propto T/(n\Lambda) \propto T^{1.7}/n \propto r^{-4/15}$, and since the advection time $t \propto r/v \propto r$, this gives $t_{\rm cool}/t_{\rm adv} \propto r^{-19/15}$, which decreases rapidly with radius. \citet{thompson16} derived a `cooling radius' for the outflow, when $t_{\rm cool} \sim t_{\rm adv}$, beyond which the flow becomes radiative: 
\begin{equation} 
r_{\rm cool} \approx 4 \, {\rm kpc} \frac{\alpha^{2.13}}{\beta^{2.92}} R_{0.3}^{1.79} \left( \frac{\Omega_{\rm 4 \pi}}{\dot{M}_{\rm SFR,10}} \right)^{0.789}, 
\label{eq:rcool} 
\end{equation} 
where $\alpha$ is the fraction of supernova energy thermalized in the hot plasma ($\dot{E}_{\rm wind} = 3 \times 10^{41} {\rm erg \, s^{-1} \alpha \dot{M}_{\rm SFR}}$), $\beta=\dot{M}_{\rm wind}/\dot{M}_{\rm SFR}$ is the mass loading factor, $R_{0.3} = R/(0.3 \, {\rm kpc})$ is a characteristic radius for the injection region, $\dot{M}_{\rm SFR,10} = \dot{M}_{\rm SFR}/(10 \, M_{\odot} {\rm yr^{-1}})$ is the star formation rate, and $\Omega_{\rm 4 \pi} = \Omega/4\pi$ is the opening angle of the outflow. This model bypasses the need to entrain cold gas, which is born comoving, and produces cold gas in abundance, since essentially the entire wind cools to $T \sim 10^4$K at $r_{\rm cool}$. From equation (\ref{eq:rcool}), it is clear that the viability of a cooling wind is sensitive to the kinetic injection factor $\alpha$ and mass loading factor $\beta$: \citet{thompson16} estimate that for $\beta \le \beta_{\rm min} \approx 0.6 \alpha^{0.6} (R_{\rm 0.3} \Omega_{4 \pi}/\dot{M}_{\rm SFR,10} )^{0.36}$, the wind remains adiabatic to large scales, since $t_{\rm cool}/t_{\rm adv}$ never falls below unity before $T < 10^5$K, when the cooling function changes slope. Indeed, in high resolution simulations with high ($\beta \sim 0.5$) mass loading rates, \citet{schneider18} found wind cooling in good agreement with equation (\ref{eq:rcool}), but no wind cooling for simulations with $\beta \sim 0.1$ \citep{schneider20}, consistent with above expectations. This sensitivity to mass loading also appears in semi-analytic models of multi-phase winds \citep{fielding22}.    

\subsection{Cold Gas Survival and Growth} 
\label{sec:survival}

\subsubsection{Condensation Enables Entrainment and Growth} 
\label{sec:hydro-cloud-wind}

We now describe condensation, which like precipitation turns hot into cold gas. A combination of cold gas `seeds' and velocity shear against the hot phase produces mixed intermediate temperature gas, which subsequently cools. The cooled gas retains its momentum, and in time the cold phase both grows in mass and comoves with the hot phase. This form of mixing-induced thermal instability thus enforces kinematic coupling between phases, and the high rate of conversion between phases has important implications for the baryon cycle. 

Condensation is closely related to the well-known `cloud entrainment problem'. 
Atomic and molecular gas is observed outflowing (in galactic winds) or inflowing (as intermediate or high velocity clouds, IVCs/HVCs), at velocities often comparable to virial velocities. How can it withstand hydrodynamic instabilities? The problem is easily stated: if cold gas is overdense by a factor $\chi = \rho_{\rm c}/\rho_{\rm h}$, the timescale for acceleration by hydrodynamic ram pressure, $t_{\rm acc} \sim p/\dot{p} \sim (\rho_{\rm c} R^{3} v)/(\rho_{\rm h} v^{2} R^{2}) \sim \chi R/v$, is longer than the timescale for the cloud to mix into its surroundings, the `cloud-crushing' time, 
$t_{\rm cc} \sim t_{\rm KH} \sim \sqrt{\chi} R/v$, by a factor $t_{\rm acc}/t_{\rm cc} \sim \chi^{1/2} \sim 10-30$, independent of cloud size or velocity \citep{klein94,FGQM12,zhang17}. Here, $\rho_h, \rho_c$ are the densities of hot and cold gas, $R$ is the cloud radius, $v$ the relative velocity between the cold and hot gas, and $p$ is the momentum. Many simulations, both with and without radiative cooling, confirmed that clouds in a wind tunnel are destroyed before they are entrained \citep{klein94,mellema02,pittard05,cooper09,scannapieco15,schneider17}. Here, we describe an emerging paradigm for cloud acceleration and survival where hot gas condenses onto cold gas, by mixing and cooling, thereby transferring mass and momentum to the cold phase. 

\begin{figure}
  \includegraphics[width=\textwidth]{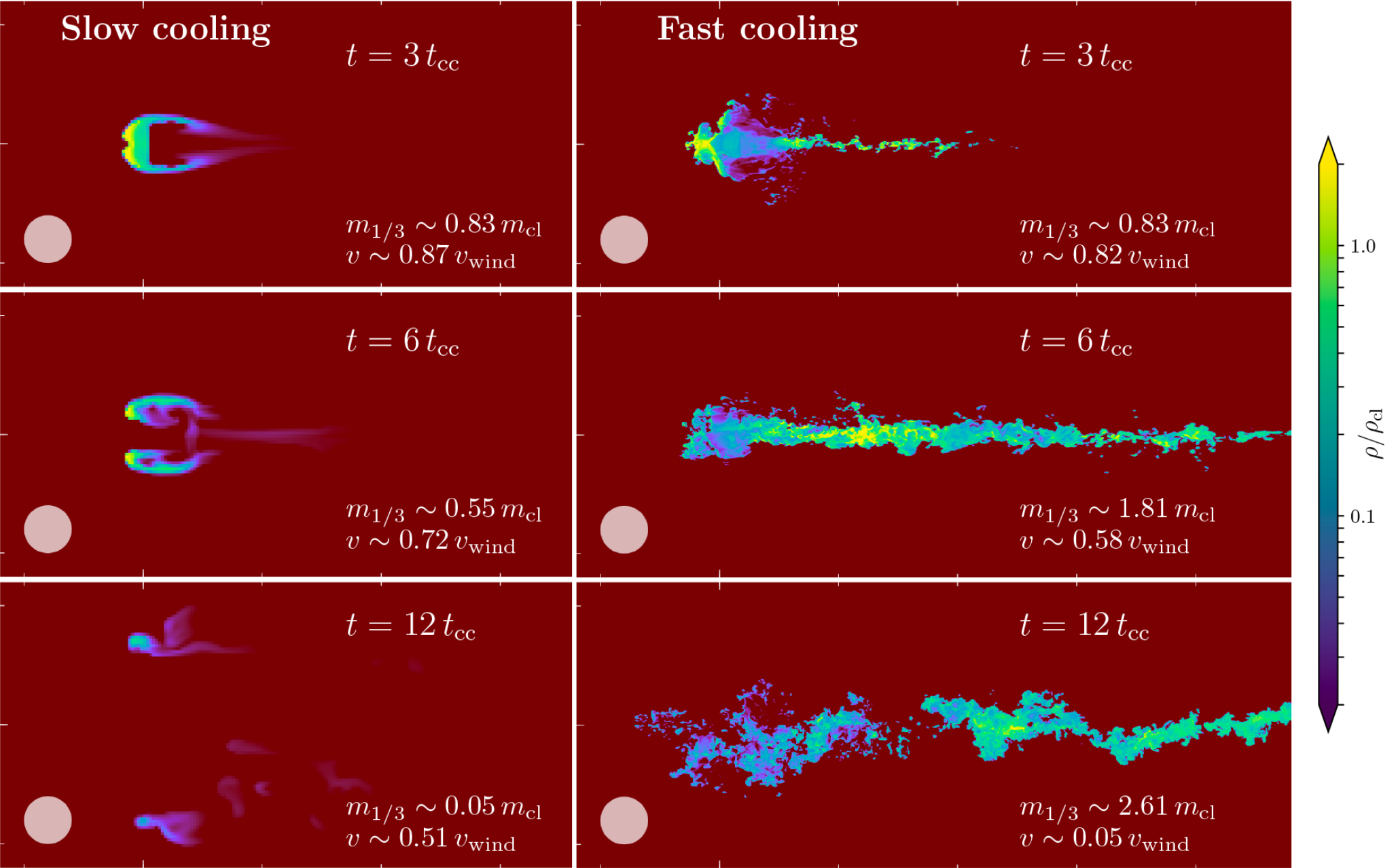}
  \parbox[c]{\linewidth}{%
    \caption{
      \small{Clouds with $t_{\rm cool,mix}/t_{\rm cc} < 1$ can entrain, survive and grow in a wind. While the simulation on the \textit{left}, where $t_{\rm cool,mix}/t_{\rm cc} \sim 7.7$ shows the expected behavior well known from previous studies, i.e., the destruction of the cloud on a few $t_{\mathtt{cc}}$, the simulation to the \textit{right}, where $t_{\rm cool,mix}/t_{\rm cc} \sim 0.077$ shows that the cloud does not get destroyed but instead grows to a mass greatly exceeding its initial mass ($m_{1/3}$ is the mass of gas with density at least 1/3 of the initial cloud density), and also gets entrained, with $\Delta v/v_{\rm wind} \ll 1$. The grey disk in each panel illustrates the original size of the cloud. From \citet{gronke18_published}.}
      \label{fig:cloud-entrainment}
    }}
\end{figure}

Cloud growth (rather than destruction) was first seen in wind tunnel simulations with radiative cooling by \citet{marinacci10} and \citet{armillotta16}, in the context of HVC survival. 
\citet{gronke18_published} found not only cloud growth, but cloud entrainment in a wind; they quantified a criterion for cloud survival, $t_{\rm cool,mix} \lsim t_{\rm cc}$, which matched simulation results well. Here, $t_{\rm cool,mix}$ is the cooling time of mixed gas at $T_{\rm mix} \sim \sqrt{T_{\rm c} T_{\rm h}}$ (\citealt{begelman90}; \S\ref{sec:TML}), where $T_{\rm c},T_{\rm h}$ are the cold and hot phase temperatures. 
This criterion sets a lower bound on cloud size \citep{gronke18_published}:
\begin{equation}
R > \frac{v_{\rm wind} t_{\rm cool,mix}}{\chi^{1/2}} \approx 2 \, {\rm pc} \ \frac{T_{\rm cl,4}^{5/2} \mathcal{M}_{\mathrm{wind}}}{P_{3} \Lambda_{\rm mix,-21.4}} \frac{\chi}{100}
\label{eq:rcrit}
\end{equation}
where $T_{\rm cl,4} \equiv (T_{\rm cl}/10^{4} \, {\rm K})$, $P_{3} \equiv nT/(10^{3} \, {\rm cm^{-3} \, K})$, $\Lambda_{\rm mix,-21.4} \equiv \Lambda(T_{\rm mix})/(10^{-21.4} \, {\rm erg \, cm^{3} \, s^{-1}})$, $\mathcal{M}_{\mathrm{wind}}$ is the Mach number of the wind, and we write $v_{\mathrm{wind}} = c_{\mathrm{s,wind}} \mathcal{M}_{\rm wind} \sim c_{\mathrm{s,cl}}\mathcal{M}_{\rm wind}\chi^{1/2}$, assuming isobaric conditions. In all cases where the cloud survives, it forms a cometary tail, similar to the observed head-tail morphology of HVCs \citep{putman11}, which grows in mass. Cloud growth and entrainment are intertwined: as hot gas condenses, it imparts its momentum, and accelerates the cloud. See Fig. \ref{fig:cloud-entrainment} for an example. 

One can quantify growth and entrainment times \citep{gronke20-cloud}; analytic models match simulation results well. Interestingly, cloud growth does {\it not} stop once the cloud is entrained, even though there is little shear to drive mixing. Instead, growth is maximized for comoving clouds! In entrained clouds, mixing is driven by cooling-driven cloud pulsations, which arise for clouds which lose sonic contact as gas condensing onto the cloud cools, $t_{\rm cool} \ll t_{\rm sc}$ (where $t_{\rm sc} \sim R/c_{\rm s,c}$ is the sound crossing time). The small loss of pressure balance causes the cloud to contract, overshoot and subsequently expand, which in turn drives more mixing and cooling, and subsequent pulsations. Cooling in a pressure confined cloud drives overstable acoustic oscillations governed by $t_{\rm cool}/t_{\rm sc}$, just as cooling in a stratified medium drives overstable buoyant oscillations governed by $t_{\rm cool}/t_{\rm ff}$ (\S\ref{sec:precipitation}). Such pulsations are also seen in clouds in initially static surroundings which are perturbed \citep{gronke20-mist}. They are the linear, small-amplitude version of `shattering' (\S\ref{sec:frag-coag}), where the entire cloud falls violently out of pressure balance due to cooling, creating highly non-linear, large amplitude oscillations which tear the cloud apart. Hot gas inflow velocities were found to be $v_{\rm mix} \sim c_{\rm s,c} (t_{\rm cool}/t_{\rm sc})^{-1/4}$, and thus growth happens on a timescale $t_{\rm grow} \sim m/\dot{m} \sim \rho_{\rm c} A r/(\rho_{\rm h} A v_{\rm mix}) \sim \chi t_{\rm sc}$. We dissect $v_{\rm mix}$, and the unintuitive scaling $v_{\rm mix} \propto (t_{\rm cool}/t_{\rm sc})^{-1/4}$, in \S\ref{sec:TML}. Surprisingly, mass growth is converged at fairly low resolution ($\sim 8$ cells per cloud radius; \citealt{gronke20-cloud,kanjilal21}), even when the mixing surface area is manifestly unconverged, and thermal diffusion lengths or Field lengths are unresolved (indeed, even without explicit conduction); these issues are also discussed in \S\ref{sec:TML}. 

The criterion $t_{\rm cool,mix} < t_{\rm cc}$, where $T_{\rm mix} \sim \sqrt{T_{\rm c} T_{\rm h}}$, is an idealization which happens to match simulations with temperature floors of $T \sim 10^4$ K (due to photoionization) well. 
In reality, mixed gas has a broad range of temperatures. A more refined criterion based on the thermodynamic history of fluid elements, modeling how mixing and cooling modify entropy, would be more accurate, at the expense of greater complexity \citep{abruzzo22}. If gas cools to lower temperatures, different criteria (still all variants of a cooling length) potentially apply \citep{farber22,abruzzo22b}. The definition of cloud survival is also important. \citet{li20} and \citet{sparre20} argued that it is the cooling time of hot gas which matters, i.e. that $t_{\rm cool,hot} < \alpha t_{\rm cc}$ (where $\alpha \sim 10$, since it takes several cloud-crushing times to destroy a cloud) is required for survival. This is indeed true if one cares about the survival of original cloud material, as appropriate in some applications. However, the abundance of cold gas is not monotonic. After an initial decline, eventually the mixed gas cools, and the cold gas mass recovers and continues to grow\footnote{The cooling time of the hot gas cannot be the rate limiting step for cloud growth, since turning cooling on or off for $T > 0.6 \, T_{\rm h}$ appears to have little impact on mass growth rates \citep{gronke18_published,abruzzo22}; emission is dominated by lower temperature gas. This suggests that mixing is responsible for the initial drop in hot gas entropy.}, even if much of the original cloud material has not survived \citep{kanjilal21}. 

The physics above has been demonstrated in 3D hydrodynamic simulations with transonic, constant winds and $\chi \sim 100-1000$. Below, we discuss how other parameter choices and physics affect outcomes, but cover turbulence (\S\ref{sec:morphology}), and cosmic rays (\S\ref{sec:CR}) elsewhere. These are all issues of active research, and by no means settled. Of these, it is most apparent that infall under gravity changes survival criteria (\S\ref{sec:gravity}), and the impact of higher overdensities and wind Mach numbers is most uncertain (\S\ref{sec:wind-cloud}).

\subsubsection{Condensation: Magnetic Fields} \label{sec:b-fields} The hot medium B-field has a much larger impact than initial cloud B-fields \citep{li20}. Unfortunately, the plasma $\beta$ of the CGM is still highly uncertain, although it is expected to lie between the ISM ($\beta \sim 1$) and ICM ($\beta \sim 100$). Effects are strongest when B-fields are perpendicular to the direction of cloud motion, and `magnetic draping' \citep{lyutikov06,dursi07} takes place: the cloud sweeps up field lines, which are amplified to rough equipartition with ram pressure ($\epsilon_{\rm B,drape} \sim \alpha \rho_{\rm wind} v_{\rm wind}^{2}$, where $\epsilon_{\rm B,drape}$ is the magnetic energy density of the drape and $\alpha \sim 2$). The amplified B-fields have two important dynamical effects. Firstly, magnetic drag couples the cold and hot gas, reducing acceleration times by a factor  
$\sim ( 1 + 2/[\beta_{\mathrm{wind}} \mathcal{M}_s^2] )^{-1}$
compared to the hydrodynamic case \citep{dursi08,mccourt15}. Secondly, B-fields suppress the Kelvin-Helmholtz (KH) instability via magnetic tension. This clearly suppresses mixing in both adiabatic \citep{jones97} and radiative \citep{ji19} KH simulations. These two effects might be expected to have a major impact on cloud survival and mass growth rates. In fact, while cloud morphology changes dramatically, becoming much more filamentary, the survival criterion (equation \ref{eq:rcrit}) and cloud mass growth rates appear relatively unchanged\footnote{\citet{gronnow18} claim significantly reduced mass growth rates in the MHD simulations, but they only followed cloud evolution for $ t \lsim 3 t_{\rm cc}$. Their results are consistent with other MHD sims where growth rates rise at later times.}  \citep{gronke20-cloud,li20,sparre20,jennings22}. For typical overdensities $\chi \sim 100$, magnetic drag only influences survival for strong background fields ($\beta_{\rm wind} \lsim 1$). Why does the strong draped field not suppress mixing completely? Although the B-field strength of the hot medium does not affect the energy density of the draping layer or the time it takes to grow ($\sim R/v_{\rm wind}$), it does affect its thickness, and thus the scale of the modes which are stabilized\footnote{The draping layer has thickness $l_{\rm drape} \sim R/(6 \alpha \mathcal{M}_{\rm A}^{2})$ and stabilizes modes $\lambda \lsim 10 l_{\rm drape} \sim R/\mathcal{M}_{\rm A}^2 \sim R/(\beta_{\rm wind} \mathcal{M}_s^{2})$ \citep{dursi07}, where $\mathcal{M}_{\rm A} =v/v_{\rm A}$, $\mathcal{M}_{\rm s} =v/c_{\rm s}$ are the Alfven and sonic Mach numbers. The total magnetic energy in the drape is $E \sim \epsilon_{\rm B,drape} A l_{\rm drape} \sim \rho v^2 A R v_A^2/v^2 \sim \epsilon_{\rm B,wind} V$, i.e. it is the magnetic energy of the wind displaced by the cloud.}. For weak B-fields (high $\beta$), only small scale modes are stabilized, and mixing can still occur, albeit at a reduced rate. Cloud entrainment still relies primarily on momentum transfer via cooling of mixed hot gas. 

Still, given that mixing is attenuated, the fact that mass growth rates are not strongly suppressed is puzzling; only for relatively strong fields ($\beta_{\rm wind} \sim 1$) is significant suppression seen \citep{jennings22}. 
One important clue is that growth rates {\it are} suppressed in MHD simulations of accreting cold streams, similar to KH simulations (N. Mandelker, B. Tan, private communication), and unlike clouds at identical Mach numbers. Evidently, clouds have additional degrees of freedom which change mass growth. This issue is still unresolved, but two points are worth noting. Firstly, morphology is very different: clouds can grow extended tails and also fragment; the significantly larger surface area measured in simulations -- due to lower cloud densities (see below) and filamentary morphology -- could boost the mass accretion rate, even if mixing rates per unit area area reduced. 
Secondly, mixing in shear layers and cold streams is via the Kelvin-Helmholtz instability -- if shear drops to zero, so does mass growth. By contrast, mass growth in clouds peaks when it is entrained, due to cooling-induced cloud pulsations. Thus, the suppression of the KH instability by B-fields may have a weaker effect. 

Note that plasma $\beta$ and density can vary dramatically in growing magnetized clouds. Compressional B-field amplification of entrained wind material can lead to low density, magnetically supported gas \citep{gronke20-cloud,nelson20}. Even though high density regions dominate by mass, these low density, low $\beta$ regions dominate by area, and could account for unexpectedly low density cloud material inferred from COS observations \citep{werk14}. In a sense, non-thermal pressure support can accelerate simulation convergence, since it reduces the scales to which gas is compressed and needs to be resolved.

\subsubsection{Condensation: Thermal Conduction} 

Conduction does not significantly affect the structure of the cloud itself, since hot electrons can only penetrate a skin depth $\lambda_{\rm skin} \sim \lambda_{\rm e,h}/\chi \sim 0.03  T_{\rm h,6}^{2} n_{\rm c,-1}^{-1} \, {\rm pc}$, where $\lambda_{\rm e,h}$ is the electron mean free path in the hot medium (equation \ref{eq:Nmfp}). However, it can create a thick boundary layer of warm gas around the cloud, which affects mixing and cooling. 
One can show that 
in the cloud survival regime, $r> r_{\rm crit}$ (equation \ref{eq:rcrit}) 
we should be in the classical diffusive (Spitzer) regime; for a hotter background medium, conduction becomes saturated \citep{li20}. Also, the ratio $r_{\rm crit}/\lambda_{\rm F} \sim 0.5 f^{-1/2} \mathcal{M} T_6^{-5/4}$ (where $\lambda_{\rm F}$ is the Field length, equation \ref{eq:Nfield}), so conduction need only be mildly suppressed for clouds which survive mixing to also evade evaporation; cloud crushing and evaporation times are comparable, for typical CGM parameters. Accordingly, simulations with isotropic Spitzer conduction show modified cloud-wind interactions \citep{bruggen16,armillotta16},  but not dramatic changes. Effects include cloud compression and slower entrainment (attributed to the smaller cross-section); whether conduction impedes or enhances destruction depends on Mach number and overdensity. 

Importantly, however, real conduction is anisotropic, since electrons gyrate around B-field lines; cross-field conduction is strongly suppressed. 
For this reason, simulations with field-aligned conduction find the effects of conduction to be much weaker; once magnetic fields drape over the cloud, conduction is strongly suppressed, and has little effect on cloud mass evolution, relative to MHD simulations \citep{li20,jennings22}.  
Conduction can also be suppressed by modified electron scattering rates, e.g. due to whistler instabilities driven by heat flux \citep{roberg16,komarov18,drake21}. Still, the influence of conduction on cloud growth is not fully mapped out, and there could still be surprises. For instance, in all simulations, conduction considerably increases the amount of warm gas in the simulation domain. It still arises in anisotropic conduction, since downstream gas is not magnetically shielded by draping. Downstream gas does not influence cloud evolution in current setups with conduction. But if there are multiple clouds, or if the velocity field is turbulent and time-varying, the abundant warm gas will interact with cold gas, with stronger effects.

\subsubsection{Condensation: Wind and Cloud Properties} \label{sec:wind-cloud} Wind tunnel simulations are of course highly idealized. They frequently assume a spherical, warm ($T\sim 10^{4}$K, $\chi \sim 100$) cloud in a constant, trans-sonic ($\mathcal{M} \sim 1$) wind. What is the effect of relaxing these assumptions? 

Begin with cloud properties. A number of simulations have considered non-spherical clouds -- e.g., with initial conditions extracted from a turbulent box \citep{schneider17,LR20_mist,banda-barragan19,gronke20-cloud,li20}. These appear to have little effect, apart from somewhat faster growth or destruction in the cloud growth/destruction regimes respectively, presumably due to larger surface area. 
Dust-laden molecular gas is often observed outflowing at high velocity \citep{menard10,fischer10,cicone14}; the higher overdensities\footnote{High overdensities $\chi$ are challenging to simulate, as clouds typically develop tails of length $\sim \chi r$, so longer simulations boxes are needed.} and different cooling regimes at lower temperatures are important modifications. Given their formation pathways, dust and molecules are often thought originate from the host galaxy; if so, they must survive the mixing process without being destroyed by high temperature gas\footnote{Though molecule formation and dust regrowth is also potentially possible in wind-driven radiative shocks \citep{richings18a,richings18b}.}. Simulations of the entrainment of molecular gas find a complex variety of outcomes, with cold gas surviving, destroyed, or transformed to $T\sim 10^{4}$K gas \citep{farber22}. Importantly, entrainment was faster than the naive expectation $t_{\rm acc} \sim \chi r/v$, due to a `cocoon' of warm $T\sim 10^{4}$K gas forming, which decreases effective overdensity. Indeed, molecular gas is almost always surrounded by atomic gas; in galactic HVCs, the atomic gas mass can be larger by a factor $\sim 10$ \citep{lehner09}. This would lead to acceleration more analogous to the `standard' $\chi \sim 100$ case; the atomic gas potentially protects the dust and molecules from exposure to hotter gas. 

For highly supersonic winds, as expected in starburst galaxies, the cloud develops a strong bow shock; it therefore interacts with higher pressure post-shock gas. Simulations of high Mach number winds show tails and entrainment times which are longer by a factor $\sim (1+ \mathcal{M})$ \citep{scannapieco15,bustard22}. This can be attributed to the compression by an oblique shock, so that $\chi \rightarrow \chi (1+ \mathcal{M})$, which also produces a streamwise pressure gradient \citep{scannapieco15}. One important outstanding issue is convergence: 3D simulations \citep{gronke20-cloud} are not converged at high Mach numbers; only 2.5D simulations\footnote{Such simulations only have a limited number of cells and periodic boundary conditions in the `0.5' direction.} are well-converged \citep{bustard22}. This is not understood, though it may be related to suppression of lateral instabilities in 2.5D. In the latter, growth times scale as $t_{\rm grow} \propto \mathcal{M}^3$ for high Mach numbers, and the survival criterion $t_{\rm cool,mix} < t_{\rm cc}$ remains surprisingly robust \citep{bustard22}. 

This points to a potentially important disconnect. The poster child for the galactic wind community is M82, where velocities are $\sim 2000 \, {\rm km \, s^{-1}}$, the expanding wind evolves rapidly, and the Mach number increases with radius (since $v \sim$const and the sound speed falls due to adiabatic cooling); for instance,  $\mathcal{M} \sim 2-8$ is possible, though the lower end of the range is likely relevant if most clouds are launched at small radii. Still, the very hot winds at the launch radius ($T\sim 10^8$K) also imply larger overdensities ($\chi \sim 10^4$) than are often simulated. This contrasts with $\mathcal{M} \sim 1-2, \chi \sim 100-1000$ in typical idealized simulations, which is more appropriate for relatively quiescent CGM conditions\footnote{E.g., the infalling cold clouds simulated in Luminous Red Galaxy (LRG) hosts in TNG50, which condense hot gas, grow, and survive over cosmological timescales \citep{nelson20}.}. 
For strong starbursts, global simulations verify some aspects of the condensation picture, but do not agree in others. Using the GPU based Cholla hydrodynamics code \citep{schneider15} for uniquely high resolution ($\sim 5$ pc) global disk simulations of galaxy outflows over a large (20 kpc) domain, cold clouds were found to accelerate out to $\sim 800 {\rm km \, s^{-1}}$ \citep{schneider20}. They attributed this to mixing of hot gas, from the linear relationship between cool gas velocity and value of a passive scalar originally injected in wind material. As the value of this passive scalar in cold gas increases -- indicating hot wind gas which has cooled out -- the cold gas velocity rises. Specific cold gas momentum and energy fluxes increase monotonically with radius, as one might expect if cold gas is being entrained. At the same time, the cold gas mass flux is not monotonic -- it rises then falls, contradicting expectations of steady cloud growth in adiabatically expanding winds \citep{gronke20-cloud}, albeit for much lower overdensities and Mach numbers\footnote{Wind tunnels can mimic time-dependent winds by appropriately modified boundary conditions.}. 
An important clue is that \citet{schneider20} find cold gas to be under-pressured by a factor $\sim 10$ relative to the hot phase, unlike the pressure balance found in idealized simulations. While its origin is not understood--it could be an artifact of under-resolving cold clouds --  underpressured clouds have different cooling times and are vulnerable to cloud-crushing shocks. More work needs to be done in the computationally challenging high overdensity and Mach number regime appropriate for starburst winds, particularly in understanding convergence.

\subsubsection{Condensation: Gravity} \label{sec:gravity} Besides outflowing in a wind, cold clouds also fall onto the central galaxy under gravity. Such clouds could originate from thermal instability in the halo, cosmological accretion, or galactic fountain recycling \citep{fraternali17}. The question of whether such clouds can survive and grow is crucial for explaining observations of intermediate and high velocity clouds, as well as fueling of star formation in the disk \citep{putman12}. There is a crucial distinction between outflowing and infalling gas clouds. The latter gradually entrain, so destruction processes become weaker until the cloud comoves with the hot gas, and hydrodynamic instabilities are quenched. By contrast, infalling clouds {\it never} become comoving -- they accelerate under the action of gravity, and have to survive unrelenting shear at the terminal velocity. HVC simulations often simulate a wind tunnel, which ignores this distinction, though some simulation include gravity \citep{heitsch09}, and find cloud growth \citep{heitsch22,gronnow22}. Condensation in the cloud's wake means that it is largely composed of cooled halo gas; the contamination increases linearly with time and from head to tail, impacting cloud metallicities \citep{heitsch22}. However, the survival criterion is more stringent \citep{tan22-gravity-subm}: $t_{\rm grow} \sim \chi r/v_{\rm mix} < t_{\rm cc}$ (rather than $t_{\rm cool,mix} < t_{\rm cc}$). Survival is only weakly sensitive to cloud size. Instead, it is sensitive to cooling; a critical pressure is required for growth and accretion induced braking. In our Galaxy, this corresponds to a distance $\sim 10$ kpc from the disk; above that, clouds should fall ballistically, and must be large enough to reach the `zone of survival'. The drag due to accretion-induced mixing $F_{\rm drag} \sim \dot{m} v$ balances gravity $F_{\rm grav} \sim m g$ when $v_{\rm T, grow} \sim g m/\dot{m} \sim g t_{\rm grow}$, a terminal velocity which can be substantially smaller than traditional terminal velocities due to hydrodynamic ram pressure $v_{\rm T, hydro} \sim \sqrt{2 R \chi g}$. Compared to the virial velocity $v_{\rm vir} \sim g t_{\rm ff}$, we have $v_{\rm T,grow} \sim v_{\rm vir} (t_{\rm grow}/t_{\rm ff}) < 1$, i.e., infall is sub-virial \citep{tan22-gravity-subm}. Sub-virial infall velocities are commonly observed in LRGs \citep{huang16,zahedy19} and galaxy clusters \citep{russell16}. These velocities are much lower than predicted terminal velocities from traditional ram-pressure drag, unless the `drop height' of colds is fine-tuned \citep{lim08}. They are in much better agreement with velocities from accretion-induced drag \citep{tan22-gravity-subm}.

\subsubsection{Condensation: Cold Streams} An important scenario for cold gas survival is supersonic cold streams in $M_{\rm halo} > 10^{12} \, {\rm M_{\odot}}$ halos at $z \gsim 2$, which are potentially crucial for gas accretion onto central galaxies (\S\ref{sec:cold_streams}). They typically have Mach numbers $\mathcal{M} \sim 0.7-2.3$, density contrasts $\chi \sim 30-300$ with respect to the hot gas, and stream radii $\sim 0.03-0.3$ times the virial radius \citep{Mandelker20_streams_radiative}. Since they are poorly resolved in cosmological simulations, idealized simulations play an important role. Their geometry gives them properties intermediate between planar shearing layers (\S\ref{sec:TML}) and finite sized clouds, supporting both surface and body modes. Surface modes are present only for $\mathcal{M} < \mathcal{M}_{\rm crit} = (1+ \chi^{-1/3})^{3/2}$ (\citealt{chandrasekhar61}; e.g., $\mathcal{M}_{\rm crit} \approx 1.3$ for $\chi=100$). Heuristically, pressure perturbations which drive the KH instability operate on a sonic time, which exceeds the flow time for supersonic flows. However, for slabs or cylinders of finite thickness, body modes -- waves reverberating between stream boundaries -- can destabilize the system, for $\mathcal{M} > 1$. Thus, the system is always unstable, and adiabatic simulations find that streams disrupt and expand; thinner streams will not make it to the halo center \citep{mandelker16,padnos18,mandelker19}. As with clouds, magnetic fields slow down but do not halt disruption \citep{berlok19}. However, in the presence of radiative cooling, streams which satisfy $t_{\rm cool, mix} < t_{\rm shear}$\footnote{Here, $t_{\rm shear}$ is defined to be the time in which the mixing layer expands to the width of the stream in the non-radiative case. For CGM parameters, this typically gives timescales comparable to the cloud-crushing time $t_{\rm cc}  \sim \chi^{1/2} R/v$, though in general it differs.}, or equivalently which are larger than a critical radius $r > r_{\rm crit}$, survive and grow in mass, analogous to the cloud case \citep{Mandelker20_streams_radiative}. 

Current idealized simulations assume an infinite stream in a uniform background; stream evolution as it encounters radially varying CGM gas has only been modeled analytically \citep{Mandelker2020_Lya}. More importantly, current setups include initial velocity shear but no gravity. Thus, streams decelerate as they grow. In reality, since they are falling under gravity, they should face unremitting shear at some terminal velocity, similar to infalling clouds (\S\ref{sec:gravity}). Indeed, unlike idealized simulations, cosmological simulations find that streams reach a roughly constant terminal velocity, in strong disagreement with free-fall velocity profiles \citep{dekel09,goerdt15}, although the role of numerical viscosity needs to be carefully assessed. The lack of deceleration could change the survival criterion for streams, as it did for clouds. 

\subsection{Cold Gas Morphology}
\label{sec:morphology}

\begin{figure}
  \includegraphics[width=\textwidth]{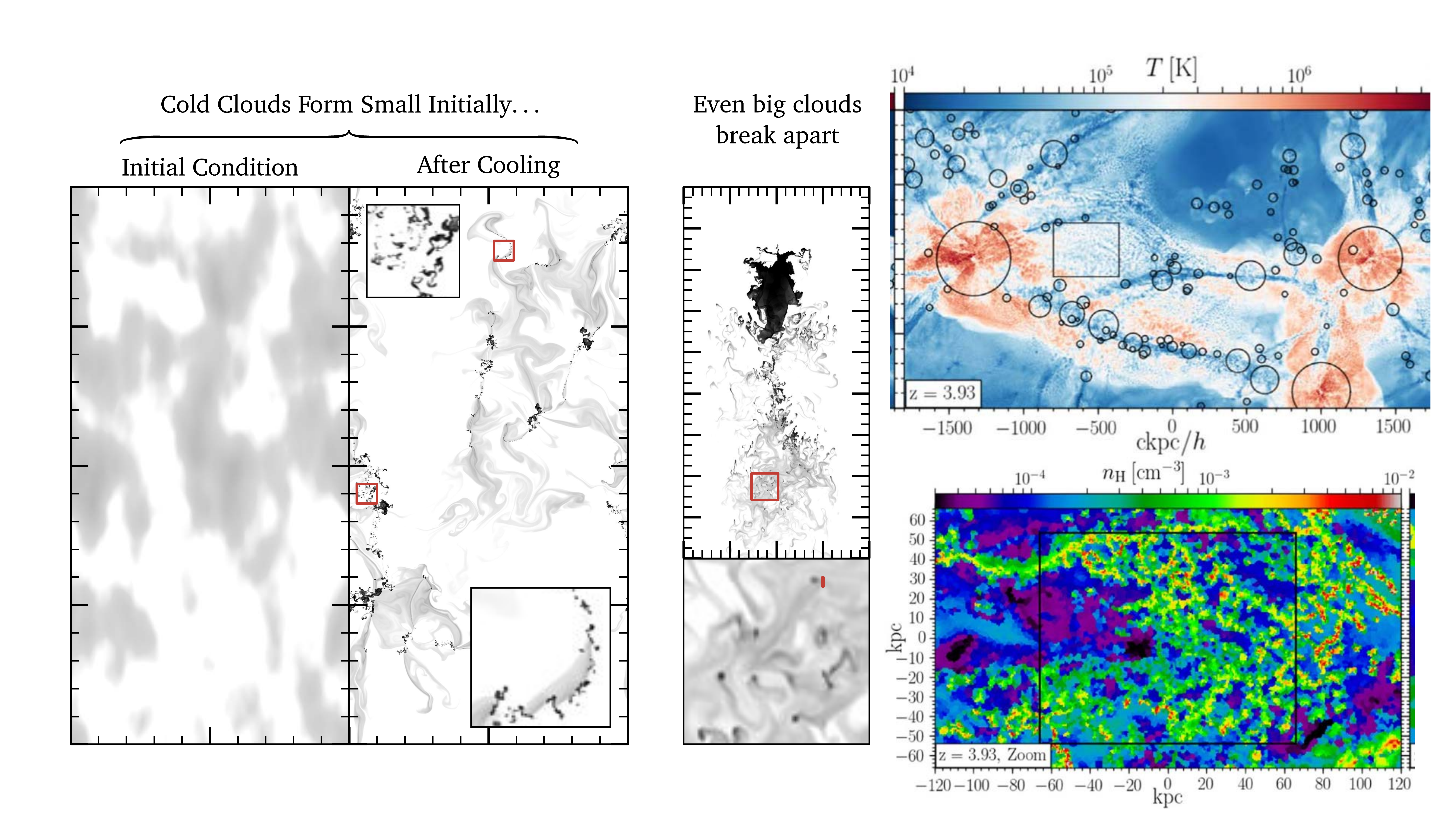}
  \parbox[c]{\linewidth}{%
    \caption{
      \small{Shattering via thermal instability and mixing. \textit{Left}: An initial perturbation
      cooling down to $10^4$\,K.  When the domain size is large
      ($\sim$\,kpc), the perturbation shatters into much smaller
      fragments with a characteristic scale
     $c_{\rm s} t_{\rm cool} \sim0.1$\,pc. \textit{Middle}: a
      $\sim10^{4}\,$ K cold cloud moving through an ambient medium at
      $\sim10^{7}$\,K.  The bottom shows a zoom-in of the red
      box in the top panel; even this apparently empty region of the
      wake is full of tiny cloudlets of cold
      gas. Adapted from \citet{McCourt18}. \textit{Right}: Shattering on large scales. The top shows temperature map of merging halos at $z\sim4$. The strong merger shock leads to thermal instabilities in the post-shock region, as in the rectangular box. Zooming in on this box (bottom) shows that cold gas ($T\sim 2 \times 10^4$ K) in the sheet has shattered into $\sim$kpc scale fragments, embedded in $T \sim 10^6$ K gas. Adapted from \citet{Mandelker19_shattering_sheet}. \copyright AAS. Reproduced with permission.}
      \label{fig:shattering}
    }}
\end{figure}

\subsubsection{Why Does Cold Gas Morphology Matter?} \label{subsect:morphology-matters} Is there a characteristic scale for cold gas, and why should we care? As we have already seen, size affects the survival and entrainment of cold gas; only clouds with $r > r_{\rm crit} \approx v t_{\rm cool,mix}/\sqrt{\chi}$ are able to withstand hydrodynamic instabilities in a wind. Cloud size also dictates mass growth rates, and the impact of thermal conduction (\S\ref{sec:survival}). It affects the interpretation of a host of observables, ranging from column densities, kinematics, cold gas mass fractions, and radiative transfer, in large part because cloud size sets the ratio of volume to surface area. For instance, cloud size/column densities affect ionizing and Ly$\alpha$ photon escape (see \S\ref{sec:cold_streams} for a discussion of spatially extended Ly$\alpha$ emission). The interpretation of kinematic line profiles changes if absorption comes from discrete `clouds', or a `fog' of droplets entrained in hot gas: non-thermal broadening would either trace small-scale cold gas turbulence or large-scale hot gas kinematics respectively. On a pragmatic note, a characteristic scale potentially sets the resolution at which numerical simulations should converge. 

In recent years, growing observational evidence for small-scale structure in CGM cold gas has emerged. From photoionization modeling, maximum cloud sizes $l \sim N_{\rm H}/n \sim 35\,$pc are inferred from observations of the CGM of $z\sim2-3$ galaxies \citep{hennawi15,lau16}; the cloud size could be smaller if the observed column is the result of intersecting many ($f_A > 1$) cloudlets along the line of sight. Another argument comes from the surprisingly large area covering factor $f_{\rm A} \sim \mathcal{O}(1)$ of CGM cold gas inferred from both absorption and emission line studies, despite the high overdensities ($\chi \sim 10^2-10^3$; \citealt{hennawi15}). Even if the mass fraction of cold gas is large $f_{\rm M} \sim \mathcal{O}(1)$, its volume fraction $f_{\rm V} \sim f_{\rm M}/\chi \sim 10^{-3}$ is tiny, and it is hard to explain its apparent abundance, unless it is somehow arranged in a thin shell. \citet{McCourt18} argued that just like terrestrial fog, this becomes explicable if cold gas is widely dispersed in tiny droplets\footnote{These arguments only apply to high overdensity gas, as suggested by photoionization modeling. For low overdensity gas, as reported in $z \sim 0.5$ COS observations \citep{werk14}, the area covering factor of large, low density clouds can be substantial. While these low densities -- about an order of magnitude lower than expected -- is still a puzzle, non-thermal pressure support could play a role. Indeed, such low densities and high covering fractions are seen in simulations where magnetic support dominates \citep{nelson20}.}: if there are $N$ droplets of size $l$ in a halo of radius $R$, $f_{\rm A} \sim N l^2/R^2$, $f_{\rm V} \sim N l^3/R^3$, so $f_A/f_V \sim R/l \gg 1$. A fog can also explain large, highly suprathermal line widths (e.g., $\sigma \sim 1000 {\rm km \, s^{-1}}$ in CII absorption; \citealt{hennawi15}). If this reflects turbulence within the cloud, they should be torn apart, but for a `mist', it just reflects background hot gas motions. Very small scale structure is also observed in HVCs, quasar BLR, BAL regions (see references in \citealt{McCourt18}), and also in the ISM \citep{stanimirovic18}. A `fog' could explain the surprising success of the uniform slab model in explaining Ly$\alpha$ spectra from galaxies \citep{gronke16,gronke17}. Radio scintillation could potentially probe the `mist' in the CGM \citep{vedantham19}, just as it probes small-scale density fluctuations in the ISM \citep{armstrong95}. 

\subsubsection{Fragmentation and Coagulation: Physical Processes} \label{sec:frag-coag} How can gas fragment to small scales? There are at least two possible mechanisms: (i) if the cooling time $t_{\rm cool}$ falls far below the sound-crossing time $t_{\rm sc} \sim R/c_s$ in a cooling cloud, causing it to become strongly underpressured relative to surrounding hot gas, $P_{\rm cloud} \ll P_{\rm hot}$, the cloud-crushing shock can `shatter' the cloud \citep{McCourt18}. These authors argued that similar to the Jeans instability, where gravitational fragmentation leads to a characteristic scale $\lambda_{\rm J} \sim c_{\rm s} t_{\rm ff}$, fragmentation in cooling, pressure-confined clouds imprints a scale \rcloudlet $\sim c_{\rm s} t_{\rm cool} \sim 0.1 \, {\rm pc} \, (n/{\rm cm^{-3}})^{-1}$, where the quantity $c_{\rm s} t_{\rm cool}$ is evaluated at its minimum at $T \sim 10^{4}$K. While this length scale depends on the ambient pressure confining the clouds, the column density through an individual fragment $N_{\rm cloudlet} \sim 10^{17} {\rm cm^{-2}}$ is essentially independent of environment.
Cooling of large ($R\gg c_{\rm s} t_{\rm cool}$) clouds which subsequently `shatter' is easily triggered by large non-linear perturbations (e.g. via turbulence). Shocks are particularly effective at `shattering' cold gas, either via compression of pre-existing cold gas \citep{mellema02}, or fragmentation in the radiative shock \citep{Mandelker19_shattering_sheet}.  
Shattering can also arise during initially isobaric linear thermal instability\footnote{By contrast, and somewhat counterintuitively, linearly unstable isochoric modes do not appear to break up into small pieces. In 1D simulations, such clouds are instead compressed to high density, and oscillate until they regain pressure balance \citep{waters19-linear,das21}, though this needs to be verified in 3D simulations.}, once the cloud loses sonic contact with its surroundings as it cools \citep{gronke20-mist,das21}.   (ii) Another avenue for producing small cloudlets is via mixing in Kelvin-Helmholtz instabilities; the tails of clouds in winds show a plethora of dense small clumps \citep{cooper09,McCourt18,sparre19}, which have a column density distribution which peaks at ${\rm N}_{\rm cloudlet} \sim 10^{17} {\rm cm^{-2}}$ \citep{LR20_mist}, corresponding to the characteristic scale $l \sim c_s t_{\rm cool}$ suggested for cooling driven fragmentation. Figure \ref{fig:shattering} shows fragmentation by these two respective processes. `Shattering' due to thermal instability in halo merger shocks can produce very low metallicity Lyman-limit systems, far from any galaxy \citep{Mandelker19_shattering_sheet}.  Numerical resolution obviously affects the onset and efficiency of shattering, but it can even affect the multi-phase nature of gas. \citet{mandelker21_whim} find that if $c_{\rm s} t_{\rm cool}$ at $T \sim 10^5$K is unresolved, gas `piles up' at $T\sim 10^5$K as further cooling becomes inefficient.

In detail, there are subtleties. 
In idealized simulations, `shattering' does not appear to take the form of hierarchical fragmentation, as in the Jeans instability. 
Instead, the cloud is strongly compressed by its surroundings, overshoots, then `explodes' into small pieces in a rarefaction wave \citep{mellema02,waters19-linear,gronke20-mist}. For break up, the density perturbation must be highly non-linear ($\delta \rho/\rho \gg 1$) before losing sonic contact with its surroundings. Strong density inhomogeneities, attributed to Raleigh-Taylor instabilities \citep{gronke20-mist} or small scale isobaric thermal instability\footnote{\citet{das21} argue from 1D simulations that clouds must be isobarically unstable but isochorically stable for fragmentation (this only happens in narrow temperature ranges; see \ref{sec:TI}). Otherwise, small scale isobaric modes within the cloud are suppressed by compression of the underpressured cloud. This interesting hypothesis needs to be tested in 3D. They found this to be true only for linear density perturbations; the distinction between isochoric and isobaric stability is irrelevant for non-linear perturbations.} \citep{das21} are crucial to fragmentation: an overpressured, uniform cloud does not fragment. Crucially, break-up is only seen when the final overdensity of the cloud is $\chi_{\rm f} \gsim 300$, otherwise the cloud recoagulates \citep{gronke20-mist}. While a density threshold is intuitive, the value $\chi_{\rm f} \gsim 300$ is not understood from first principles. If this high density threshold is robust, break-up via cloud crushing instabilities only happens in very hot winds and ICM-like conditions ($T_{\rm h} \sim 10^{7}$K, $\chi \sim 1000$), rather than the CGM ($T_{\rm h} \sim 10^{6}$K, $\chi \sim 100$), unless background turbulence can disperse droplets before they recoagulate.   
As for the production of cloudlets via mixing, this is only important for clouds in the destruction regime. There, given enough time, the entire cold gas mass (including the cloudlets) would mix into the hot gas. For larger clouds in the entrainment/survival regime, after an initial period of fragmentation, cloudlets in laminar winds are `focused' onto the cometary tail, and the cloud remains monolithic. 
Indeed, there appears to be a fundamental contradiction between `clouds' (which have to be large, $r > r_{\rm crit} \sim 10 c_s t_{\rm cool}$ to survive) and a `mist' of tiny clouds. In simulations, cold gas in a mist only survives by coalescing to form larger clouds. 

Thus, one cannot neglect coagulation, which competes with fragmentation. There are at least two kinds of coagulation. The most obvious is direct collisions, whereby two colliding cloudlets stick to one another, similar to dust grain growth in protoplanetary disks. It is heavily influenced by turbulence. While there is an enormous literature on turbulent coagulation (beginning with the seminal paper by \citealt{saffman56}), given its relevance to terrestrial cloud formation, there have not been studies for CGM-like conditions; more work is needed. A second mechanism is coagulation due to the advective flow generated by hot gas condensing onto a cold cloud \citep{elphick91,elphick92,koyama04,waters19-merge}. The inflow velocities in these studies was set by thermal conduction (or numerical diffusion). If inflow is instead set by hot gas cooling in turbulent mixing layers, from mass conservation the inflow velocity is $v_{\rm in} \sim v_{\rm mix} (r/r_{\rm cl})^{-\alpha}$, where $\alpha \approx 0,1,2$ for plane-parallel (e.g., a semi-infinite turbulent mixing layer), cylindrical (e.g., the cometary tail behind a cloud), and spherical geometry respectively. Cloudlets feel a gentle `breeze' from this advective flow, and entrain, just as in standard  wind tunnels \citep{gronke22-coag}. At first blush, it seems this should be utterly negligible: since $v_{\rm mix} \sim c_{\rm s,c}$, the corresponding Mach number is $ 
{\mathcal M} \sim ({v_{\rm mix}}/{c_{\rm s,h}}) \left( {r_{\rm cl}}/{d}\right)^{2} \lsim 10^{-2}$, which should be easily overwhelmed by turbulence. Yet, it is undeniably observed in simulations: cloud debris coagulates onto a cometary tail in a wind, and small droplets coagulate into clouds. The geometric dilution can be offset by a large increase in surface area\footnote{There is an interesting analogy between the coagulation force, $F_{12} \sim \rho v_{\rm mix}^2 A_1 A_2/(4 \pi r^2)$, and gravitational forces $F_{12} \sim G M_1 M_2/r^2$ \citep{gronke22-coag}. While both have a $1/r^2$ scaling, the former can be enhanced by breaking up into small droplets, which greatly increases the surface area and speeds up coagulation. By contrast, mass is conserved under fragmentation.}, and when the mass fraction of cold gas is high, increasing inertia, the local rms velocity dispersion can be comparable to $c_{\rm s,c}$. 

\subsubsection{Multi-Phase Turbulence: Structure on all Scales?} Individual clouds, and `mist' in particular, do not have to be long-lived; they can be continuously created and destroyed, {\it if} there is a long-lasting supply of cold gas. This appears to be what happens in turbulent flows, where large growing clouds continually shed small clouds, which then mix into the hot medium \citep{gronke22_survival}. 
Crudely, cold gas condenses at large scales, which then cascades to small scales. The survival criterion $r > r_{\rm crit}$ is similar to wind tunnel simulations, although turbulence makes survival close to $r_{\rm crit}$ highly stochastic; only clouds with $r \gg r_{\rm crit}$ are assured of survival. The velocity structure function of cold and hot gas are similar, which implies cold gas can be used to trace hot gas kinematics. The mass spectrum is a scale-free power law, $dn/dm \propto m^{-2}$, with roughly equal mass per logarithmic interval; similar power-law tails are also seen in other contexts \citep{li14a}. The entire system grows exponentially, on a timescale $t_{\rm grow} \sim \sqrt{(L/v_{\rm turb}) t_{\rm cool}}$ -- similar to that of turbulent mixing layers (\S\ref{sec:TML}; here, $L$ is the stirring length and $v_{\rm turb}$ is the rms turbulent velocity), for good reason: -- it is a TML writ large, with very similar physics, including self-similarity.  
While the small-scale `mist' and large-scale `clouds' have comparable mass, `clouds' dominate mass growth and `mist' dominates areal covering fractions -- and hence are much more likely to be observed along a random line of sight, in absorption line spectroscopy. Viewed in this light, the question of whether cold gas is `cloudy' or `foggy' is misguided. It is both; there is structure on all scales, and the most relevant scale depends on properties we probe (column density, mass, emission measure, covering fraction, etc). If the mass distribution is a roughly scale-free power-law, as in the stellar IMF, cut-offs (such as $r_{\rm crit}$ or $c_{\rm s} t_{\rm cool}$) can still reflect important physical processes. Just as for the IMF, the cloud-size distribution is set by a complex interplay between accretion, fragmentation and coalescence, except clouds are pressure confined rather than gravitationally confined. The origin of the IMF is still debated in the star formation community, and is by no means a settled question. Given that example, and weak observational constraints on cloud sizes, we likely have a long way to go towards a robust theory of cloud masses in the CGM community. For example, magnetic fields could substantially change cloud morphology and the mass PDF, as they do in laminar flows (\S\ref{sec:b-fields}). The interaction between turbulence, shocks and gravitational infall can produce a complex, rain-like morphology \citep{banda-barragan21}. 

\subsection{Cold Gas Interactions: Turbulent Mixing Layers}
\label{sec:TML}

\subsubsection{Physics of TMLs} Gas phases exchange mass, momentum and energy at phase boundaries, via microscopic (thermal conduction, viscosity) and macroscopic (turbulent) diffusive processes. The mean free path of turbulent eddies is much larger than electron or ion mean free paths. Thus, turbulent diffusion tends to be more efficient, as demonstrated in its role in stellar heat transport (turbulent convection), or momentum transport (turbulent viscosity) in accretion disks. At phase boundaries where there is velocity shear or differential acceleration, turbulence is driven by Kelvin-Helmholtz or Rayleigh-Taylor instabilities. In recent years, significant progress has been made in understanding turbulent mixing layers (TMLs), particularly the rich interplay between turbulence and radiative cooling, usually in idealized planar shearing setups \citep{kwak10, kwak11, henley12, kwak15, ji18, fielding20,tan21,yang22,chen22}. This physics lies at the heart of condensation, which enables cold gas survival and entrainment (\S\ref{sec:survival}). It has close parallels with turbulent combustion fronts \citep{zeldovich69, zeldovich85}. 

In the TML, hot and cold gas mix to form intermediate temperature gas, which subsequently cools. The net result is that hot gas is converted to cold gas. An early analytic paper suggested that TMLs are characterized by a temperature\footnote{\label{footnote:Tmix} We can derive this result from a slightly different viewpoint. Suppose the hot and cold gas have a turbulent velocities $v_{\rm h},v_{\rm c}$ respectively. The cold and hot phases share kinetic energy, $\rho_c v_c^{2} \sim \rho_{\rm h} v_{\rm h}^{2}$, so that $v_c \sim v_h/\chi^{1/2}$, where the density contrast $\chi = \rho_{\rm c}/\rho_{\rm h}$. TMLs are isobaric, $P_{\rm h} \sim P_{\rm c}$. This implies that whenever different phases mix, the hot phase dominates the enthalpy flux: $\dot{E}_{\rm h}/\dot{E}_{\rm c} \sim P_{\rm h} v_{\rm h}/P_{\rm c} v_{\rm c} \sim \chi^{1/2} \gg 1$. Conversely, the cold phase dominates the mass flux $\dot{m}_{\rm h}/\dot{m}_{\rm c} \sim \rho_{\rm h} v_{\rm h}/\rho_{\rm c} v_{\rm c} \sim \chi^{-1/2} \ll 1$. These ratios have been confirmed in numerical simulations \citep{ji19}. If hot gas provides enthalpy and cold gas provides heat capacity, mixing results in a mean temperature $\bar{T} \sim \dot{m}_{\rm h} T_{\rm h}/\dot{m}_{c} \sim \rho_{\rm h} v_{\rm h} T_{\rm h}/\rho_{\rm c} v_{\rm c} \sim (T_{\rm h} T_{\rm c})^{1/2}$.} $T_{\rm mix} \sim (T_{\rm h} T_{\rm c})^{1/2}$ and width $l \sim v_{\rm t} t_{\rm cool,mix}$, where $v_{\rm t}$ is the turbulent velocity and $t_{\rm cool,mix}$ is the cooling time of mixed gas \citep{begelman90}. Despite its crude formulation, $t_{\rm cool,mix}$ appears to determine thresholds for cold cloud survival (equation \ref{eq:rcrit}) and governs the boundary between single and multi-phase TMLs (equation \ref{eq:Da}). Suprisingly, however, the commonsense `cooling length' prescription $l \sim v  t_{\rm cool}$, which works well for radiative shocks, and predicts column densities $N \sim n l_{\rm cool}$ which do not depend on density (since $t_{\rm cool} \propto 1/n$) turns out to be incorrect for TMLs \citep{ji19}.  

\begin{figure*}
    \centering
    \includegraphics[width=\linewidth]{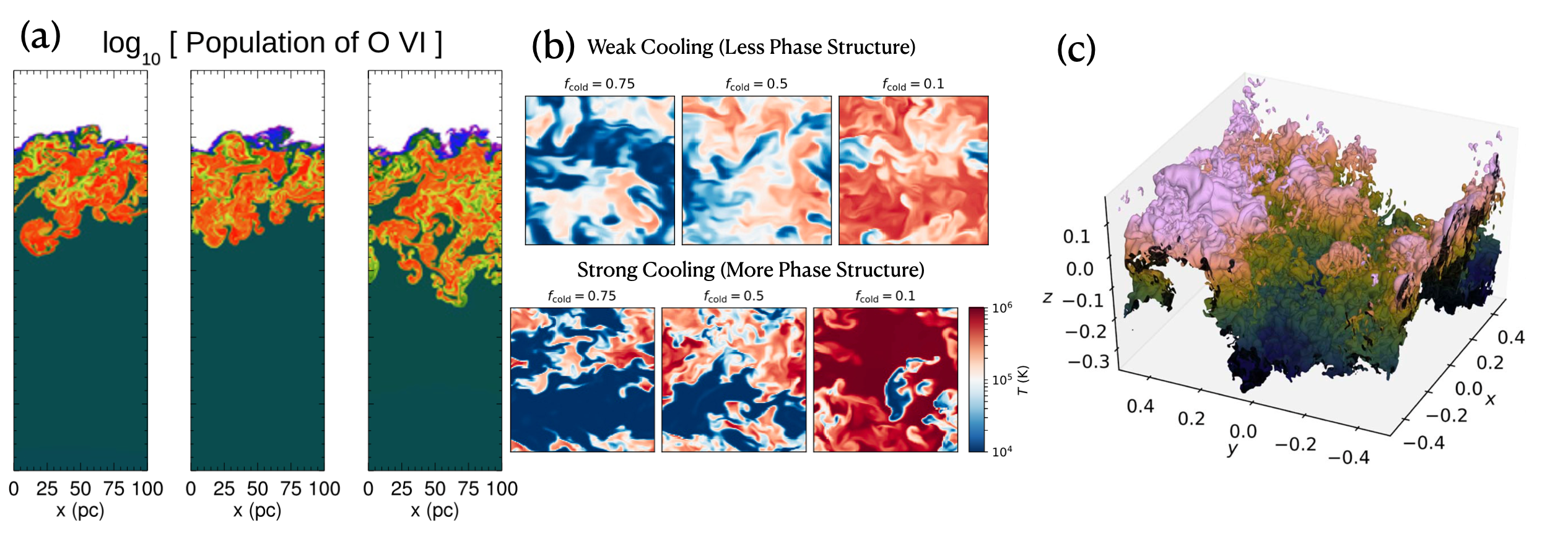} 
    \hspace{1cm}
    \caption{\small{Some properties of turbulent mixing layers (TMLs). (a) TMLs have abundant high ions, such as OVI. Adapted from \citet{kwak10}. \copyright AAS. Reproduced with permission.} (b) The \Da number (equation \ref{eq:Da}) determines if the TML is single-phase (Da $<$1; weak cooling regime) or multi-phase (Da$>1$; strong cooling regime). Adapted from \citet{tan21}. (c) In the fast cooling regime, the TML phase boundary has a highly wrinkled, fractal structure. Adapted from \citet{fielding20}. \copyright AAS. Reproduced with permission.}
    \label{fig:TML-properties}
\end{figure*}

A key quantity is the inflow velocity $v_{\rm mix}$ of hot gas into the mixing layer. This determines the mass flux $\dot{m} \sim \rho_{\rm h} v_{\rm mix}$ and enthalpy flux $\dot{e} \sim (5/2) P v_{\rm mix} [1 + \mathcal{M}^{2}]$ into the TML. The factor $\mathcal{M}^2$ represents kinetic energy of the hot gas, which is eventually thermalized. Since the inflowing hot gas mixes and cools, in steady state $\dot{m}$ gives the cold gas mass growth rate, and $\dot{e}$ gives the TML surface brightness. 
Recent work has found the following $v_{\rm mix}$ properties (some of which are shown in Fig \ref{fig:TML-properties}): 

\begin{itemize} 

\item{{\it TMLs are roughly isobaric.} Pressure fluctuations are small, and tend to decline with increasing resolution. 

\item{{\it The inflow velocity is of order the cold gas sound speed $v_{\rm mix} \sim c_{\rm s,c}$}. Thus, for $T_c \sim 10^{4}$ K, to order of magnitude $v_{\rm mix} \sim 10 \, {\rm km \, s^{-1}}$. This may seem surprising, since this velocity is low compared to typical shear velocities, $v_{\rm h} \sim 100-1000 \, {\rm km \, s^{-1}}$. The fundamental reason is that the inertia of the cold gas limits the rate at which the hot/cold interface deforms and the two phases mix. For a Kelvin-Helmholtz time $t_{\rm KH} \sim \sqrt{\chi} l/v_{\rm h}$, the mixing velocity is $v_{\rm mix} \sim l/t_{\rm KH} \sim v_{\rm h}/\sqrt{\chi} \sim {\mathcal M}_{\rm h} c_{\rm s,c}$. 
Of course, in detail, $v_{\rm mix}$ must also depend on cooling rates, as we discuss below.} 

\item{{\it The inflow velocity has different scalings in the fast ($v_{\rm mix} \propto t_{\rm cool}^{-1/4}$; multi-phase) or slow ($v_{\rm mix} \propto t_{\rm cool}^{-1/2}$; single phase) cooling regimes.} Just as $t_{\rm cool}/t_{\rm ff}$ determines whether a stratified medium is single or multi-phase, the \Da number \citep{damkohler40}: 
\begin{equation}
{\rm Da} = \frac{t_{\rm turb}}{t_{\rm cool}} = \frac{L}{v_{\rm t,c} t_{\rm cool}(T_{\rm mix})}, 
\label{eq:Da} 
\end{equation} 
where $L$ is the outer scale of turbulence, determines if the TML is single (Da$<1$) or multi-phase (Da$>1$) \citep{tan21}. Turbulent heat diffusion can be characterized by a diffusion coefficient $D_{\rm turb} \sim v_{\rm t,c} L$. For a single phase TML, $v_{\rm mix} \sim (D_{\rm turb}/t_{\rm cool})^{1/2} \propto t_{\rm cool}^{-1/2}$ (here, we abbreviate $t_{\rm cool,mix}$ with $t_{\rm cool}$). On the other hand, when the TML is multi-phase, 
only a small fraction of the volume is actively cooling. Most gas is either in the hot or cold stable phase; only their relative fractions change with depth in the TML. The effective cooling time is the geometric mean\footnote{This is a common random walk result: the effective optical depth for a photon in a medium with scattering and absorption optical depths $\tau_{\rm S},\tau_{\rm A}$ respectively is $\tau_{*} \sim \sqrt{\tau_{\rm S} \tau_{\rm A}}$. Another example is the Field length $\lambda_{\rm F} \sim \sqrt{\lambda_{\rm e} v_{\rm e} t_{\rm cool}}$, which is the effective mean free path of a thermal electron. It is set by the geometric mean of the electron's elastic ($\sim \lambda_{\rm e}$, set by Coulomb interactions) and inelastic ($\sim v_{\rm e} t_{\rm cool}$, where $v_{\rm e} \sim \sqrt{3 kT/m_e}$ is the thermal velocity) mean free paths.} of the cooling time and eddy turnover time $\tilde{\tau}_{\rm cool} \sim (t_{\rm eddy} t_{\rm cool})^{1/2}$; the expected scalings in emissivity $\tilde{\epsilon} \sim P/\tilde{\tau}_{\rm cool} \propto (v_{\rm  t,c}/t_{\rm cool})^{1/2}$ have been directly verified in numerical simulations \citep{tan21}. In this strong cooling case, which is most relevant to condensation (\S\ref{sec:entrainment}),
\begin{equation}
v_{\rm mix} \sim \left( \frac{D_{\rm turb}}{\tilde{\tau}_{\rm cool}} \right)^{1/2} \sim v_{\rm c,t}^{3/4} \left( \frac{L}{t_{\rm cool}} \right)^{1/4}.
\end{equation}
Since they scale with $v_{\rm mix}$, the mass growth rates and surface brightness follow the same $t_{\rm cool}^{-1/2},t_{\rm cool}^{-1/4}$ scalings. The fast cooling regime is the main regime of interest for us, since the criterion for cloud survival $t_{\rm cc}/t_{\rm cool,mix} > 1$  (\S\ref{sec:survival}) is equivalent to ${\rm Da}_{\rm mix} > 1$.}

\item{{\it Convergence in net cooling and mass transfer rates only requires resolving large-scale motions.} In the multi-phase strong cooling regime, the interface is highly wrinkled, greatly increasing the turbulent front surface area $A_{\rm T}$ from its laminar value $A_{\rm L}$. \citet{fielding20} showed from their simulations that it is a fractal, with 
${A_{\rm T}}/{A_{\rm L}} = ( {\lambda}/{L} )^{2-D}$,
where $\lambda$ is the smoothing scale and $D=2.5$ is the fractal dimension measured in their simulations.  
Since the area (and volume) of the cooling region is resolution dependent, one might expect the total cooling rate to be resolution dependent. However, simulation surface brightness and mass entrainment rates converge at surprisingly low resolution \citep{ji19,fielding20,tan21}, even when the cooling length $c_{\rm s} t_{\rm cool}$ is unresolved, and without explicit thermal conduction\footnote{If thermal conduction is implemented, it does not affect cooling rates and $v_{\rm mix}$, as long as $D_{\rm turb} \sim v_{\rm t,c} L > D_{\rm conduct} = \kappa/(\rho c_{\rm P})$ \citep{tan21}.}, when the cooling interface is one cell thick. Note that for many practical applications (e.g., cloud entrainment, \S\ref{sec:entrainment}), convergence in these quantities is equivalent to convergence in mass, momentum and energy transfer between phases. 

As \citet{tan21} argue, we are already familiar with the resolution of this conumdrum. When we stir cream into coffee, the mixing rate is independent of molecular diffusion rates, even though the latter is ultimately responsible for fine-grained mixing. In Kolmogorov turbulence, turbulent dissipation rates $\epsilon \approx \rho v^{3}/l$ are independent of viscosity. Microscopic diffusion (or viscous) times are extremely long at macroscopic scales. Instead, the cream (or kinetic energy) cascades from large to small scales, until the timescale for microscopic processes becomes short enough to take over. Similarly, turbulence increases the area of the fractal cold/hot interface until (numerical or physical) thermal diffusion takes over. However, the mixing rate, and thus overall cooling rate, is independent of the details of the fractal surface. Instead, as in the coffee cup, the rate of mixing is set by the eddy turnover time at the outer scale. {\it This} is the scale which needs to be resolved, which is relatively undemanding. Thus, for instance, cloud growth and entrainment is converged for clouds which are resolved by $\sim 8$ cells; the TML is completely unresolved \citep{gronke18_published}. 
}}
\end{itemize} 

These properties assume $\mathcal{M} \lsim 1$, hydrodynamic TMLs. Supersonic TMLs have been studied in the slow cooling regime \citep{yang22}. The TML separates into two zones: a Mach number independent zone (similar to what we have discussed), plus an expanding turbulent zone with large velocity dispersion. Turbulent dissipation dominates over enthalpy advection, and reverses the sign of mass flux: cold gas evaporates. In MHD simulations, B-fields amplify to quasi-equipartition with turbulence, at which point magnetic tension suppresses mixing \citep{ji19}. These effects potentially strongly curtail cold gas growth. However, due to the influence of finite size effects, they may be better examined in wind tunnel simulations of a macroscopic cloud or filament, rather than planar mixing layers. In supersonic flows, a standoff bow shock forms, and the cloud interacts with subsonic postshock gas (\S\ref{sec:wind-cloud}). Wind tunnel simulations with B-fields show cloud growth which is surprisingly similar between MHD and hydro simulations, unlike TML simulations (\S\ref{sec:b-fields}).

\subsubsection{TMLs: Observational Predictions} \label{sec:TML-observations} A potential observational diagnostic of TMLs are high ions like OVI, which peak at $T \sim 10^{5}$K in collisionally ionized gas. Such gas should have short cooling times ($\sim 10$ Myr in CGM conditions), but is nonetheless seen in abundance in QSO sightlines \citep{tumlinson11,prochaska11} and HVCs \citep{savage14}. High ions naturally arise in steady state TMLs. Moreover, the surprising alignment of velocity centroids in low and high ions \citep{tripp08,rudie19,haislmaier21} arises naturally in TMLs, since different portions of a mixing layer participate in the same large scale turbulence. 

However, predicted column densities of individual TMLs are significantly lower than observations, typically by two orders of magnitude \citep{slavin93,kwak10,ji19}. Non-equilibrium ionization \citep{kwak10,ji19} and photoionization \citep{ji19} can only increase column densities by moderate amounts. Thus, if TMLs are responsible for observations, sightlines must pierce hundreds of mixing layers-- which may be plausible if the CGM exists as a `fog' of tiny cloudlets (see \S\ref{sec:morphology}). 

Another frequently used discriminant between theoretical models (conductive interfaces, cooling flows, etc; e.g., \citealt{wakker12}) is observed lines ratios such as N(SiIV)/N(CIV) or N(CIV)/N(OVI). It is important to realize that line ratios are {\it not} uniquely predicted by TML simulations. Instead, they arise from the temperature dependence of thermal conduction, $\kappa(T)$. Although macroscopic quantities like mass, momentum and energy transfer rates between phases are independent of thermal conduction, line ratios depend explicitly on the temperature PDF $P(T)$ of the TML, which depends on small-scale interface structure and hence {\it does} depend on thermal conduction \citep{tan21-lines}. Neglecting thermal conduction is equivalent to assuming temperature-independent thermal (numerical) diffusion. A simple 1D semi-analytic model for conduction fronts reproduces simulated temperature PDFs and line ratios for different conduction laws remarkably well\footnote{Just as conduction models probe the {\it fine-grained} thermodynamic temperature distribution, the {\it coarse-grained} mean temperature profile $\bar{T}(x) = f_{\rm c} (x) T_{\rm c} + (1-f_{c}(x) T_{\rm h}$, where $f_c(x)$ is the spatially varying mass fraction of cold gas, can be reproduced by a mixing length model \citep{tan21,chen22}.} \citep{tan21-lines}.

\subsection{Cold Gas Interactions: Cosmic Rays}
\label{sec:CR}

Although they only represent a billionth of all particles, cosmic rays have an energy density comparable to turbulence, thermal gas and magnetic fields in the Milky Way ISM \citep{grenier15}. They are accelerated primarily at supernova shocks, where $\sim 10\%$ of supernova kinetic energy is converted to cosmic rays, although secondary acceleration in the halo via termination shocks \citep{jokipii87,dorfi12,bustard17} or turbulence \citep{drury17,bustard22-accel} is also possible. Since Coulomb interaction timescales are long, radiative losses by electrons do not affect CR protons. Thus, CRs are unaffected by radiative cooling. CR scatter orders of magnitudes more than photons (e.g., $\sim 10^{6}$ times before leaving our Galaxy), enforcing tight coupling and making them a better candidate for mediating feedback \citep{socrates08}. Although the possibility of CR driven winds was noted early on \citep{ipavich75,breitschwerdt91}, in recent years there has been a veritable explosion of work on CR driven winds (e.g., \citealt{everett08,uhlig12,booth13,salem14,pakmor16,zweibel17,wiener17-transport,ruszkowski17, mao18,chan19, buck20, quataert22-diffusion}).  
If the halo becomes CR dominated, the pressure support provided by CRs can potentially allow the cool, photoionized ($T \sim 10^4$K) phase to become volume-filling, with abundant OVI \citep{ji20}, and suppression of virial shocks \citep{ji21}. Even if the gas remains multi-phase, CR pressure support in cold clouds reduces their density and increases buoyancy, increasing free-fall times \citep{butsky20} and altering kinematic absorption-line signatures \citep{butsky22}. Heating by CRs can also offset radiative cooling, though this has mostly been modeled in the ICM \citep{guo08-CR,jacob17a,ruszkowski17}. Regrettably, this body of work is simply too enormous to survey here. Instead, we briefly review CR hydrodynamics. Then, in keeping with the theme of this section (\S\ref{sec:small_scales}), we consider the specific impact of multi-phase gas structure on CR transport. 

\subsubsection{CR Hydrodynamics}
Since they travel at the speed of light, CRs should zip across our Galaxy in $\sim 30,000$ years. Instead, spallation and radioactive decay products indicate CRs have residence times a thousand times longer. CRs originate in discrete, transient sources (supernovae) and should be highly directional on the sky. Instead, CRs are remarkably isotropic, to 1 part in $\sim 10^{4}$ (at $\sim$GeV energies, where the energy density peaks). These observations make sense if the galaxy is `optically thick' to CRs, so that they scatter frequently ($\sim 10^{6}$ times before leaving, with $\sim$pc mean free paths) and slowly random walk out. Scattering slows the bulk CR propagation speed, generating macroscopic CR pressure gradients that push on the gas. 

Magnetic irregularities on scales of order the CR gyroradius ($\sim 1$ AU for a GeV CR) can resonantly scatter CRs. These perturbations can originate from two sources: extrinsic turbulence, whereby the fluctuations are part of an externally driven cascade, or from the CRs themselves (`self-confinement'), which can amplify Alfv\'en waves through a resonant streaming instability \citep{kulsrud69,wentzel74}. While turbulent scattering could be important for high energy CRs, theory and observations alike point to a shift towards self-confinement for CRs with $E < 300$GeV \citep{amato18}. The CR energy density $E_c$, which determines CR influence on surroundings, is dominated by low energy ($\sim$GeV) CRs at the peak of the energy spectrum.  For this reason, we focus on the physics of self-confined CRs. We emphasize that CR transport is still quite uncertain. For instance, there are troubling discrepancies between `standard' models of CR transport and observations in the Milky Way \citep{kempski22,hopkins21-CR-problems}, though these mostly arises at higher energies. The issue is similar to that with conduction: CR scattering rates\footnote{While these are usually calculated in quasi-linear theory, there has been recent encouraging progress in calculating scattering rates in Particle-in-Cell and hybrid codes \citep{bai19,holcomb19,bai22}.} remain uncertain. Particularly important is the influence of chaotic small-scale tangled B-fields, known as `Field Line Wandering' (FLW); see \citet{mertsch20} for a recent review. This renders CR transport effectively diffusive, or even superdiffusive \citep{yan04,sampson22}, even if CRs `stream' along field lines. Diffusive CR transport can be viewed as a subgrid model for this tangled field structure. We describe canonical `cosmic ray hydrodynamics' \citep{skilling71,zweibel17}, but it is on shakier ground than say, MHD.

When tightly self-confined or ``coupled", CRs are trapped by the scattering Alfv\'en waves, which reach an equilibrium amplitude (e.g., $\delta B/B \sim 10^{-3}$ for Galactic parameters; see \citealt{farmer04,wiener13a}) between growth and damping processes.\footnote{Sources of damping include non-linear Landau damping \citep{cesarsky81}, ion-neutral damping \citep{depontieu01}, turbulent damping \citep{farmer04}, and dust damping \citep{squire21}.} Their small mean free path renders their collective behavior fluid-like, and their bulk flow is described as a sum of advection with the gas, `streaming' relative to the gas with waves at the local Alfv\'en velocity ${\mathbf v}_{\rm A}$, and a second order correction from CR diffusion\footnote{Although this is also referred to as `diffusion', it is distinct from diffusion due to FLW. } relative to the wave frame, due to the finite scattering rate \citep{skilling71}. Streaming transport only proceeds {\it down} the CR gradient (i.e., to regions with lower CR energy density $E_{\rm c}$) along magnetic field lines. CRs propagating up their gradient damp the streaming instability, and do not couple to the gas.

CRs also \emph{push} the gas with a force $\nabla P_{\rm c}$ and \emph{heat} the gas at a rate $v_{A} \cdot \nabla P_{\rm c}$. These momentum and energy transfer rates can be considerable, but they also depend sensitively on local plasma conditions. This last point is important. One reason why CR feedback is considered so attractive is that it can act on large scales and suffers weaker losses than say, thermal gas. This assumption, which generally comes from low-resolution galaxy-scale simulations, or when CR streaming is neglected, deserves continued scrutiny in higher-resolution calculations. For instance, in the tight-coupling limit, CR pressure traces variations in velocity, density, and B-field strength \citep{breitschwerdt91}: 
\begin{equation}
P_{\rm c} \propto (v + v_{\rm A})^{-4/3} = (v + B/\sqrt{4\pi \rho})^{-4/3}. 
\label{eq:CR-adiabat}
\end{equation}
Small scale structure in these quantities can thus significantly modulate CR profiles. 

Complicating matters further, CRs are not always coupled to the gas! Coupling is weak if wave damping is strong, or if there are insufficient CRs to power the streaming instability. CRs also only scatter and couple to the gas if they are anisotropic. As with a radiation field, spatial anisotropy implies that CRs with energy density $E_{\rm c}$ have a net flux $F_{c}$, with some effective drift speed $v_{\rm D} \sim F_{c}/E_{c}$. For the streaming instability to be excited, the drift speed must exceed the local Alfv\'en velocity $v_{\rm A}$. In regions where the CRs are isotropic ($\nabla P_{\rm c}=0 $), or have small drift speed, $v_{\rm D} < v_{\rm A}$, CRs will not scatter; they decouple from the gas and free stream out of these `optically thin' regions at the speed of light. Effectively, CRs behave like a radiation field with a very unusual opacity, which depends on the anisotropy of the radiation field. This important property, which strongly influences CR transport and dynamics, has traditionally been difficult to simulate, as it leads to a rapidly growing grid-scale numerical instability, which can only be offset by regularization \citep{sharma10-stream}, which is numerically costly (quadratic timestep scaling, $\Delta t \propto (\Delta x)^2$, where $\Delta x$ is the grid cell size) . In recent years, two-moment methods adapted from radiative transfer, which solve time-dependent equations for both CR pressure $P_c$ and flux $F_c$, have allowed for fast, stable and accurate solution of CR hydrodynamics with $\Delta t \propto \Delta x$ scaling \citep{jiang18,thomas19,chan19}.

\subsubsection{CR transport in a multi-phase medium} 
\begin{figure}
  \includegraphics[width=\textwidth]{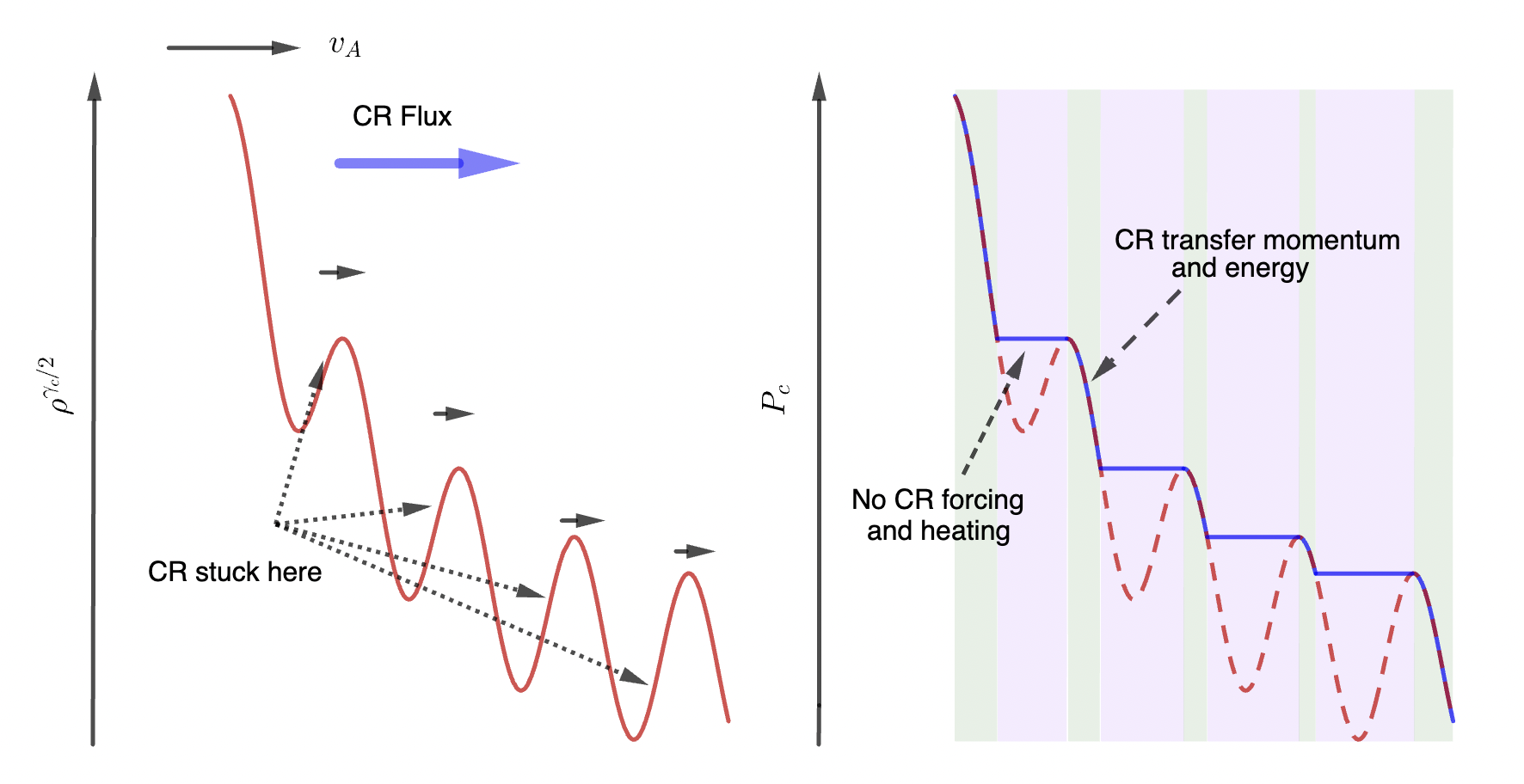}
  \parbox[c]{\linewidth}{%
    \caption{
      \small{The CR bottleneck effect. {\it Left:} in steady state, and if CR streaming dominates, $P_{\rm c} \propto (v_A + v)^{-\gamma_c} \propto \rho^{\gamma_c/2}$, where the last step assumes $v\ll v_{\rm A}$ and $B \approx$ const. However, in the presence of density bumps, this requires CRs stream up their gradient. {\it Right:} Instead, CRs `bottleneck' there, and produce a CR staircase structure consisting of plateaus, where CRs are decoupled, and sharp gradients, which undergo intense heating and forcing. Thus, density fluctuations (e.g., from multi-phase structure) drastically changes the spatial footprint of CRs. Figure credit: Navin Tsung.}
      \label{fig:staircase}
    }}
\end{figure}

The CR `bottleneck' is a dramatic example of how sharp density (or velocity) jumps introduce significant complexities into CR transport \citep{skilling71,begelman95,wiener17-cold-clouds}. Since $v_{\rm D} \sim v_A \propto \rho^{-1/2}$, a cloud of warm ($T\sim 10^{4}$K) ionized gas embedded in hot ($T\sim 10^{6}$K) gas results in a minimum in drift speed. This produces a `bottleneck' for the CRs: CR density is enhanced as CRs are forced to slow down, akin to a traffic jam. Since CRs cannot stream up a gradient, the system readjusts to a state where the CR profile is flat up to the minimum in $v_A$; thereafter the CR pressure falls again. If there are multiple bottlenecks, this produces a staircase structure in the CR profile (Fig \ref{fig:staircase}). A convex hull construction (connecting the highest peaks in $(v+v_A)^{-1}$ with horizontal ridgelines) agrees with numerical simulations \citep{tsung22-staircase}.  Importantly, since $\nabla P_c = 0$ in the plateaus, CRs there are no longer coupled to the gas, and can no longer exert pressure forces or heat the gas. Instead, momentum and energy deposition is focused at the CR `steps'. Small-scale density contrasts can thus have global influence on CR driving and heating. Staircase structures appear when CR streaming dominates over advection or diffusion. They are seen in 1D simulations of CR driven winds \citep{quataert22-streaming}, CR driven acoustic instability\footnote{This is an instability in low $\beta$ plasmas where CRs drive sound waves unstable, creating a series of closely spaced weak shocks \citep{begelman94}. The density jump at the shock serves as a propagating bottleneck.} \citep{tsung22-staircase}, and in 2D/3D simulations of thermal instability \citep{tsung22-TI} and cloud acceleration (see below). The importance of bottlenecks depends on the covering fraction of overdense gas, and whether they are threaded by B-field lines.  

Cosmic rays can accelerate cold clouds by exerting a direct force. Recall that $P_c (v_A + v)^{\gamma_c}$ is conserved in the tight-coupling limit (equation \ref{eq:CR-adiabat}). Suppose $v_{\rm A} \gg v$, so that CR streaming dominates. As the Alfven velocity drops within the cloud, CRs bottleneck and build up. The resulting steep pressure gradient at the cloud interface accelerates the cloud \citep{wiener17-cold-clouds,wiener19,bruggen20,bustard21,huang22}. Simulations (mostly in 2D) have found efficient acceleration (although it must be extrapolated to observed velocities), and that radiative cooling wards off the destructive effects of CR heating. In detail, there are caveats. In warm ($T \sim 10^4$K) clouds, the CRs exert their force on the {\it rear} end, which is the final bottleneck where they accumulate. Thus, they stretch the cloud. Depending on field strength, this differential acceleration can shred the cloud \citep{huang22}. This mechanism requires field lines to thread the cloud, which does happen for clouds which form via thermal instability \citep{huang22}, but does not happen for clouds which undergo magnetic draping. All simulations adopt a setup where the hot and cold gas are initially at rest. The bottleneck diminishes as their relative velocity $v$ increases, and goes to zero if $v$ is super-Alfvenic (relative to the hot gas Alfven speed, $v_A \sim c_{s,h}/\sqrt{\beta}$). Thus, this mechanism alone cannot accelerate clouds significantly in a high $\beta$ plasma. Still, even if CRs accelerate the hot gas alone, the latter could transfer momentum to the cold gas, producing indirect acceleration. CRs will pressurize and broaden the interface; \citet{huang22} find that the reduction in cooling suppresses cool gas growth. 

Self-shielding and modulation of ionized fraction can also impact CR coupling. Ion-neutral damping attenuates the MHD waves which couple CR to the gas, and increase CR diffusion \citep{farber18,bustard21}. Also, since $v_A = B/(4 \pi \rho_i)^{1/2}$, where $\rho_i$ is the {\it ion} density, the Alfven speed can rise, rather than fall, in the interiors of self-shielded clouds. \citet{bustard21} find that this varying Alfven speed profile creates bottlenecks at both the cloud front and back, with modest effects on cloud acceleration compared to the single bottleneck case.

\subsubsection{Outlook} This is fast-moving field; our discussion barely scrapes the surface. The primary challenge is to settle on the correct equations for CR transport. At present, the standard procedure is to consider limits where either CR streaming or diffusion dominates, using constant diffusion coefficients calibrated to Galactic constraints. In reality, the latter almost certainly vary with environment. Researchers are starting to use models with diffusion coefficients calculated from quasi-linear theory; differing assumptions (e.g., for wave damping), compounded by non-linear effects, can result in very different outcomes in the CGM \citep{hopkins21-CR-test}. For instance, the low CGM B-fields (and hence Alfven speeds) in FIRE simulations mean that CR streaming is unimportant; strong CR diffusion {\it must} take place to achieve sufficiently rapid CR transport to satisfy $\gamma$-ray constraints \citep{chan19}, but this may not be true in simulations with stronger B-fields (e.g., \citealt{voort21}). Even if diffusion is simply an effective subgrid model for field line wandering, how it changes with turbulence and plasma parameters, and the rate of cross-field transport, is still unclear. FLW is usually studied in the test particle approximation in a static tangled B-field, which ignores the backreaction of CRs (from CR forces and heating) on B-field structure, evolution of the B-field due to turbulence, and the constraint that self-confined CRs can only propagate down their gradient.  
Even with the current set of equations, unresolved structure (such as multi-phase gas) in simulations can significantly change outcomes, as we have seen in our discussion of bottlenecks, and this may also have to be handled in a subgrid manner. Finally, increased contact with observations is crucial. CRs are currently constrained by non-thermal radio (synchrotron emission from cosmic rays electrons) and gamma-ray emission (from pions produced by CR protons); the latter more directly constrain CR energy densities. In starburst galaxies, there are well-known correlations between radio and gamma-ray luminosities and far-infrared luminosity, which probe the star formation rate \citep{lacki10,abdo10a,abdo10b,ackermann12}. However, these generally probe conditions at the base of the wind rather than the CGM, and constraints on CR transport suffer from degeneracies (e.g., between CR diffusion coefficient and halo size; \citealt{trotta11}). Comparison with observations requires codes with spectral energy resolution \citep{armillotta21,krumholz22,girichidis22,hopkins22-spectral}.  Progress in CR transport could also come from more local, high resolution observations \citep{thomas20}. Given the recent pace of activity in this field, there are ample grounds for optimism.

\section{RESOLVING SMALL SCALES IN LARGE-SCALE SIMULATIONS}
\label{sec:cosmo_resolving}

This review has emphasized the multi-scale nature of the multi-phase CGM. The large dynamic range on which important physical processes operate means that achieving numerical convergence in cold gas properties -- which generally have smaller characteristic length and mass scales than hot gas -- can be extremely challenging in large-scale simulations. In this section, we compile some key length scales, discuss what is currently achievable in cosmological simulations, and indicate potential paths toward progress. When discussing convergence, it is important to be specific about the metric of interest. For instance, convergence in cold gas morphology is much more demanding than convergence in cold gas mass, as we explain below.
We expect that numerical convergence in column densities, which is relevant for observational predictions, is intermediate between convergence in cold gas mass and convergence in cold gas morphology, given that it requires convergence in the PDF of the 2D projected density field, rather than the full 3D density field of cold gas. In particular, for high covering fractions (where a line of sight passes through multiple clouds), small scale structure should be washed out in projection. However, exact convergence criteria have yet to be quantified (see further discussion in \S\ref{sec:cosmo_res}). 

\subsection{Some Characteristic Scales}
\label{sec:char_scales}
We begin by summarizing some key length scales relevant to multi-phase gas in the CGM, many of which have been discussed earlier in this article (see also \citealt{li20}). 
Since many of these are collisional length scales, which scale as $\lambda \propto n^{-1}$ (for instance, cooling lengths $l \sim v t_{\rm cool}$ are proportional to $t_{\rm cool} \propto n^{-1}$), it is useful to define them in terms of column density, which is a density invariant quantity. 
Besides quoting spatial resolution, it is useful for simulators to quote resolved cold phase column densities, as this quickly gives an idea of which processes can be resolved. 
We note that while the collisional length scales are density invariant, the minimum resolvable columns in simulations in general do depend on density (see \S \ref{sec:cosmo_res}), so this must be taken into account in estimating the range of columns resolved in different regions of a simulation.

Firstly, there are variants of the cooling length $l \sim v t_{\rm cool}$, (where $v$ is a characteristic velocity and $t_{\rm cool}$ is a characteristic cooling time), which appear to be necessary to have converged cold gas mass (equation \ref{eq:Ncrit}) or converged cold gas morphology (equation \ref{eq:Nsonic}). We caution that this is an area of active research, and researchers still debate {\it which} velocity and {\it which} cooling time is most appropriate (e.g., see \S\ref{sec:hydro-cloud-wind}), but some version of these length scales is likely to remain relevant when the dust settles. In our opinion, of all the length scales listed below, for converged cold gas masses as well as rates of mass, momentum and energy transfer between phases, the cold gas survival length scale is the most important one to target. 

\begin{itemize}
\item{{\it Cold gas survival.} The survival length is the minimal size for a cloud of overdensity $\chi$ to survive when there is velocity shear relative to the hot gas with Mach number $\mathcal{M}= v/c_{\rm s,h}$. Adopting the criterion $t_{\rm cool,mix} < t_{\rm cc}$ as in equation (\ref{eq:rcrit}), so that $l \sim v_{\rm shear} t_{\rm cool,mix}/\sqrt{\chi}$, we obtain: 
\begin{equation}
N_{\rm H}^{\rm crit} \sim 10^{18} \, {\rm cm^{-2}} \, \frac{\mathcal{M} \chi_{100} T_{\rm cl,4}^{3/2}}{\Lambda_{\rm mix,21.4}} \sim 10^{18} \, {\rm cm^{-2}} \,\mathcal{M} T_{h,6}^{5/4},
\label{eq:Ncrit} 
\end{equation}
where in the second step we assume isobaric conditions $\chi = \rho_c/\rho_h = T_h/T_c$, $\Lambda(T_{\rm mix}) \propto T_{\rm mix}^{-0.5}$ (appropriate for $10^5 < T_{\rm mix} < 10^{7.5}$ K), and $T_{\rm cl}=10^4$ K. Clouds with $N_{\rm H} > N_{\rm H}^{\rm crit}$ can survive and grow in a wind. This criterion is still somewhat uncertain for $\mathcal{M}, \chi_{100} \gg 1$ conditions ($\chi_{100}=\chi/100)$.} 

\item{{\it Sonic length.} The sonic length (sometimes referred to as the `shattering' length, see \S\ref{sec:frag-coag}) is the minimum value of the cooling length $c_s t_{\rm cool}$, at which the sound crossing time $t_{\rm sc} \sim l/c_s$ and the cooling time $t_{\rm cool}$ are equal, so $l \sim c_s t_{\rm cool}$ For solar metallicity, collisionally ionized cooling curves, it arises at $T\sim 10^4$ K, giving 
\begin{equation} 
N_{\rm H}^{\rm sonic} \sim 10^{17}~{\rm cm^{-2}},
\label{eq:Nsonic}
\end{equation}
although it can be larger if cooling is inefficient, e.g. at low metallicity or due to photoionization.
}
\end{itemize}

To put these length scales in context, it is useful to compare them against some others. For instance, there are length scales associated with the diffusive transport of momentum and energy, i.e. viscosity (mediated by ions) and thermal conduction (mediated by electrons). Viscosity tends to be less important, given long viscous times: the thermal velocity of ions is $(m_{\rm p}/m_{\rm e})^{1/2} \sim 40$ times smaller than the thermal velocity of electrons. Thus, simulations which implement Braginskii viscosity in cloud-wind interactions see little effect \citep{li20}. There are two length scales associated with thermal conduction: 
\begin{itemize} 
\item{{\it Electron mean free path.} The Coulomb mean free path 
\begin{equation} 
N_{\rm H}^{\rm mfp} \sim 10^{16}~{\rm cm^{-2}}~f_{\rm mfp} T_6^2
\label{eq:Nmfp} 
\end{equation} 
determines the nature of thermal conduction. For $N_{\rm H} > N_{\rm H}^{\rm mfp}$, classical diffusive (Spitzer) conduction is applicable. Here, $f_{\rm mfp}<1$ is a factor which set the reduction of the electron mean free path by  scattering mechanisms other than Coulomb interactions.}
\item{{\it Field length.} The Field length is the length scale at which classical diffusive conduction and radiative cooling balance. If the heat flux is given by $F = - \kappa(T) \nabla T$, then $\nabla \cdot F \sim \kappa(T) T/\lambda_F^2 \sim n^2 \Lambda(T)$ gives $\lambda_{\rm F} \sim (\kappa(T) T/n^2 \Lambda(T))^{1/2}$. Conduction (cooling) dominates on smaller (larger) length scales. Evaluated in the hot medium, 
\begin{equation}
N_{\rm H}^{\rm Field} \sim 10^{18} \, {\rm cm^{-2}} \, f_{\rm cond}  T_6^2,
\label{eq:Nfield} 
\end{equation}
where $f_{\rm cond}<1$ is a suppression factor (e.g., due to tangled B-fields, or additional scattering). Static cold clouds with $N_{\rm H} < N_{\rm H}^{\rm Field}$ embedded in a hot medium evaporate, while larger ones will condense if pressure exceeds a critical value \citep{mckee90}. Note that the Field length decreases rapidly with temperature. At small scale interfaces between different phases, it sets the local temperature scale height, even in the presence of turbulence \citep{tan21-lines}. Thus, to obtain the correct temperature PDF at phase interfaces (e.g., for computing collisional ionization), the Field length must be resolved. 
If not resolved, this can be implemented at the subgrid level.} 
\end{itemize}
As of this writing, given their computational expense, we do not believe the electron mean free path and the Field length are important length scales to be targeted for typical purposes, and if required, they can be incorporated via subgrid prescriptions. 

Finally, there are length scales associated with self-gravity and radiative transfer: 

\begin{itemize}
\item{{\it Jeans length.} The Jeans length $\lambda_{\rm J} \sim c_s t_{\rm ff}$ is the critical size for clouds to become self-gravitating. Evaluated at $T \sim 10^4$ K, it equivalent to a column density: 
\begin{equation}
N_{\rm H}^{\rm J} \sim 5\times10^{20} \, {\rm cm^{-2}} \, P_{2}^{1/2},
\label{eq:NJeans} 
\end{equation}
where $P_{2} = n T /(10^{2}~{\rm K~cm}^{-3}) = n_{-2} T_{4}$ ($n_{-2}=n/(10^{-2}~{\rm cm^{-3}}$)) is the pressure of the cold gas.  Clouds with $N_{\rm H} > N_{\rm H}^{\rm J}$ become self-gravitating. Most clouds in the CGM are pressure confined rather than self-gravitating.}
\item{{\it Self-shielding.} Most photoionization modeling in the CGM context assumes that cold gas is optically thin. However, clouds start to self-shield from hydrogen ionizing photons once their column densities start to exceed $N_{\rm HI} \sim \sigma_{\rm HI}^{-1} \sim 10^{17} \, {\rm cm^{-2}}$, where $\sigma_{\rm HI}$ is the photoionization cross section at the Lyman edge (13.6 eV). However, the total associated hydrogen column density is larger, $N_{\rm H} \sim N_{\rm HI,crit} x_{\rm HI,crit}^{-1}$, where $x_{\rm HI, crit} \ll 1$ is the hydrogen neutral fraction when self-shielding starts to kick in \citep{zheng02}. Thus, the column density associated with cooling to $T \sim 10-100 \, {\rm K} \ll 10^{4}$ K, via fine-structure and molecular lines \citep{robertson08,krumholz11}, is larger: 
\begin{equation}
N_{\rm H}^{\rm shield} \sim \, 10^{22} \, {\rm cm^{-2}} \, Z_{0.1}^{-1}, 
\label{eq:Nshield} 
\end{equation}
where $Z_{0.1} = Z/(0.1 Z_{\odot})$. 
Such large columns are typically only found in the ISM, though at high redshift they can be reached also in the inner CGM \citep[e.g.][]{Stern2021_DLAs}. 
Thus, most clouds in the CGM will stay at warm (atomic) temperatures $T \sim 10^4$ K.}
\end{itemize}

\subsection{Achievable Resolutions in Cosmological Simulations}
\label{sec:cosmo_res}
It is useful to compare the above physical scales for cold gas to the resolution achieved by different types of cosmological simulations. 
We caution upfront that the simple scalings below are only crude guides to the scales resolved by different simulations, because this depends on the details of the specific simulation and numerical method, including the order-of-accuracy of the solver in addition to the cell size or particle mass. 
As mentioned above, which cold gas length scales should be resolved depends on the question asked. 
For example, for many purposes it is likely not necessary to resolve the small sonic/shattering length (see discussion in \S\ref{sec:morphology} and \S\ref{sec:TML}). 

Most cosmological galaxy formation simulations use a Lagrangian (or quasi-Lagrangian) method, in which the mass of resolution is fixed and the spatial resolution adapts according to where the mass ends up. 
This approach has the advantage that it concentrates most of the resolution in galaxies, while making it possible to simulate large boxes in which most of the volume has low density. 
A simple estimate of the spatial resolution in the gas for a Lagrangian simulation is 
\begin{equation}
\label{eq:hgas_Lagrangian}
h_{\rm gas} \approx \left( \frac{m_{\rm gas}}{\rho} \right)^{1/3}\approx 160~{\rm pc}~m_{\rm gas, 3}^{1/3} n_{-2}^{-1/3} ,
\end{equation}
where $m_{\rm gas}$ is the mass of gas cells and $m_{\rm gas,3}=m_{\rm gas}/({\rm 10^{3}~M_{\odot}})$. 
In terms of column density, the resolution is
\begin{equation}
\label{eq:Nres_Lagrangian}
N_{\rm H}^{\rm res} \approx h_{\rm gas} n \approx 5 \times 10^{18}~{\rm cm^{-2}} \, m_{\rm gas,3}^{1/3} n_{-2}^{2/3}.
\end{equation}
This shows that lower columns have more stringent resolution requirements (lower $m_{\rm gas}$) and also that the minimum resolved column depends on density in Lagrangian codes. 
We express resolution in terms of a fiducial density $n=10^{-2}$ cm$^{-3}$ because this is representative of cold streams in simulations \citep[e.g.,][]{Mandelker20_streams_radiative}, but in reality cold gas has a range of densities. 
The scalings in equations (\ref{eq:hgas_Lagrangian}) and (\ref{eq:Nres_Lagrangian}) apply to moving-mesh codes \citep[e.g.,][]{Springel10}, mesh-free finite mass (MFM) or mesh-free finite volume (MFM) methods \citep[][]{Hopkins15_GIZMO}, and adaptive mesh refinement (AMR) codes using quasi-Lagrangian refinement \citep[e.g.,][]{Kravtsov97_ART, Bryan14_ENZO, Teyssier02_RAMSES}.  
SPH codes are also widely used for cosmological simulations but the spatial resolution for these is better approximated by the size of the smoothing kernel, which typically contains $N_{\rm nbg}\sim 32-100$ particles. 
Adopting a fiducial neighbor number $N_{\rm nbg}\approx 60$ \citep[e.g.,][]{Schaye15}, the spatial resolution of an SPH simulation can be estimated as $\sim N_{\rm ngb}^{1/3} \approx 4\times$ coarser than implied by equation (\ref{eq:hgas_Lagrangian}), with the cell mass replaced by the SPH particle mass. 
The minimum resolved column densities are correspondingly larger. 
The fiducial cell mass $m_{\rm gas}\approx1,000$ M$_{\odot}$ in the equations above are representative of today's state-of-the-art zoom-in simulations for Milky Way-mass galaxies \citep[e.g.,][]{Grand21}. 
However, full cosmological boxes with side lengths $\sim 100$ Mpc typically have a much larger cell mass $m_{\rm gas}\sim 10^{6}$ M$_{\odot}$ \citep[e.g.,][]{Dubois14, Springel18_TNG}, corresponding to a $\sim10\times$ coarser spatial resolution, i.e. $\sim 1.6$ kpc or $N_{\rm H} \sim 5\times 10^{19}$ cm$^{-2}$ in $n = 10^{-2}$ cm$^{-3}$ gas. 

Comparing the above resolution scalings with characteristic scales in \S \ref{sec:char_scales} suggests that state-of-the-art zoom-in simulations can resolve the Jeans scale of cold gas (eq. \ref{eq:NJeans}) and may possibly start to marginally resolve the cold gas survival scale (eq. \ref{eq:Ncrit}), but the sonic/shattering scale is generally not resolved. 
This implies that the highest resolution zoom-in simulations are approaching resolutions where the cold gas masses are converged, but that the cold gas morphology (e.g., the amount of small-scale structure) remains unconverged. 
This is consistent with the moving-mesh, zoom-in simulations of \cite{vandevoort19}, who find that the total HI mass in the CGM is converged across the resolution levels investigated, but that the amount of small-scale structure in HI maps increases with increasing CGM resolution. 
Consistent with the varying small-scale structure, these authors find that the covering factors of HI above different column density thresholds also depend on resolution. 
Large-volume cosmological simulations resolve the Jeans scale but neither the cold gas survival scale nor the sonic scale. 
Thus, cold gas masses and morphology may both be unconverged in such simulations, although the degree of convergence likely depends on subgrid models and whether they are ``recalibrated'' at different resolutions \citep[e.g.,][]{Schaye15}.

In the last few years, multiple groups have noted that a Lagrangian approach may not be ideal for CGM studies because the concentration of resolution in galaxies implies a relatively low spatial resolution in dilute gas outside galaxies. 
To address this, new refinement schemes have been devised to enforce either a minimum or an uniform spatial resolution in a specified region extending into the CGM or beyond. 
This approach has been implemented in AMR \citep[][]{hummels19, peeples19} and moving-mesh codes \citep[][]{vandevoort19}, and it could in principle also be used in mesh-free codes that allow splitting of resolution elements. 
In these simulations, the spatial resolution $\Delta x$ is fixed while the mass resolution depends on density as 
\begin{equation}
m_{\rm res} = (\Delta x)^{3} \rho \approx 2.5\times10^{5}~{\rm M_{\odot}}~(\Delta x)_{1}^{3} n_{-2}
\end{equation}
and the minimum resolved column density is
\begin{equation}
\label{eq:Nres_Eulerian}
N_{\rm H}^{\rm res} \approx (\Delta x) n \approx 3\times 10^{19}~{\rm cm}^{-2}~(\Delta x)_{1} n_{-2},
\end{equation}
where $(\Delta x)_{1}=\Delta x/(1~{\rm kpc})$. 
In published simulations with uniform or enhanced CGM refinement, the maximum cell size is $(\Delta x)_{1} \approx 0.5-1$ (quoted here at $z=0$, since some simulations prescribed the maximum cell size in proper units and others in comoving units).
Therefore, for the fiducial cold gas density $n=10^{-2}$ cm$^{-3}$, the minimum resolved column is comparable to (or larger than) what is achievable with Lagrangian zoom-ins (eq. \ref{eq:Nres_Lagrangian}). 
These studies focused on CGM resolution have been influential in demonstrating clearly that finer structures are resolved as the spatial resolution is increased, especially in the cold gas (see an example of a resolution study in Fig. \ref{fig:hummels}). 
However, in agreement with the analytic estimates here, the cold gas morphology is not converged even at the highest spatial resolutions simulated to date. 
An important limitation of simulations that limit the size of resolution elements to a prescribed maximum is that while this produces very fine mass resolution in low-density gas, this tends to be predominantly in hot gas rather than in the cold, dense gas.  
Unfortunately, it is not clear how much higher resolution in hot gas helps, relative to Lagrangian refinement schemes that achieve comparable resolution in cold gas but coarser resolution in hot gas, since the hot gas tends to be much smoother. 
A promising variant may be to adopt refinement criteria that are tailored to specific processes that one wishes to resolve, such as refining on the local cooling length \citep[e.g.,][]{mandelker21_whim}. 
The insights on relevant physical scales from small-scale studies will play an important role in designing such refinement criteria.

\subsection{Incorporating Small Scales via Subgrid Recipes} 

Another approach is to distill the insights from small-scale simulations as subgrid models. There has been some progress on this front for cloud-wind interactions. For instance, PhEW (Physically Evolved Winds; \citealt{huang20}) calculates the exchange of mass and metals between an SPH cloud `particle' and its surroundings, incorporating isotropic thermal conduction, hydrodynamic instabilities, and the compressive and elongational effects of a bow shock, calibrated to the high resolution cloud crushing simulations of \citet{bruggen16}, albeit only in the cloud destruction regime. PhEW has been implemented in cosmological simulations \citep{huang22}. It improves numerical convergence, since small-scale interactions are handled in subgrid. 

While omitting some of the physics in PhEW, such as conduction and the effects of a bow shock, the semi-analytic multiphase wind model of \citet{fielding22} {\it does} include source terms for cloud entrainment and growth (as discussed in \S\ref{sec:hydro-cloud-wind}), which reproduce the results of idealized wind tunnel and shearing layer simulations in this regime\footnote{As noted in \S\ref{sec:wind-cloud}, cloud survival and growth in the high overdensity or Mach number regimes is still not completely clear, so their extrapolations in these regimes is necessarily uncertain.}. \citet{fielding22} include a wind thermalization term (see also \citealt{nguyen21}) which matches the rising entropy seen in high-resolution wind simulations \citep{schneider20}. Their model enables fast exploration of parameter space in semi-analytic models of galaxy formation, and there are plans to incorporate it into hydrodynamic simulations. 

A related approach is to use a two-fluid approximation, similar to that used in cosmic ray or radiation hydrodynamics simulations, where hot and cold gas exchange mass, energy and momentum via prescribed source terms \citep{weinberger22}. Of course, all the magic lies in these source terms, which must encode our growing understanding of multi-phase interactions. These prescriptions have yet to be formulated in this approach. 

These developments illustrate how small and large scale simulations can have a fruitful synergy, similar to treatments of the ISM and star formation in zoom-in or cosmological scale simulations. While this approach holds considerable promise, these are early days. More work is needed, both in understanding the physics of small-scale interactions and encoding these insights into subgrid recipes. Perhaps the mark of a mature subgrid recipe would be the ability to match high-resolution meso-scale simulations in disparate environments, such as multi-phase galactic winds \citep{schneider20} or `chaotic cold accretion' onto black holes \citep{gaspari13,gaspari18}. Such a model could then be used with confidence in large-scale simulations. 

\section{OUTLOOK AND FUTURE DIRECTIONS}
\label{sec:outlook}

While galactic halo gas has been studied for decades, 
the advent of the HST Cosmic Origins Spectrograph in the late 2000s transformed the field. Many researchers (present authors included) voted with their feet and stampeded to work on the CGM. \citet{tumlinson17} reviewed the remarkable ensuing observational progress in this journal. Their Figure 1, featured in innumerable conference talks, is a wildly influential cartoon of gas flows in the CGM, showing filamentary accretion from the IGM, bipolar winds, and gas recycling. In their memorable formulation, ``[the CGM] is potentially the gas fuel tank, waste dump, and recyling center all at the same time."  Gas flows indeed lie at the heart of the CGM's influence on galaxy formation. The review concluded that 
the `missing' baryons/metal problem, which had plagued researchers for years, could largely be laid to rest. 

This review partly serves as a theoretical counterpart, summarizing theoretical progress on understanding gas flows -- both on halo scales, as well as small-scale mass transfer between phases. While the \citet{tumlinson17} cartoon picture is still informative, the correct physical picture for the CGM likely evolves with both redshift and halo mass, with additional processes contributing. For instance, besides accretion via cold streams, we have reviewed the potential impact of hot mode accretion, thermal instability and precipitation, cooling-induced condensation onto cold gas `seeds', and the contribution of `intergalactic gas transfer' between galaxies. 
Our new cartoon picture in Figure \ref{fig:multiscale_cartoon} illustrates some of this complexity, and in particular emphasizes the multiscale structure expected in cold gas.

What are potential agenda items for continued theoretical progress? They include: 
\begin{itemize}
\item{{\it Clarity on cold vs. hot mode accretion.} Despite more than two decades of work, the role of cold vs. hot mode accretion is not yet fully settled. A key issue we have reviewed is progress on the survival of cold gas streams and clouds, processes which are not well resolved in existing cosmological simulations, using idealized simulations. 
Another issue is the effects of different accretion modes on the properties and evolution of central galaxies. 
While there has been much attention devoted to the role of cold streams in feeding high-redshift disk galaxies (which are highly turbulent and may be considered ``thick''), recent work suggests that hot-mode accretion may be important for the formation of thin disks (see \S \ref{sec:ICV_implications}). 
Future work should aim to clarify how the mode of gas accretion affects galaxy formation, and how this depends on mass and redshift. 
One specific area where a more systematic analysis of simulations would be beneficial would be the mechanisms that exchange angular momentum in the CGM, and how these exchanges affect the delivery of angular momentum to galaxies (\S \ref{sec:accretion_angular momentum}).} 

\item{{\it Understanding the structure of cold gas.} Most work on small-scale structure in cold gas is based on high-resolution simulations with idealized initial and boundary conditions. What are the properties of the cold gas in realistic galaxy halos? Are halos filled with a quasi-uniform cold `mist', or composed of discrete cloud complexes, as seen in HVCs?  If there is structure on all scales, what is the physics that determines the cloud mass function, including its low and high mass cut-offs? And what mass range is most germane to particular observational probes?}

\item{{\it Connecting large and small scales.} How do we build on recent theoretical progress on large and small scales to develop models of the CGM that realistically incorporate the cosmological environment, and still have enough physics/resolution to model fine-scale structure in the cold gas? Converged cold gas masses and column densities at CGM scales is one goal. Another would be models that can be meaningfully compared with high-resolution observations of kinematic profiles, which are sensitive to detailed structure along the line of sight.
Arguably, neither existing cosmological nor idealized simulations simultaneously have all the physics and resolution to do this well at present.} 

\item{{\it Progress on non-thermal physics.} B-field strengths affect MHD forces, turbulence, and CR streaming speeds. The MHD equations are in principle straightforward to solve. However, in practice, B-fields on CGM scales vary greatly between different simulation groups, likely because magnetic amplification and advection depend on the feedback model.  
Subgrid models for unresolved tangling might be needed, given its effects on CR transport and thermal conduction.
Unlike MHD, it is as yet unclear whether we are even solving the correct equations for CR transport. Progress will likely come with continued confrontation of observations with physical models.}

\end{itemize} 

What observational inputs would most benefit theorists? A partial inventory includes: 
\begin{itemize}
\item{{\it Better characterization of the hot CGM.} In most galaxies at $L_*$ and below, X-ray emission from coronal gas is too faint to be observed, and we are confined to observing cold gas ($T \sim 10^{4}$ K). Even if pressure balance holds between phases, the unknown contribution of non-thermal pressure support and large uncertainties in cold gas properties inferred from ionization model modeling \citep[e.g.,][]{stern16,chen17} imply considerable uncertainties in hot gas pressures. Direct constraints on coronal gas properties -- which affect mass, momentum and energy exchange between phases -- and their radial profiles are sorely needed. FRB dispersion measure measurements (which probe the total columns of free electrons), as well as a combination of thermal and kinetic Sunyaev-Zeldovich measurements (which together can reveal density and temperature profiles; e.g., \citealt{Schaan2021}), could potentially shed light on this.} 

\item{{\it Better characterization of non-thermal components.} We have reviewed the significant impact that B-fields, cosmic rays, and turbulence play on physical processes in the CGM. At the same time, depending on underlying assumptions, simulations can give widely disparate predictions for non-thermal energy content, from energetically insignificant to magnetically or CR dominated halos (e.g., \citealt{hopkins20,ji20}). 
FRBs could shed more light on B-fields and turbulence, as well as small-scale clumping \citep{prochaska19,chawla22}. The CR component may be the most difficult -- while in principle it is constrained by gamma-ray or radio observations, in practice these are only detectable near the host galaxy, rather than far out in the CGM.

\item{{\it Spatially resolved observations.} Most information about the CGM comes from quasar absorption line data which suffers from poor spatial coverage (typically just one pencil beam per halo); what is needed is data that combines high spatial sampling and high spectral resolution \citep{peeples19-white-paper}.  
Gravitational lensing has been used fruitfully to enhance the density of background quasars \citep[e.g.,][]{chen14,bowen16, Lopez2018}, and future studies using 30m-class telescopes will take spectra of multiple fainter background galaxies per halo \citep[][]{steidel10}}. 
Even better, emission lines could provide detailed spatial and kinematic maps. At high redshift, halos are already seen glowing in Ly$\alpha$ emission (\S \ref{sec:lya}) but the complexity of Ly$\alpha$ radiative transfer precludes simple interpretation. Optically thin metal lines, although faint, are potentially much more informative \citep{Sravan16,corlies20,piacitelli22}. Directly imaging the CGM was highlighted as a key discovery area in the Astro 2020 decadal survey \citep{astro2020}. 
Although CGM imaging has recently been enabled by new integral field spectrographs on 8-10m telescopes, imaging diffuse gas is presently limited either to the brightest, inner regions of the CGM, or to special cases where the gas is illuminated by a quasar. 
Order-of-magnitude improved sensitivities will be needed to map the CGM of normal galaxies on larger scales.}

\end{itemize}

A potential preview of the future of CGM studies comes from their more massive sibling, the ICM. All the same physical ingredients are present: radiative cooling, turbulence, stratification, multi-phase gas, B-fields and cosmic-rays. However, we see the hot medium directly, in spatially resolved X-ray observations. We also see cold filaments directly in resolved H$\alpha$ and CO emission. Despite short central cooling times, cooling flow models \citep{fabian94} were shown by Chandra and XMM observations to be untenable in clusters \citep{peterson06}. Instead, the ICM is understood to be in approximate hydrostatic and thermal equilibrium -- not too dissimilar from a star, with feedback from the central AGN playing the role of nuclear reactions \citep{mcnamara07}. We see the feedback directly, in the form of spatially resolved AGN blown bubbles. Non-thermal components are much better characterized: $\sim \mu$G magnetic fields (leading to $\beta \sim 100$ plasmas) are constrained by Faraday rotation; cosmic rays ($P_{\rm c}/P_{\rm g} \lsim {\rm few} \%$) by lack of gamma-ray emission\footnote{This was considerably smaller than expected \citep{pinzke10}, given the expected high efficiency of CR acceleration at strong accretion shocks, and the fact that the ICM is largely a closed box -- a sobering reminder of surprises and uncertainties in CR physics.}; and kinematics of highly subsonic $\mathcal{M} \sim 0.1-0.2$ turbulence by emission line broadening \citep{hitomi16}. Bulk sloshing motions are constrained by observed cold fronts \citep{markevitch07}. Which is not to say that matters are fully resolved. For instance, theorists still debate how AGN energy is transmitted and isotropized in the ICM; conduction and viscosity, and the role of kinetic instabilities in modulating them, are highly uncertain; the origin of cold filaments, as well as their kinematic and energetic coupling with the hot phase, is still debated \citep{DV22_review}. But the field is certainly in a significantly more mature state than CGM studies, due to the transformative power of spatially resolved observations. 

One possible lesson for CGM theorists is, wherever possible, to use systems with a wealth of high resolution data as laboratories for models of key physical processes, before sallying forth into terra incognita. 
If, after multiple decades there is no direct evidence for thermal conduction in the ISM or ICM, it is hard to imagine we could detect its effects in the CGM anytime soon.
A powerful laboratory for physical models of CR transport is our Galaxy \citep{kempski22,hopkins21-CR-problems}. 
Models of small scale structure in photoionized gas should confront $\sim 1-10^{4}$ AU scale atomic structures and extreme scattering events in the ISM \citep{stanimirovic18}, which may also constrain MHD turbulence on CR gyro-scales. Models of turbulent mixing layers and cold gas condensation onto cometary tails must confront resolved observations of HVCs and cluster filaments, jellyfish galaxies \citep{muller21}, and AGB winds such as Mira A \citep{martin07,li19}.  

It's also good to remember that progress is often fueled by cross-fertilization from other fields. Precision cosmology was greatly aided by the adoption of Markov Chain Monte Carlo, and machine learning techniques are playing a similar role today \citep[][]{Dvorkin2022_ML}. Similarly, the scale problem that CGM theorists confront is not too dissimilar from the problem of thermo-nuclear burning on white dwarfs \citep{schmidt06}, which is even more difficult to treat due to the high temperature sensitivity of nuclear burning rates. It is also similar to what engineers working on turbulent mixing face \citep{sreenivasan19}. Climate science simulations tackle another complex multi-scale problem: indeed, clouds are the largest source of uncertainty and usually have to be handled in a subgrid fashion \citep{huang22-clouds}. Even if the stakes for humanity are not quite as high, our fledging efforts to model galactic atmospheres can draw hope and inspiration from the progress in climate science simulations over the past few decades. 

\section*{ACKNOWLEDGEMENTS}
We are grateful to our many collaborators, who have contributed greatly to our understanding of the CGM. 
We thank Chad Bustard, Drummond Fielding, Max Gronke, Eliot Quataert, and Jonathan Stern for comments on a draft of this review article.

\bibliography{references_CA,master_references}

\section*{ACRONYMNS AND DEFINITIONS}
\noindent {\bf Virialized gas:} Gas with thermal velocities set by the gravitational potential. In massive halos, this corresponds to a hot phase ($T_{\rm vir} \gtrsim 10^{6}$ K).\\

\noindent {\bf Cold flows/cold streams:} Cold halo gas, typically at a temperature $T \sim 10^{4}$ K and filamentary in morphology. Associated with “cold mode” accretion.\\ 

\noindent {\bf Cooling flow:} Model for hot gas accretion in halos driven by radiative cooling and neglecting feedback. Associated with “hot mode” accretion.\\

\noindent {\bf Precipitation:} Formation of cold gas via linear thermal instability in a gravitationally stratified hot medium. The overdense cold gas eventually falls, like rain.\\

\noindent {\bf Condensation:} Conversion of hot to cold gas when initially hot gas mixes with cold gas, and the mixed gas cools. The hot gas `condenses' onto cold gas seeds.\\

\noindent {\bf Turbulent mixing layer:} A turbulent interface between hot and cold phases where the interaction between turbulence and radiative cooling governs the exchange of mass, momentum and energy between phases.\\

\end{document}